\documentclass[%
reprint,
superscriptaddress,
bibnotes,
amsmath,amssymb,
aps,
longbibliography
]{revtex4-1}

\usepackage{qcircuit}
\usepackage[dvipdfmx]{graphicx}
\usepackage{graphicx}
\usepackage{amsmath,amssymb,amsthm,mathrsfs,amsfonts,dsfont}
\usepackage{subfigure, epsfig}
\usepackage{braket}
\usepackage{bm}
\usepackage{enumerate}
\usepackage[colorlinks,linkcolor=blue,citecolor=blue]{hyperref}
\usepackage{multirow}

\newcommand{\black}[1]{\textcolor{black}{#1}}

\newcommand{\mathL}{\mathcal{L}}
\newcommand{\mathH}{\mathcal{H}}

\newcommand{\mathW}{\mathcal{W}}
\newcommand{\mathR}{\mathcal{R}}
\newcommand{\mathRL}{\mathcal{R}_{\mathcal{L}}}
\newcommand{\mathM}{\mathcal{M}}

\newcommand{\prepare}{{\tt PREPARE}}
\newcommand{\unitprep}{{\tt UNIT}-{\tt PREP}}
\newcommand{\select}{{\tt SELECT}}

\newcommand{\order}{O}

\newcommand{\balpha}{\boldsymbol{\alpha}}
\newcommand{\bmu}{\boldsymbol{\mu}}
\newcommand{\bp}{\boldsymbol{p}}

\usepackage{hyperref}
\hypersetup{colorlinks=true,linkcolor=blue,citecolor=blue,urlcolor=blue}

\graphicspath{{./fig/}}  
\begin{document}
\title{Hunting for quantum-classical crossover in condensed matter problems}

\author{Nobuyuki Yoshioka}
\email{nyoshioka@ap.t.u-tokyo.ac.jp}
\affiliation{Department of Applied Physics, University of Tokyo, 7-3-1 Hongo, Bunkyo-ku, Tokyo 113-8656, Japan}
\affiliation{Theoretical Quantum Physics Laboratory, RIKEN Cluster for Pioneering Research (CPR), Wako-shi, Saitama 351-0198, Japan}
\affiliation{JST, PRESTO, 4-1-8 Honcho, Kawaguchi, Saitama, 332-0012, Japan}

\author{Tsuyoshi Okubo}
\email{t-okubo@phys.s.u-tokyo.ac.jp}
\affiliation{Institute for Physics of Intelligence, University of Tokyo, 7-3-1 Hongo, Bunkyo-ku, Tokyo 113-0033, Japan}
\affiliation{JST, PRESTO, 4-1-8 Honcho, Kawaguchi, Saitama, 332-0012, Japan}

\author{Yasunari Suzuki}
\email{yasunari.suzuki.gz@hco.ntt.co.jp}
\affiliation{NTT Computer and Data Science Laboratories, Musashino 180-8585, Japan}
\affiliation{JST, PRESTO, 4-1-8 Honcho, Kawaguchi, Saitama, 332-0012, Japan}

\author{Yuki Koizumi}
\affiliation{NTT Computer and Data Science Laboratories, Musashino 180-8585, Japan}

\author{Wataru Mizukami}
\email{mizukami.wataru.qiqb@osaka-u.ac.jp}
\affiliation{Center for Quantum Information and Quantum Biology, Osaka University, 1-2 Machikaneyama, Toyonaka, Osaka, 560-0043, Japan}
\affiliation{Graduate School of Engineering Science, Osaka University, 1-3 Machikaneyama, Toyonaka, Osaka 560-8531, Japan.}
\affiliation{JST, PRESTO, 4-1-8 Honcho, Kawaguchi, Saitama, 332-0012, Japan}

\begin{abstract}
The intensive pursuit for quantum advantage in terms of computational complexity has further led to a modernized crucial question: {\it When and how will quantum computers outperform classical computers?} 
The next milestone is undoubtedly the realization of quantum acceleration in practical problems. 
Here we provide a clear evidence and arguments that the primary target is likely to be condensed matter physics. Our primary contributions are summarized as follows:  1) Proposal of systematic error/runtime analysis on state-of-the-art classical algorithm based on tensor networks; 2) Dedicated and high-resolution analysis on quantum resource performed at the level of executable logical instructions; 3) Clarification of quantum-classical crosspoint for ground-state simulation to be within runtime of hours using only a few hundreds of thousand physical qubits for 2d Heisenberg and 2d Fermi-Hubbard models, \black{assuming that logical qubits are encoded via the surface code with the physical error rate of $p=10^{-3}$.}
To our knowledge, we argue that condensed matter problems offer the earliest platform for demonstration of practical quantum advantage that is order-of-magnitude more feasible than ever known candidates, in terms of both qubit counts and total runtime.
\end{abstract}

\maketitle

\let\oldaddcontentsline\addcontentsline
\renewcommand{\addcontentsline}[3]{}


\section*{Introduction}
When and how will quantum computers outperform classical computers? 
This pressing question drove the community to perform random sampling in quantum devices that are fully susceptible to noise~\cite{arute2019quantum, zhong_quantum_2020, zhong_phase_2021}. 
We anticipate that the precedent milestone after this quantum transcendence is to realize quantum acceleration for practical problems.
In this context, an outstanding question remains: {\it in which problem next?} This encompasses research across a range of fields, including natural science, computer science, and, notably, quantum technology.

Research on quantum acceleration is predominantly focused on two areas: cryptanalysis and quantum chemistry.
In the realm of cryptanalysis, there has been a substantial progress since Shor introduced a polynomial time quantum algorithm for integer factorization and finding discrete logarithms~\cite{shor1999polynomial,fowler2012surface, gheorghiu2019benchmarking, Gidney2021howtofactorbit}. 
Gidney et al. have estimated that a fully fault-tolerant quantum computer with 20 million ($2\times10^7$) qubits could decipher a 2048-bit RSA cipher in eight hours, \black{and a 3096-bit cipher in approximately a day}~\cite{Gidney2021howtofactorbit}. 
\black{This represents an almost hundred-fold enhancement in the the spacetime volume of the algorithm compared to similar efforts, which generally require several days~\cite{fowler2012surface, gheorghiu2019benchmarking}.}
Given that the security of nearly all asymmetric cryptosystems is predicated on the classical intractability of integer factoring or discrete logarithm findings~\cite{rivest1978method, kerry2013digital}, the successful implementation of Shor's algorithm is imperative to safeguard the integrity of modern and forthcoming  communication networks.

\begin{figure}[t]
    \begin{center}
    \includegraphics[width=1.05\linewidth]{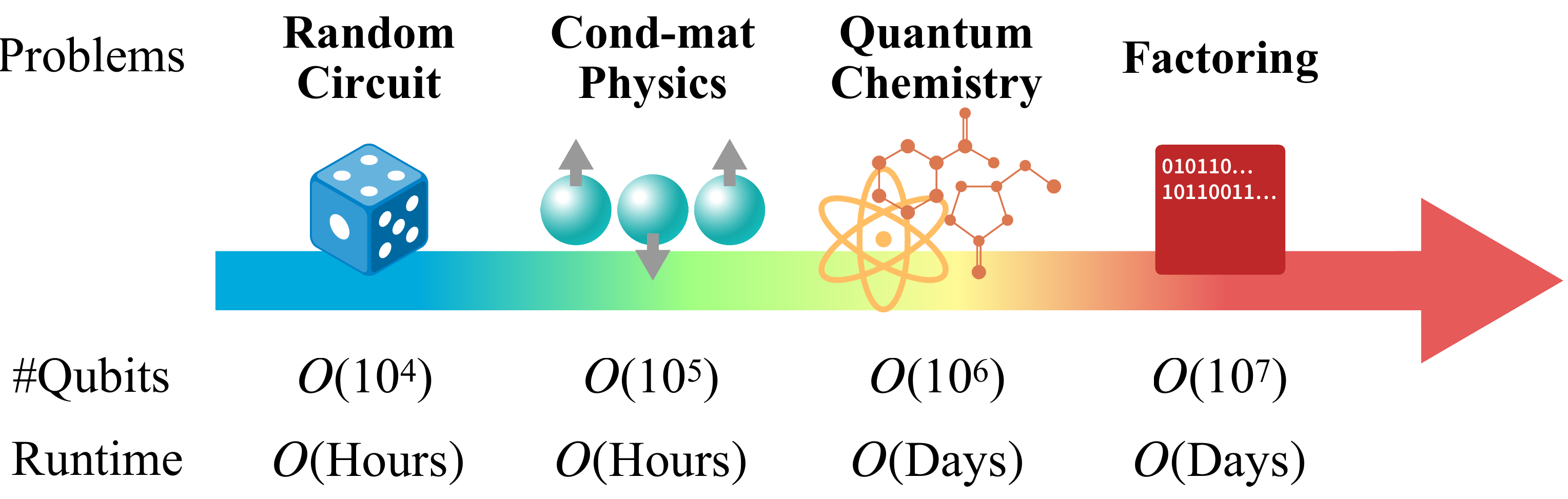}
    \caption{Schematic diagram for scaling of computational resource required to achieve quantum advantage using fault-tolerant quantum computers.
    }\label{fig:fig1}
    \end{center}
\end{figure}

The potential impact of accelerating quantum chemistry calculations, including first-principles calculations, is immensely significant as well.
Given its broad applications in materials science and life sciences, it is noted that computational chemistry, though not exclusively quantum chemistry, accounts for  40\% of HPC resources in the world~\cite{sherrill_jcp_2020}.
Among numerous benchmarks, a notable target with significant impact is quantum advantage in simulation of energies of a molecule called FeMoco, found in the reaction center of a nitrogen-fixing enzyme~\cite{Reiher201619152}. 
\black{According to the resource estimation that employs the state-of-the-art quantum algorithm,} calculation of the ground-state energy of FeMoco requires about four days on a fault-tolerant quantum computer equipped with four million $(4\times10^6)$ physical qubits~\cite{lee_evenmore_2021}.
Additionally, Goings et al. conducted a comparison between quantum computers and the contemporary leading heuristic classical algorithm for cytochrome P450 enzymes, suggesting that the quantum advantage is realized only in computations extending beyond four days~\cite{goings_pnas_2022}.

A practical quantum advantage in both domains has been proposed to be achievable within a timescale of days with millions of physical qubits. Such a spacetime volume of algorithm may not represent the most promising initial application of fault-tolerant quantum computers.
This paper endeavors to highlight condensed matter physics as a novel candidate~(See Fig.~\ref{fig:fig1}). 
We emphasize that while models in condensed matter physics encapsulate various fundamental  quantum many-body phenomena, their structure is simpler than that of quantum chemistry Hamiltonians. 
Lattice quantum spin models and lattice fermionic models serve as nurturing grounds for strong quantum correlations, facilitating phenomena such as quantum magnetism, quantum condensation, topological order, quantum criticality, and beyond. 
Given the diversity and richness of these models, coupled with the difficulties of simulating large-scale systems using classical algorithms, even with the most advanced techniques, it would be highly beneficial to reveal the location of the crosspoint between quantum and classical computing based on runtime analysis.

Our work contributes to the community's knowledge in  three primary ways: 1) Introducing a systematic analysis method to estimate runtime to simulating quantum states within target energy accuracy using the extrapolation techniques, 2) Conducting an end-to-end runtime analysis of quantum resources at the level of executable logical instructions, 3) Clearly identifying the quantum-classical crosspoint for ground-state simulation to be within the range of {\it hours} using physical qubits on the order of $10^5$.
To the best of our knowledge, this suggests the most imminent practical and feasible platform for the crossover.

We remark that there are some works that assess the quantum resource to perform quantum simulation on quantum spin systems~\cite{childs2018toward, beverland2022assessing}, while the estimation is done solely regarding the dynamics; they do not involve time to extract information on any physical observables. Also, there are existing works on phase estimation for Fermi-Hubbard models~\cite{Kivlichan2020improvedfault, campbell2021early} that do not provide estimation on the classical runtime. In this regard, there has been no clear investigation on the quantum-classical crossover prior to the current study that assesses end-to-end runtime. 

\section*{Results} \label{sec:algorithms}
Our argument on the quantum-classical crossover is based on the runtime analysis needed to compute the ground state energy within desired total energy accuracy, denoted as $\epsilon$. 
The primal objective in this section is to provide a framework that elucidates the quantum-classical crosspoint for systems whose spectral gap is constant or polynomially-shrinking.
\black{In this work, we choose two models that are widely known due to their profoundness despite the simplicity: the 2d $J_1$-$J_2$ Heisenberg model and 2d Fermi-Hubbard model on a square lattice (see the Method section for their definitions).} Meanwhile, it is totally unclear whether a feasible crosspoint exists at all when the gap closes exponentially. 

\black{
It is important to keep in mind that condensed matter
physics often entails extracting physical properties beyond
merely energy, such as magnetization, correlation
function, or dynamical responses.
Therefore, in order to assure that expectation value estimations can done consistently (i.e. satisfy $N$-representability), we demand that we have the option to measure the physical observable after computation of the ground state energy is done.
In other words, for instance the classical algorithm, we perform the variational optimization up to the desired target accuracy $\epsilon$; we exclude the case where one calculates less precise quantum states with energy errors $\epsilon_i \geq \epsilon$ and subsequently perform extrapolation. The similar requirement is imposed on the quantum algorithm as well.
}


\begin{figure}[t]
    \begin{center}
    \includegraphics[width=0.95\linewidth]{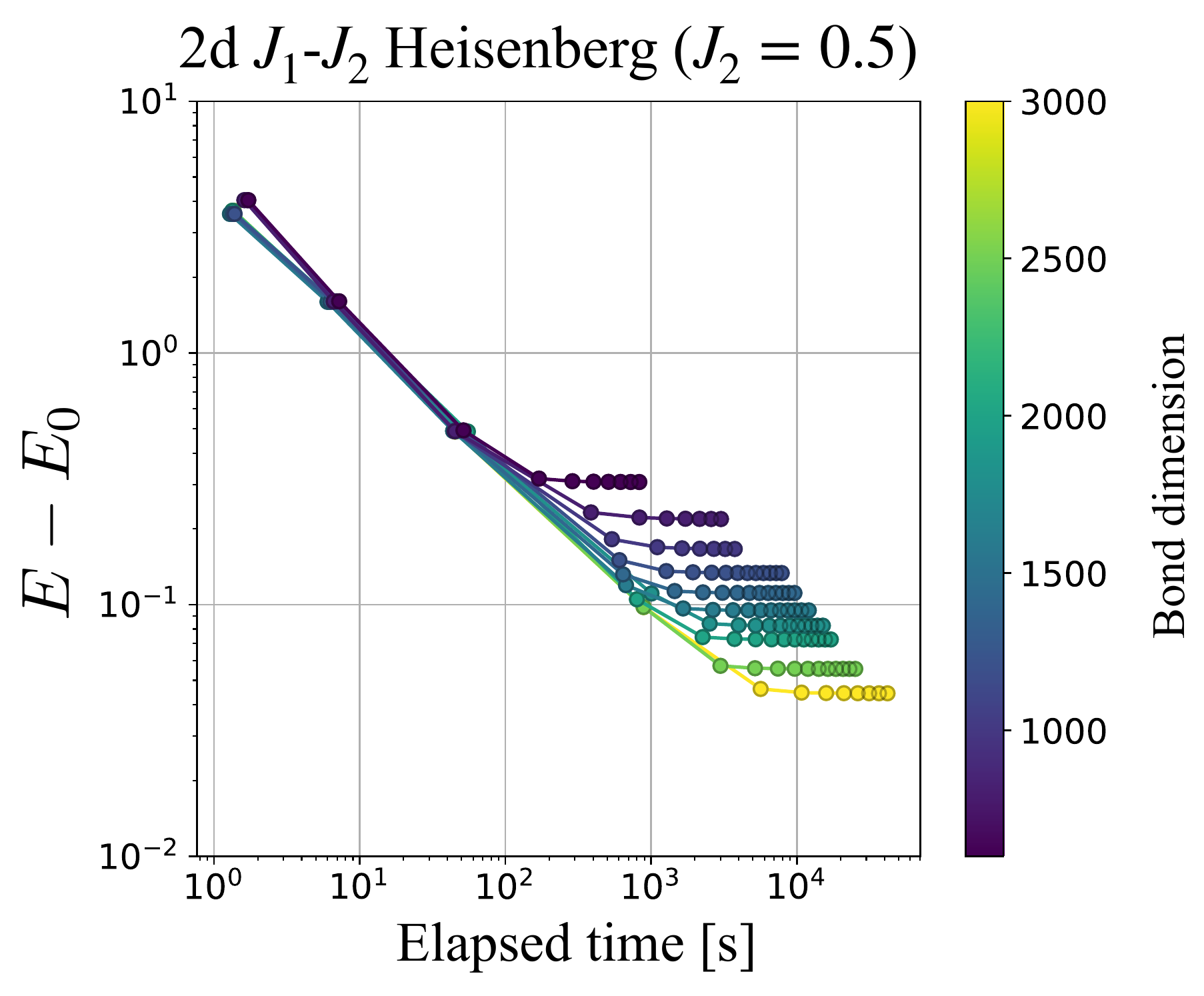}
    \caption{Elapsed time scaling of DMRG algorithm in $J_1$-$J_2$ Heisenberg model at $J_2=0.5$ with lattice size $10\times10$. Although the simulation itself does not reach $\epsilon=0.01$, learning curves for different bond dimensions ranging from $D=600$ to $D=3000$ collapse into a single curve, which implies the adequacy to estimate runtime according to the obtained scaling law. All DMRG simulations are executed using ITensor library~\cite{itensor}.
    }\label{fig:dmrg_scaling_J1J2}
    \end{center}
\end{figure}

\subsection*{Runtime of classical algorithm}
Among the numerous powerful classical methods available, we have opted to utilize the DMRG algorithm, which has been established as one of the most powerful and reliable numerical tools to study strongly-correlated quantum lattice models especially in one dimension~(1d)~\cite{White1992, WhiteH1993}.
In brief, the DMRG algorithm performs variational optimization on tensor-network-based ansatz named Matrix Product State (MPS)~\cite{OstlundR1995, DukelskyMNS1998}.
Although MPS is designed to efficiently capture 1d area-law entangled quantum states efficiently~\cite{EisertCP2010}, the efficacy of DMRG algorithm allows one to explore quantum many-body physics beyond 1d, including quasi-1d and 2d systems, and even all-to-all connected models, as considered in quantum chemistry~\cite{wouters2014density, baiardi2020density}.

A remarkable characteristic of the DMRG algorithm is its ability to perform systematic error analysis.
This is intrinsically connected to the construction of ansatz, or the MPS, which compresses the quantum state by performing site-by-site truncation of the full Hilbert space. 
The compression process explicitly yields a metric called ``truncation error," from which we can extrapolate the truncation-free energy, $E_0$, to estimate the ground truth.
By tracking the deviation from the zero-truncation result $E - E_0$, we find that the computation time and error typically obeys a scaling law~(See  Fig.~\ref{fig:dmrg_scaling_J1J2} for an example of such a scaling behavior in 2d $J_1$-$J_2$ Heisenberg model). The resource estimate is completed by combining the actual simulation results and the estimation from the scaling law. [See Sec.~\ref{sec:dmrg} in Supplementary Materials (SM) for detailed analysis.]

\black{We remark that it is judicious to select the DMRG algorithm for 2d models, even though the formal complexity of number of parameters in MPS is expected to increase exponentially with system size $N$, owing to its intrinsic 1d-oriented structure. }
Indeed, one may consider another tensor network states that are designed for 2d systems, such as the Projected Entangled Pair States (PEPS)~\cite{NishinoHOMAG2001, VerstraeteC2004}. \black{When one use the PEPS, the bond dimension is anticipated to scale as $D=O(\log(N))$ for gapped or gapless non-critical systems and $D=O({\rm poly}(N))$ for critical systems~\cite{verstraete2006criticality, haghshenas2018u1, rader2018finite} to represent the ground state with fixed total energy accuracy of $\epsilon=O(1)$~(it is important to note that the former would be $D=O(1)$ if considering a fixed energy {\it density}).
Therefore, in the asymptotic limit, the scaling on the number of parameters of the PEPS is exponentially better than that of the MPS.
Nonetheless, regarding the actual calculation, the overhead involved in simulating the ground state with PEPS is substantially high, to the extent that there are practically no scenarios where the runtime of the \black{variational PEPS} algorithm outperforms that of DMRG algorithm for our target models. }

\begin{figure}[t]
    \begin{center}
    \includegraphics[width=0.95\linewidth]{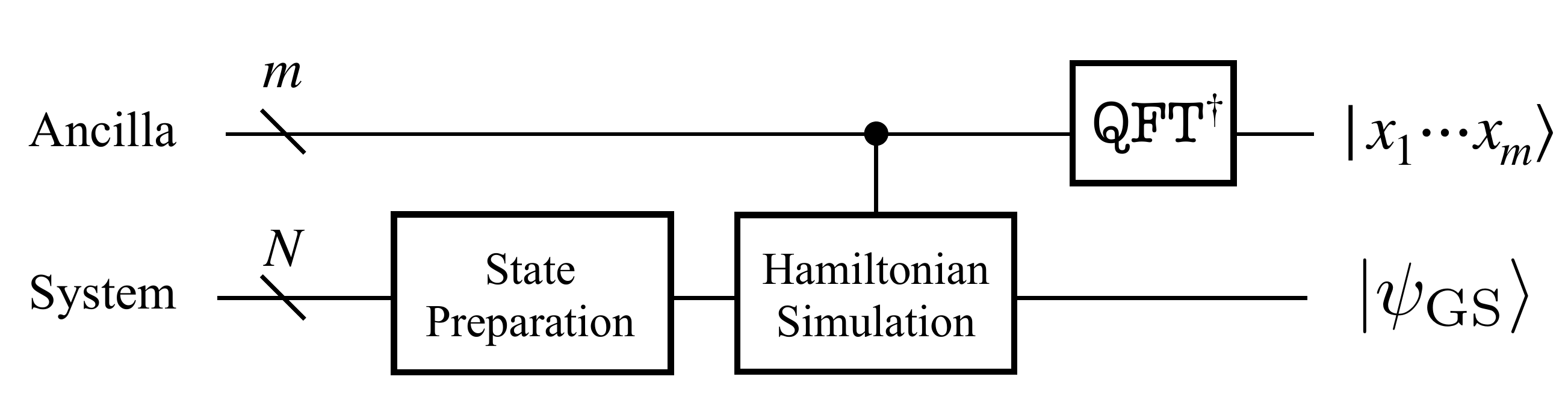}
    \caption{\black{Schematic description of quantum circuit for the QPE algorithm to compute the eigenenergy of the ground state of quantum many-body system.}
    }\label{fig:phase_estimation_circuit_main}
    \end{center}
\end{figure}

\subsection*{Runtime of quantum algorithm}
\subsubsection{Overview of quantum resource estimation}
Quantum phase estimation (QPE) is a quantum algorithm designed to extract the eigenphase $\phi$ of a given unitary $U$ by utilizing ancilla qubits to indirectly read out the complex phase of the target system.
More concretely, given a trial state $\ket{\psi}$ whose fidelity with the $k$-th eigenstate  $\ket{k}$ of the unitary is given as $f_k = \|\braket{k | \psi}\|^2$, a single run of QPE projects the state to $\ket{k}$ with probability $f_k$, and yields a random variable $\hat{\phi}$ which corresponds to a $m$-digit readout of $\phi_k$.

It was originally proposed by Ref.~\cite{Abrams99} that eigenenergies of a given Hamiltonian can  be computed  efficiently via QPE by taking advantage of quantum computers to perform Hamiltonian simulation, e.g., $U = \exp(-i H \tau)$.
\black{To elucidate this concept, it is beneficial to express the gate complexity for the QPE algorithm as schematically shown in Fig.~\ref{fig:phase_estimation_circuit_main} as
\begin{eqnarray}
    C \sim  C_{\rm SP} + C_{\rm HS} + C_{{\rm QFT}^{\dagger}},
\end{eqnarray}
where we have defined $C_{\rm SP}$ as the cost for  state preparation, $C_{\rm HS}$ for the controlled Hamiltonian simulation, and $C_{{\rm QFT}^{\dagger}}$ for the inverse quantum Fourier transformation, respectively (See Sec.~\ref{sec:phase_estimation} in SM).
The third term $C_{{\rm QFT}^\dagger}$ is expected to be the least problematic with $C_{{\rm QFT}^{\dagger}} = O(\log (N))$, while the second term is typically evaluated as $C_{\rm HS}=O({\rm poly} (N))$ when the Hamiltonian is, for instance, sparse, local, or constituted from polynomially many Pauli terms.
Conversely, the scaling of the third term $C_{\rm SP}$ is markedly nontrivial.
In fact, the ground state preparation of local Hamiltonian generally necessitates exponential cost, which is also related to the fact that the ground state energy calculation of local Hamiltonian is categorized within the complexity class of QMA-complete~\cite{kitaev2002classical, kempe2006complexity}. 
}

\begin{figure}[t]
    \begin{center}
    \includegraphics[width=0.9\linewidth]{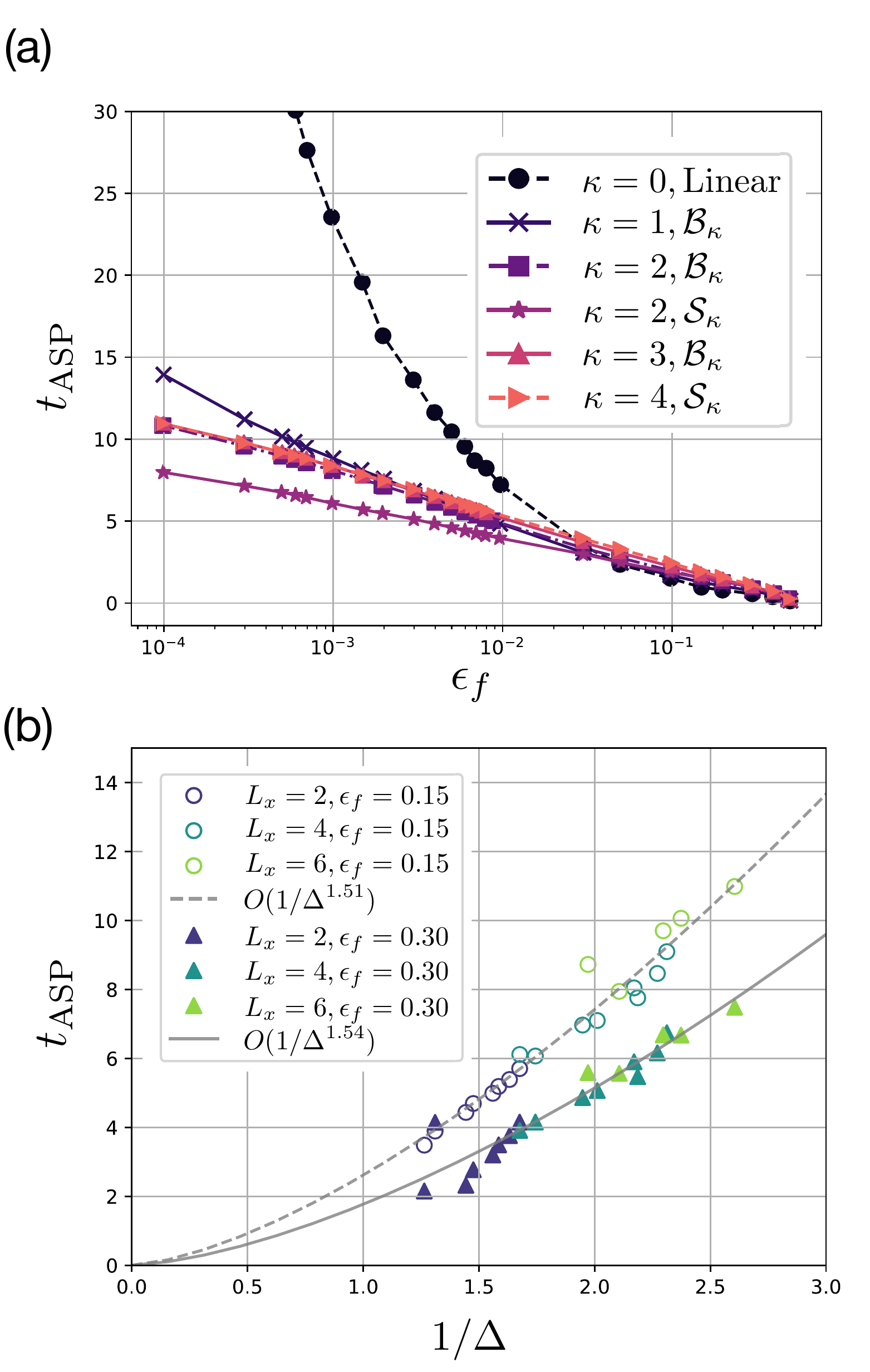}
    \caption{Scaling of the ASP time $t_{\rm ASP}$ for 2d $J_1$-$J_2$ Heisenberg model with $J_2$=0.5. (a) Scaling with the target \black{infidelity} $\epsilon_f$ for system size of $4\times 4$ lattice. The interpolation function is taken so that the derivative up to $\kappa$-th order is zero at $t=0, t_{\rm ASP}$. Here we consider the linear interpolation for $\kappa=0$, and for smoother ones we take $\mathcal{S}_\kappa$ and $\mathcal{B}_\kappa$ that are defined from sinusoidal and incomplete Beta functions, respectively (see Sec.~\ref{sec:state_preparation} in SM). 
    While smoothness for higher $\kappa$ ensures logarithmic scaling for smaller $\epsilon_f$, for the current target model, we find that  it suffices to take $s(t)$ whose derivative vanishes up to $\kappa=2$ at $t=0, t_{\rm ASP}$.
    (b) Scaling with the spectral gap $\Delta$. Here we perform the ASP using the MPS state for system size of $L_x \times L_y$, where results for $L_x=2,4,6$ is shown in cyan, blue, and green data points. We find that the scaling exhibits $t_{\rm ASP} \propto 1/\Delta^{\beta}$ with $\beta \sim 1.5$.
    }\label{fig:2dJ1J2_ASP}
    \end{center}
\end{figure}

\subsubsection{State preparation cost}
Although the aforementioned argument seems rather formidable, it is important to note that the QMA-completeness pertains to the worst-case scenario. Meanwhile, the average-case hardness in translationally invariant lattice Hamiltonians remains an open problem, and furthermore we have no means to predict the complexity under specific problem instances.
In this context, it is widely believed that a significant number of ground states that are of substantial interest in condensed matter problems can be readily prepared with a polynomial cost~\cite{deshpande2022importance}. \black{In this work, we take a further step to argue that} the state preparation cost can be considered negligible as $C_{\rm SP} \ll C_{\rm HS}$ for our specific target models, namely the gapless spin liquid state in the $J_1$-$J_2$ Heisenberg model or the antiferromagnetic state in the Fermi-Hubbard model.
\black{Our argument is based on numerical findings combined with upper bounds on the complexity, while we leave the theoretical derivation for scaling (e.g. Eq.~\eqref{eq:asp_bound_combined}) as an open problem.}

For concreteness, we focus on the scheme of the Adiabatic State Preparation (ASP) as a deterministic method to prepare the ground state through a time evolution of period $t_{\rm ASP}$. 
We introduce a time-dependent interpolating function  $s(t):\mathbb{R} \mapsto [0, 1]~(s(0)=0, s(t_{\rm ASP})=1)$ such that the ground state is prepared via time-dependent Schr\"{o}dinger equation given by
\begin{eqnarray}
    i\frac{\partial }{\partial t} \ket{\psi(t)} = H(t)\ket{\psi(t)},
\end{eqnarray}
where $H(t)=H(s(t)) = sH_f + (1-s)H_0$ for the target Hamiltonian $H_f$ and the initial Hamiltonian $H_0$. We assume that the ground state of $H_0$ can be prepared efficiently, and take it as the initial state of the ASP.
Early studies suggested a sufficient (but not necessary) condition for preparing the target ground state scales as $t_{\rm ASP}=O(1/\epsilon_f \Delta^3)$~\cite{kato1950adiabatic} where $\epsilon_f = 1 - |\braket{\psi_{\rm GS}| \psi(t_{\rm ASP})}|$ is the target infidelity and $\Delta$ is the spectral gap. This has been refined in recent research as
\begin{eqnarray}
    t_{\rm ASP} = 
    \begin{cases}
        O(\frac{1}{\epsilon_f^2 \Delta^2}|\log(\Delta)|^{\zeta})~(\zeta>1)~\text{\cite{elgart2012note}}    \\
        O(\frac{1}{\Delta^3} \log (1/\epsilon_f))~\text{\cite{ge2016rapid}}    \\
    \end{cases}.\label{eq:asp_bounds}
\end{eqnarray}
Two conditions independently achieve the optimality with respect to $\Delta$ and $\epsilon_f$.
Evidently, the ASP algorithm can prepare the ground state efficiently if the spectral gap is constant or polynomially small as $\Delta = O(1/N^{\alpha})$.

For both of our target models, numerous works suggest that $\alpha=1/2$~\cite{imada1998metal, wang_critical_2018, nomura_dirac_2021}, which is one of the most typical scalings in 2d gapless/critical systems such as the spontaneous symmetry broken phase with the Goldstone mode and critical phenomena described by 2d conformal field theory.
With the polynomial scaling of $\Delta$ to be granted, now we ask what the scaling of $C_{\rm SP}$ is, and how does it compare to other constituents, namely $C_{\rm HS}$ and $C_{\rm QFT}^\dagger$.

In order to estimate the actual cost, we have numerically calculated $t_{\rm ASP}$ required to achieve the target fidelity (See Sec.~\ref{sec:state_preparation} in SM for details) up to 48 qubits. 
With the aim of providing a quantitative way to estimate the scaling of $t_{\rm ASP}$ in larger sizes, we reasonably consider the combination of the upper bounds provided in Eq.~\eqref{eq:asp_bounds} as
\begin{eqnarray}
    t_{\rm ASP} = O\left(\frac{1}{\Delta^\beta}\log(1/\epsilon_f)\right).\label{eq:asp_bound_combined}
\end{eqnarray}
Figures~\ref{fig:2dJ1J2_ASP}(a) and (b) illustrate the scaling of $t_{\rm ASP}$ concerning $\epsilon_f$ and $\Delta$, respectively. Remarkably, we find that Eq.~\eqref{eq:asp_bound_combined} with $\beta=1.5$ gives an accurate prediction for 2d $J_1$-$J_2$ Heisenberg model.
This implies that the ASP time scaling is $t_{\rm ASP}=O(N^{\beta/2}\log(1/\epsilon_f))$, which yields gate complexity of $O(N^{1+\beta/2} {\rm polylog}(N/\epsilon_f))$ under optimal simulation for time-dependent Hamiltonians~\cite{low2016hamiltonian, dong2021efficient}. Thus,  $C_{\rm SP}$ proves to be subdominant in comparison to $C_{\rm HS}$ if $\beta<2$, which is suggested in our simulation. 
Furthermore, under assumption of Eq.~\eqref{eq:asp_bound_combined}, we can estimate $t_{\rm ASP}$ to at most a few tens for practical system size of $N\sim 100$ under infidelity of $\epsilon_f \sim 0.1$. This is fairly negligible compared to the controlled Hamiltonian simulation that requires dynamics duration to be order of tens of thousands in our target models~\cite{Note2}. This outcome stems from the fact that the controlled Hamiltonian simulation \black{for the purpose of eigenenergy extraction} obeys the Heisenberg limit as $C_{\rm HS} =O(1/\epsilon)$, a consequence of time-energy uncertainty relation. This is in contrast to the state preparation, which is not related to any quantum measurement and thus there does not exist such a polynomial lower bound.

\begin{table*}
\setlength{\leftskip}{-0.25cm}
\begin{tabular}{l|c|c|c|c|c|c|c|c|c|}
                  & \begin{tabular}[c]{@{}c@{}}Formal scaling\\ (Lattice system)\end{tabular} & \multicolumn{4}{c|}{2d $J_1$-$J_2$ Heisenberg~($J_2$=0.5)}                                                                              & \multicolumn{4}{c}{2d Fermi-Hubbard~($U$=4)}                                                                                            \\ \hline
                  & $N$:\#Total qubits                                                        & \multicolumn{1}{c|}{$6\times 6$}       & \multicolumn{1}{c|}{$10\times 10$}     & \multicolumn{1}{c|}{$20\times 20$}     & $100 \times 100$  & \multicolumn{1}{c|}{$6\times 6$}       & \multicolumn{1}{c|}{$10\times 10$}     & \multicolumn{1}{c|}{$20\times 20$}     & $100 \times 100$  \\ \hline
qDRIFT            & $O\left( N^2/\epsilon^2\right)$                                           & 5.88e+12  & 5.31e+13  & 9.82e+14  & 7.59e+17 & 1.94e+12  & 1.68e+13  & 3.03e+14  & 2.30e+17 \\          
Random Trotter (2nd) & $O\left( N^2/\epsilon^{3/2}\right)$                                       & 3.64e+09  & 3.00e+10  & 5.28e+11  & 3.89e+14 & 1.47e+10  & 1.21e+11  & 2.11e+12  & 1.57e+15 \\ 
Taylorization     & $O\left( N^2 W/\epsilon\right)$                                           & 4.24e+09  & 2.59e+10  & 3.55e+11  & 2.34e+14 & 2.59e+09  & 1.86e+10  & 2.86e+11  & 2.04e+14 \\ 
Qubitization (seq) & $O\left( N^2/\epsilon\right)$                                             & 1.17e+08  & 8.00e+08  & 1.21e+10  & 7.51e+12 & 8.47e+07  & 6.31e+08  & 9.97e+09  & 6.26e+12 \\
 Qubitization (product) & $O\left( N^2/\epsilon\right)$ & {\bf 9.38e+07}  & {\bf 5.33e+08}  & {\bf 6.85e+09}  & {\bf 3.82e+12} & {\bf 5.57e+07}  & {\bf 3.87e+08}  & {\bf 5.71e+09}  & {\bf 3.49e+12} \\ 
\end{tabular}
\caption{$T$-count required to perform the quantum phase estimation on lattice Hamiltonians based on various Hamiltonian simulation algorithms. Here, we denote the total qubit count by $N$, target energy accuracy by $\epsilon(=0.01)$, and $W=\log(N/\epsilon)/\log\log(N/\epsilon)$ as Taylorization order.
Note that the post-Trotter methods, namely the Taylorization and qubitization algorithms, consume ancillary qubits of $O(\log N)$ to block-encode the action of Hamiltonian simulation into truncated Hilbert space. This is comparable to those required by the quantum Fourier transformation that require ancillary qubits of $O(\log(N/\epsilon))$. 
\black{In practice, the newly proposed product-wise construction of the qubitization algorithm consumes twice as much ancilla logical qubits with nearly halved $T$-count for the target models in this work (see Sec.~\ref{sec:post_trotter_oracles} for details).}
}
\label{tab:tcount_comparison}
\end{table*}

\begin{figure}[ht]
    \centering
    \includegraphics[width=0.85\linewidth]{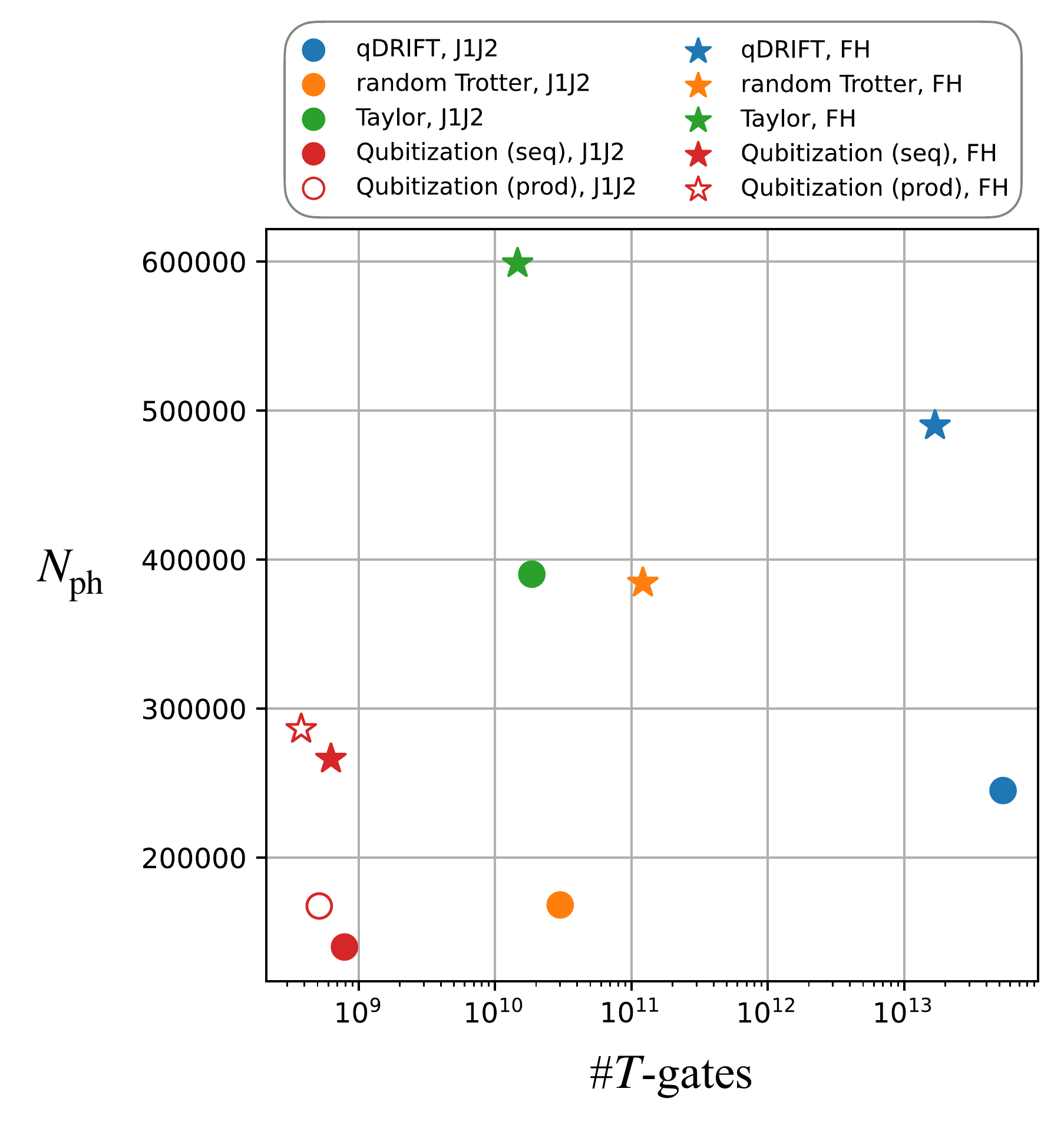}
    \caption{\black{Spacetime cost of the phase estimation algorithm, namely the $T$-count and physical qubit counts $N_{\rm ph}$. Here, we estimate the ground state energy up to target accuracy $\epsilon=0.01$ for 2d $J_1$-$J_2$ Heisenberg model ($J_2/J_1=0.5$) and 2d Fermi-Hubbard model ($U/t=4$), both with lattice size of $10\times 10$. The blue, orange, green, and orange points indicate the results that employ qDRIFT, 2nd-order random Trotter, Taylorization, and qubitization, where the circle and star markers denote the spin and fermionic models, respectively. Two flavors of the qubitization, the sequential and newly proposed product-wise construction (see Sec.~\ref{sec:post_trotter_oracles} for details), are discriminated by filled and unfilled markers. Note that  $N_{\rm ph}$ here does not account for the magic state factories, which are incorporated in Fig.~\ref{fig:heatmap}.}} \label{fig:spacetime_volume}
\end{figure}

\subsubsection{Main quantum resource}
As we have seen in the previous sections, the dominant contribution to the quantum resource is $C_{\rm HS}$, namely the controlled Hamiltonian simulation from which the eigenenergy phase is extracted into the ancilla qubits.
Fortunately, with the scope of performing quantum resource estimation for the QPE and digital quantum simulation, numerous works have been devoted to analyzing the error scaling of various Hamiltonian simulation techniques, in particular the Trotter-based methods~\cite{suzuki1985general,childs_theory_2021, campbell_random_2019}.
Nevertheless, we point out that crucial questions remain unclear; (A) which technique is the best practice to achieve the earliest quantum advantage for condensed matter problems, and (B) at which point does the crossover occur?

Here we perform resource estimation under the following common assumptions:
(1) logical qubits are encoded using the formalism of surface codes~\cite{kitaev1997quantum};
(2) quantum gate implementation is based on Clifford+$T$ formalism; 
Initially, we address the first question (A) by comparing the total number of $T$-gates, or $T$-count, across various Hamiltonian simulation algorithms, as the
 application of a $T$-gate involves a time-consuming procedure known as magic-state distillation. 
 Although not necessarily, this procedure is considered to dominate the runtime in many realistic setups. Therefore, we argue that $T$-count shall provide sufficient information to determine the best Hamiltonian simulation technique. 
Then, with the aim of addressing the second question (B), we further perform high-resolution analysis on the runtime. We in particular consider concrete quantum circuit compilation with specific physical/logical qubit configuration compatible with the surface code implemented on a square lattice.

Let us first compute the $T$-counts to compare the state-of-the-art Hamiltonian simulation techniques: (randomized) Trotter product formula~\cite{suzuki_general_1991, childs_fasterquantum_2019}, qDRIFT~\cite{campbell_random_2019}, Taylorization~\cite{berry_exponential_2014, berry_simulating_2015, Meister2022tailoringterm}, and qubitization~\cite{low2016hamiltonian}. The former two commonly rely on the Trotter decomposition to approximate the unitary time evolution with sequential application of (controlled) Pauli rotations, while the latter two, dubbed as ``post-Trotter methods," are rather based on the technique called {\it block-encoding}, which utilize ancillary qubits to encode desired (non-unitary) operations on target systems (See Sec.~\ref{sec:post_trotter_oracles} in SM). 
While post-Trotter methods are known to be exponentially more efficient in terms of gate complexity regarding the simulation accuracy~\cite{berry_exponential_2014}, it is nontrivial to ask which is the best practice in the crossover regime, where the prefactor plays a significant role. 

We have compiled quantum circuits based on existing error analysis to reveal the required $T$-counts (See Sec.~\ref{sec:phase_estimation},~\ref{sec:cost_basic},~\ref{sec:imperfect_prepare} in SM).
From results presented in Table~\ref{tab:tcount_comparison}, we find that the qubitization algorithm provides the most efficient implementation in order to reach the target energy accuracy $\epsilon = 0.01$.
Although the post-Trotter methods, i.e., the Taylorization and qubitization algorithms require additional ancillary qubits of $O(\log (N))$ to perform the block encoding, we regard this overhead as not a serious roadblock, since the target system itself and the quantum Fourier transformation requires qubits of $O(N)$ and $O(\log(N/\epsilon))$, respectively.
\black{In fact, as we show in Fig.~\ref{fig:spacetime_volume}, the qubitization algorithms are efficient at near-crosspoint regime in physical qubit count as well, due to the suppressed code distance (see Sec.~S9 in SM for details).}

We also mention that, for 2d Fermi-Hubbard model, there exists some specialized Trotter-based methods that improve the performance significantly~\cite{Kivlichan2020improvedfault, campbell2021early}. For instance, the $T$-count of the QPE based on the state-or-the-art PLAQ method proposed in Ref.~\cite{campbell2021early} can be estimated to be approximately $4\times10^{8}$ for $10\times 10$ system under $\epsilon=0.01$, which is slightly higher than the $T$-count required for the qubitization technique. Since the scaling of PLAQ is similar to the 2nd order Trotter method, we expect that the qubitization remains the best for all system size $N$.

\begin{figure*}[tb]
    \centering
    \includegraphics[width=0.99\linewidth]{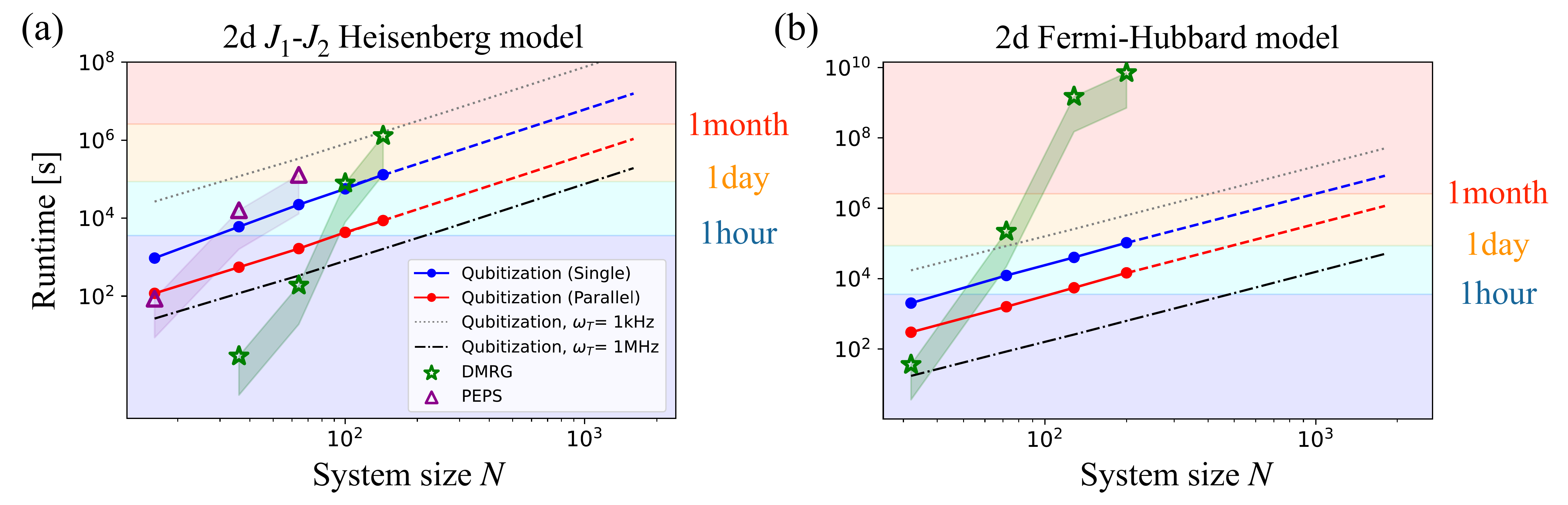}
    \caption{Quantum-classical crossover in (a) 2d $J_1$-$J_2$ Heisenberg model of $J_2/J_1=0.5$ and (b) 2d Fermi-Hubbard model of $U/t=4$. The blue and red circles are the runtime estimate for the quantum phase estimation using the qubitization technique as a subroutine, whose analysis involves quantum circuit compilation of all the steps (I), (II), and (III). All the gates are compiled under the Clifford+$T$ formalism with each logical qubits encoded by the surface code with code distance $d$ around 17 to 25 assuming physical error rate of $p=10^{-3}$~(See Sec.~\ref{sec:total_resource} in SM). Here, the number of magic state factories $n_F$ and number of parallelization threads $n_{\rm th}$ are taken as $(n_F, n_{\rm th}) = (1, 1)$ and $(16, 16)$ for ``Single" and ``Parallel," respectively. The dotted and dotted chain lines are estimates that only involve the analysis of step (II); calculation is based solely on the $T$-count of the algorithm with realistic $T$-gate consumption rate of 1kHz and 1MHz, respectively. \black{The green stars and purple triangles are data obtained from the actual simulation results of classical DMRG and \black{variational PEPS} algorithms, respectively, with the shaded region denoting the potential room for improvement  by using the most advanced computational resource~(See Sec.~\ref{sec:dmrg} in SM)}. Note that the system size is related with the lattice size $M\times M$ as $N=2M^2$ in the Fermi-Hubbard model.
    }\label{fig:crossover}
\end{figure*}

The above results motivate us to study the quantum-classical crossover entirely using the qubitization technique as the subroutine for the QPE.
As is detailed in Sec.~\ref{sec:distselect_analysis} in SM~\cite{Note1}, our runtime analysis involves the following steps:
\begin{enumerate}
    \item[(I)] {\it Hardware configuration.} Determine the architecture of quantum computers (e.g., number of magic state factories, qubit connectivity etc.).
    \item[(II)] {\it Circuit synthesis and transpilation.} Translate high-level description of quantum circuits to Clifford+$T$ formalism with the provided optimization level. 
    \item[(III)] {\it Compilation to executable instructions.} Decompose logical gates into the sequence of executable instruction sets based on lattice surgery.
\end{enumerate}
It should be noted that the ordinary runtime estimation only involves the step (II);  simply multiplying the execution time of $T$-gate to the $T$-count as $N_{T} t_{T}$. However, we emphasize that this estimation method loses several vital factors in time analysis which may eventually lead to deviation of one or two orders of magnitude.
In sharp contrast, our runtime analysis comprehensively takes all steps into account to yield reliable estimation under realistic quantum computing platforms.

\begin{figure*}[tbp]
    \centering
    \includegraphics[width=0.99\linewidth]{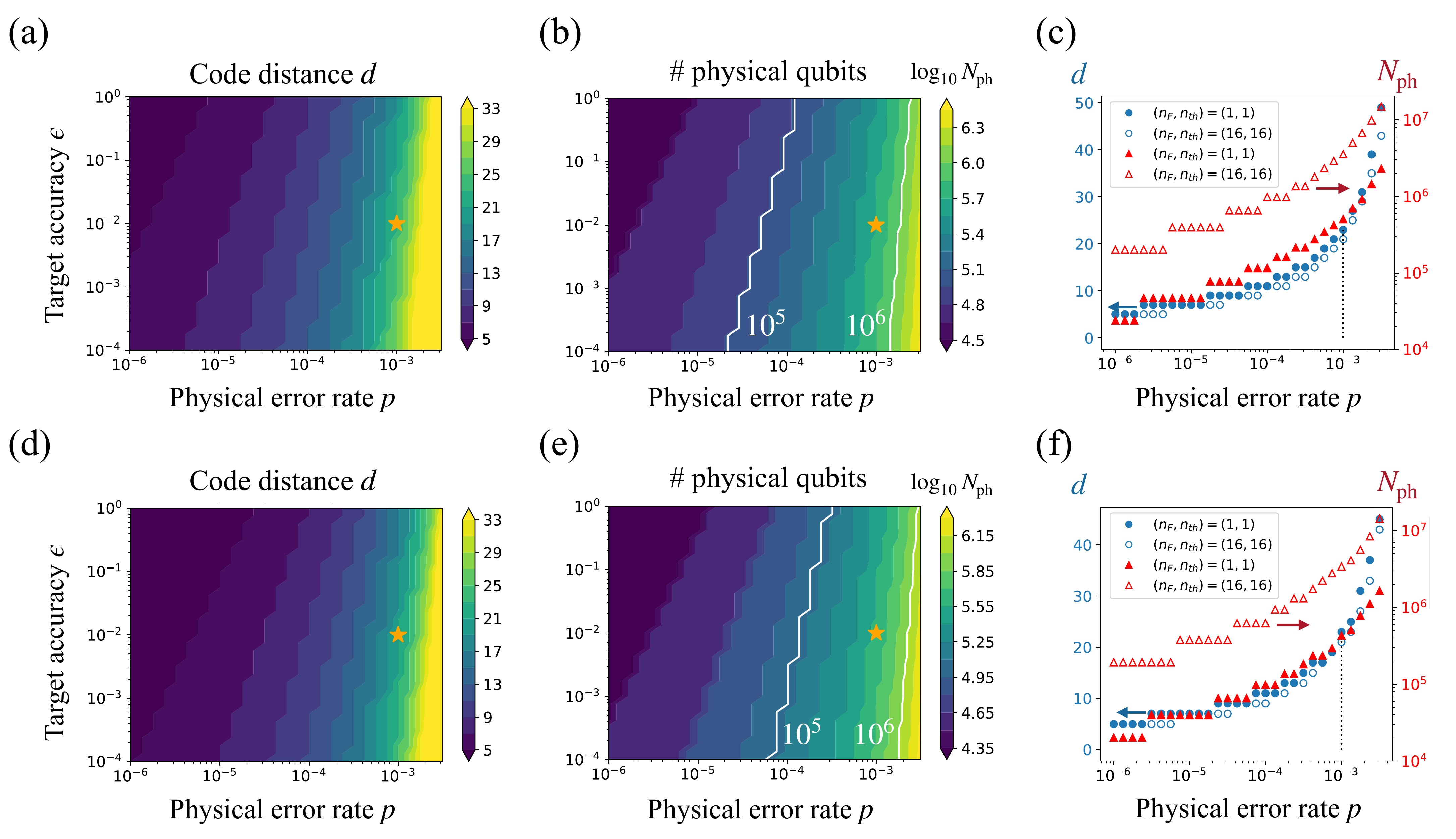}
    \caption{
    Requirements for logical and physical qubits by the phase estimation algorithm based on the qubitization to achieve target accuracy $\epsilon$ with physical error rate $p$. The panels denote (a) code distance $d$ and (b) number of physical qubits $N_{\rm ph}$  required to simulate the ground state of 2d $J_1$-$J_2$ Heisenberg model with lattice size of $10 \times 10$ with $J_2 = 0.5$. Here, the qubit plane is assumed to be organized as $(n_F, \# \text{thread})=(1,1)$. The setup used in the maintext, $\epsilon=0.01$ and $p=10^{-3}$, is indicated by the orange stars. (c) Focused plot at $\epsilon=0.01$. Blue and red points show the results for code distance $d$ and $N_{\rm ph}$, respectively, where the filled and empty markers correspond to floor plans with $(n_{F}, \#\text{thread})=(1,1)$ and $(16, 16)$, respectively. (d-f) Plots for 2d Fermi-Hubbard model of lattice size $6\times 6$ with $U=4$, corresponding to (a-c) for the Heisenberg model.
    }\label{fig:heatmap}
\end{figure*}

\begin{figure*}[t]
    \centering
    \includegraphics[width=0.7\linewidth]{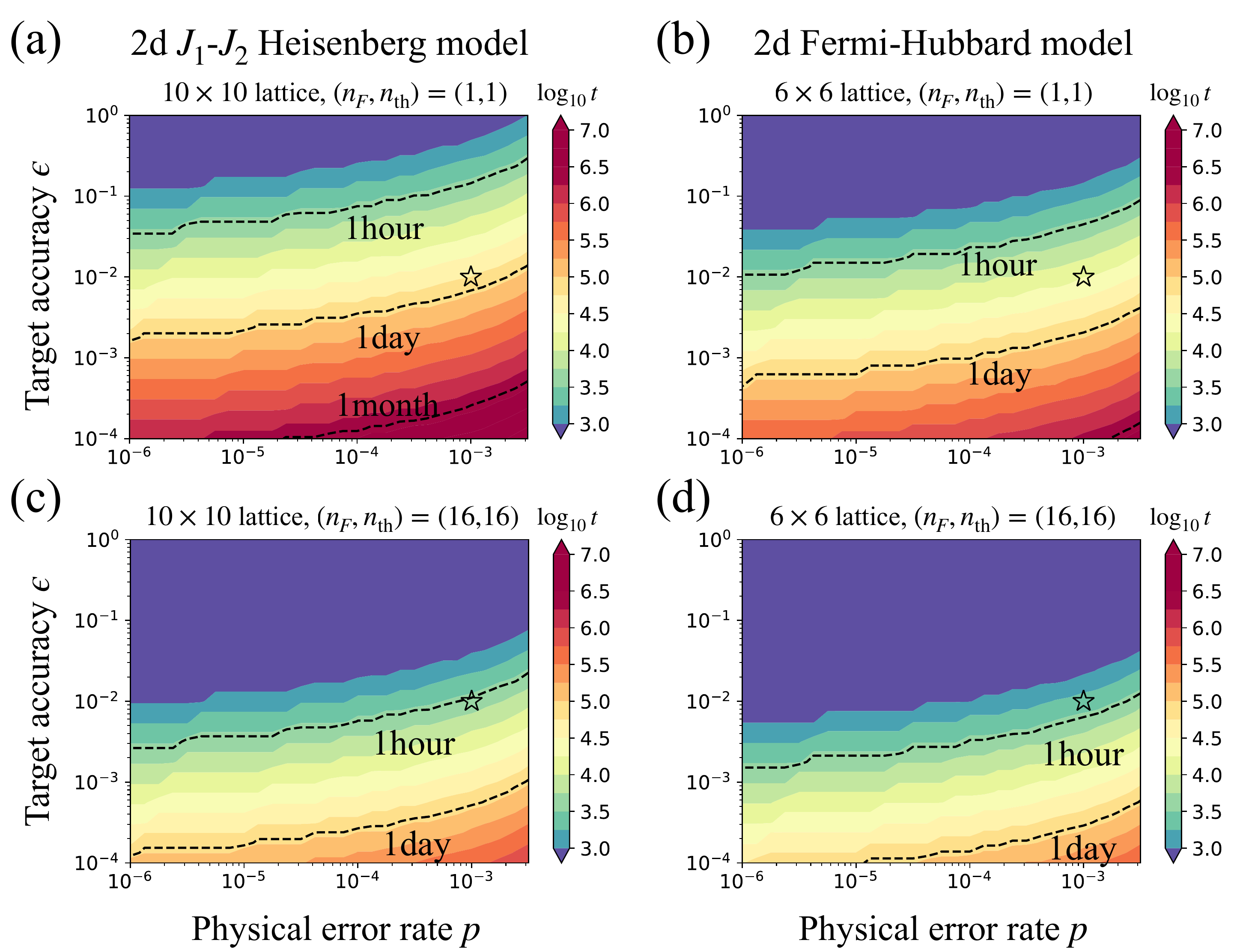}
    \caption{
    Estimated runtime for various simulation setups for (a), (c) 2d $J_1$-$J_2$ Heisenberg model of lattice size $10 \times 10$ with $J_2=0.5$ and (b), (d) 2d Fermi-Hubbard model of lattice size $6 \times 6$ with $U=4$. The floor plan of the qubit plane is assumed as $(n_F, \# \text{thread})=(1, 1)$ and $(16, 16)$ for (a),(b) and (c),(d), respectively. The setup $\epsilon=0.01$ and $p=10^{-3}$, employed in Fig.~\ref{fig:crossover}, is shown by the black open stars.
    }\label{fig:runtime_heatmap_main}
\end{figure*}

\subsection*{Crossover under $p=10^{-3}$} \label{subsec:crossover_main}
Figure~\ref{fig:crossover} shows the runtime of classical/quantum algorithms simulating the ground state energy in 2d $J_1$-$J_2$ Heisenberg model and 2d Fermi-Hubbard model. 
In both figures, we observe clear evidence of quantum-classical crosspoint below a hundred-qubit system (at lattice size of $10\times10$ and $6\times6$, respectively) within plausible runtime. 
Furthermore, a significant difference from ab initio quantum chemistry calculations is highlighted in the feasibility of system size $N\sim 1000$ logical qubit simulations, especially in simulation of 2d Heisenberg model that utilizes the parallelization tehcnique for the oracles~(See Sec.~\ref{sec:distselect_analysis} in SM for details).

For concreteness, let us focus on the simulation for systems with lattice size of $10\times 10$, where we find the quantum algorithm to outperform the classical one.
Using the error scaling, we find that the DMRG simulation is estimated to take about $10^5$ and $10^9$ seconds in 2d Heisenberg and 2d Fermi-Hubbard models, respectively.
On the other hand, the estimation based on the dedicated quantum circuit compilation with the most pessimistic equipment (denoted as ``Single" in Fig.~\ref{fig:crossover}) achieves runtime below $10^5$ seconds in both models. This is further improves by an order when  we assume a more abundant quantum resource. 
Concretely, using a quantum computer with multiple magic state factories ($n_F$=16) that performs multi-thread execution of the qubitization algorithm ($n_{\rm Th}=16)$, the quantum advantage can be
achieved within a computational time frame of several hours.
We find it informative to also display the usual $T$-count-based estimation; it is indeed reasonable to assume a clock rate of 1-10kHz for single-thread execution, while its precise value fluctuates depending on the problem instance.

We note that the classical algorithm (DMRG) experiences an exponential increase in the runtime to reach the desired total energy accuracy $\epsilon=0.01$. This outcome is somewhat expected, since one must enforce the MPS to represent 2d quantum correlations into 1d via cylindrical boundary condition~\cite{LeBlanc2015, WangS2018}.
\black{Meanwhile, the prefactor is significantly lower than that of other tensor-network-based methods, enabling its practical use in discussing the quantum-classical crossover.
For instance, although the formal scaling is exponentially better in \black{variational PEPS} algorithm, the runtime in 2d $J_1$-$J_2$ Heisenberg model exceeds $10^4$ seconds already for the $6\times 6$ model, while the DMRG algorithm consumes only $10^2$ seconds (See Fig.~\ref{fig:crossover}(a)). 
Even if we assume that the bond dimension of PEPS can be kept constant for larger $N$, the crossover between DMRG and \black{variational PEPS} occurs only above the size of $12\times 12$. 
As we have discussed in Sec.~\ref{sec:algorithms}, we reasonably expect $D=O(\log(N))$ for simulation of fixed total accuracy, and furthermore expect that the number of variational optimization also scales polynomially with $N$. This implies that the scaling is much worse than $O(N)$; in fact, \black{we have used constant value of $D$ for $L=4, 6, 8$ and observe that the scaling is already worse than cubic in our setup. 
Given such a scaling,}
 we conclude that DMRG is better suited than the \black{variational PEPS} for investigating the quantum-classical crossover,
}
\black{and also that quantum algorithms with quadratic scaling on $N$ runs faster in the asymptotic limit.}

\subsection*{Portfolio of crossover under various algorithmic/hardware setups} \label{subsec:different_setup}
It is informative to modify the hardware/algorithmic requirements to explore the variation of quantum-classical crosspoint.
For instance, the code distance of the surface code depends on $p$ and  $\epsilon$ as (See Sec.~\ref{sec:total_resource} in SM)
\begin{eqnarray}
d &=& O\left(\frac{\log(N/\epsilon)}{\log(1/p)}\right). \label{eq:code_distance_formula}
\end{eqnarray}
Note that this also affects the number of physical qubits via the number of physical qubit per logical qubit $2d^2$.
We visualize the above relationship explicitly in Fig.~\ref{fig:heatmap}, which considers the near-crosspoint regime of 2d $J_1$-$J_2$ Heisenberg model and 2d Fermi-Hubbard model. 
It can be seen from Fig.~\ref{fig:heatmap}(a),(b),(d),(e) that the improvement of the error rate directly triggers the reduction of the required code distance, which results in s significant suppression of the number of physical qubits. 
This is even better captured by Fig.~\ref{fig:heatmap}(c) and (f). By achieving a physical error rate of $p=10^{-4}$ or $10^{-5}$, for instance, one may realize a 4-fold or 10-fold reduction of the number of physical qubits.

The logarithmic dependence for $\epsilon$ in Eq.~\eqref{eq:code_distance_formula} implies that the target accuracy does not significantly affect the qubit counts; 
it is rather associated with the runtime, since the total runtime scaling is given as 
\begin{eqnarray}
t = O\left(\frac{N^2\log(N/\epsilon)}{\epsilon \log(1/p)}\right),
\end{eqnarray}
which shows {\it polynomial} dependence on $\epsilon$.
Note that this scaling is based on multiplying a factor of $d$ to the gate complexity, since we assumed that the runtime is dominated by the magic state generation, of which the time is proportional to the code distance $d$, rather than by the classical postprocessing (see Sec.~\ref{sec:distselect_analysis}, \ref{sec:total_resource} in SM).
As is highlighted in Fig.~\ref{fig:runtime_heatmap_main}, we observe that in the regime with higher $\epsilon$, the computation is completed within minutes. 
However, we do not regard such a regime as an optimal field for the quantum advantage. 
The runtime of classical algorithms typically shows higher-power dependence on $\epsilon$, denoted as $O(1/\epsilon^{\gamma})$, with $\gamma \sim 2$ for $J_1$-$J_2$ Heisenberg model and $\gamma \sim 4$ for the Fermi-Hubbard model (see Sec.~\ref{sec:dmrg} in SM), which both implies that classical algorithms are likely to run even faster than quantum algorithms under large $\epsilon$ values. 
We thus argue that the setup of $\epsilon = 0.01$ provides a platform that is both plausible for the quantum algorithm and challenging by the classical algorithm.

\section*{Discussion}\label{sec:discussion}
Our work has presented a detailed analysis of the quantum-classical crossover in condensed matter physics, specifically, pinpointing the juncture where the initial applications of fault-tolerant quantum computers demonstrate advantages over classical algorithms.
Unlike previous studies, which primarily focused on exact simulation techniques to represent classical methods, we have proposed utilizing error scaling to estimate runtime using one of the most powerful variational simulation method---the DMRG algorithm. 
We have also scrutinized the execution times of quantum algorithms, conducting a high-resolution analysis that takes into account the topological restrictions on physical qubit connectivity, the parallelization of Hamiltonian simulation oracles, among other factors.
This rigorous analysis has led us to anticipate that the crossover point is expected to occur within feasible runtime of a few hours when the system size $N$ reaches about a hundred. 
Our work serves as a reliable guiding principle for establishing milestones across various platform of quantum technologies.

Various avenues for future exploration can be envisioned.
We would like to highlight primary directions here.
Firstly, expanding the scope of runtime analysis to encompass a wider variety of classical methods is imperative.
In this study, we concentrated on the DMRG and \black{variational PEPS} algorithms due to their simplicity in runtime analysis.
However, other quantum many-body computation methods such as quantum Monte Carlo (e.g. path-integral Monte Carlo, variational Monte Carlo etc.), coupled-cluster techniques, or other tensor-network-based methods hold equal importance. 
In particular, devising a systematic method to conduct estimates on Monte Carlo methods shall be a nontrivial task.

Secondly, there is a pressing need to further refine quantum simulation algorithms that are designed to extract physics beyond the  eigenenergy, such as the spacial/temporal correlation function, nonequilibrium phenomena, finite temperature properties, among others. Undertaking error analysis on these objective could prove to be highly rewarding.

Thirdly, it is important to survey the optimal method of state preparation. 
While we have exclusively considered the ASP, there are numerous options including the Krylov technique~\cite{kirby2022exact}, recursive application of phase estimation~\cite{zhao2019state}, and sparse-vector encoding technique~\cite{zhang2022quantum}.
Since the efficacy of state preparation methods heavily relies on individual instances, it would be crucial to elaborate on the resource estimation in order to discuss quantum-classical crossover in other fields including high-energy physics, nonequilibrium physics, and so on.

Fourthly, it is interesting to seek the possibility of reducing the number of physical qubits by replacing the surface code with other quantum error-correcting codes with a better encoding rate~\cite{bravyi2023high}. 
For instance, there have been suggestions that the quantum LDPC codes may enable us to reduce the number of physical qubits by a factor of tens to hundreds~\cite{bravyi2023high}. Meanwhile, there are additional overheads in implementation and logical operations, which may increase the runtime and problem sizes for demonstrating quantum advantage.

Lastly, exploring the possibilities of a classical-quantum hybrid approach is an intriguing direction. 
This could involve twirling of Solovey-Kitaev errors into stochastic errors that can be eliminated by quantum error mitigation techniques originally developed for near-future quantum computers without error correction
~\cite{suzuki2020quantum, piveteau2021error}.

\section*{Methods}

\subsection*{Target models}\label{sec:target_model}
Condensed matter physics deals with intricate interplay between microscopic degrees of freedom such as spins and electrons, which, in many cases, form translationally symmetric structures.
Our focus is on lattice systems that not only reveal the complex and profound nature of quantum many-body phenomena, but also await to be solved despite the existing intensive studies (See Sec.~\ref{sec:target_hamiltonian} in SM):\vspace{3mm}

\noindent
{\it (1) Antiferromagnetic Heisenberg model.} Paradigmatic quantum spin models frequently involve frustration between interactions as source of complex quantum correlation. 
One highly complex example is the spin-1/2 $J_1$-$J_2$ Heisenberg model on the square lattice, whose ground state property has remained a persistent problem over decades:
\begin{eqnarray}
    H = J_1\sum_{\langle p, q\rangle} 
    \sum_{\alpha \in \{X, Y, Z\}} S_p^\alpha S_q^\alpha + 
    J_2 \sum_{\langle \langle p, q \rangle \rangle} 
    \sum_{\alpha \in \{X, Y, Z\}} S_p^\alpha S_q^\alpha,\nonumber
\end{eqnarray}
where $\braket{\cdot}$ and $\braket{\braket{\cdot}}$ denote pairs of (next-)nearest-neighboring sites that are coupled via Heisenberg interaction with amplitude $J_{1(2)}$, and $S_p^\alpha$ is the $\alpha$-component of spin-1/2 operator on the $p$-th site.
Due to the competition between the $J_1$ and $J_2$ interaction, tje formation of any long-range order is hindered at $J_2/J_1 \sim 0.5$, at which a quantum spin liquid phase is expected to realize~\cite{zhang_valence_2003, jiang_spin_2012, nomura_dirac_2021}.
In the following we set $J_2=0.5$ with $J_1$ to be unity, and focus on cylindrical boundary conditions.\vspace{3mm}

\noindent
{\it (2) Fermi-Hubbard model.} One of the most successful fermionic models that captures the essence of electronic and magnetic behaviour in quantum materials is the Fermi-Hubbard model. 
Despite the concise construction, it exhibits a variety of features such as the unconventional superfluidity/superconductivity, quantum magnetism, and interaction-driven insulating phase (or Mott insulator)~\cite{esslinger_hubbard_2010}.
With this in mind, we consider the following half-filled Hamiltonian:
\begin{eqnarray}
H = -t \sum_{\langle p, q \rangle, \sigma} (c_{p, \sigma}^\dag c_{q, \sigma}  + {\rm h.c.}) + U\sum_{p}c^\dag_{p, \uparrow}c_{p, \uparrow} c^\dag_{p, \downarrow} c_{p, \downarrow},
\nonumber
\end{eqnarray}
where $t=1$ is the hopping amplitude and $U$ is the repulsive onsite potential for annihilation~(creation) operators $c_{p, \sigma}^{(\dag)}$, defined for a fermion that resides on site $p$ with spin $\sigma$. 
Here the summation on the hopping is taken over all pairs of nearest-neighboring sites $\braket{p, q}$.
Note that one may further introduce nontrivial chemical potential to explore cases that are not half-filled, although we leave this for future work.

\section*{Data Availability} All study data are included in this article and Supplementary Materials.

\section*{Acknowledgements}
The authors are grateful to the fruitful discussions with Sergei Bravyi, Keisuke Fujii, Zongping Gong, Takuya Hatomura, Kenji Harada, Will Kirby, Sam McArdle, Takahiro Sagawa, Kareljan Schoutens, Kunal Sharma,  Kazutaka Takahashi,  Zhi-Yuan Wei, and Hayata Yamasaki.
N.Y. wishes to thank JST PRESTO No. JPMJPR2119 and the support from IBM Quantum. 
T.O. wishes to thank JST PRESTO Grant Number JPMJPR1912, JSPS KAKENHI Nos.~22K18682, 22H01179, and \black{23H03818}, and support by the Endowed Project for Quantum Software Research and Education, The University of Tokyo (https://qsw.phys.s.u-tokyo.ac.jp/).
Y. S. wishes to thank JST PRESTO Grant Number JPMJPR1916 and JST Moonshot R\&D Grant Number JPMJMS2061.
W.M. wishes to thank JST PRESTO No.\ JPMJPR191A, JST COI-NEXT program Grant No. JPMJPF2014 and MEXT Quantum Leap Flagship Program (MEXT Q-LEAP) Grant Number JPMXS0118067394 and JPMXS0120319794.
This work was supported by JST Grant Number JPMJPF2221. 
\black{This work was supported by JST ERATO Grant Number  JPMJER2302 and JST CREST Grant Number JPMJCR23I4, Japan. } 
A part of computations were performed using the Institute of Solid State Physics at the University of Tokyo.

\section*{Author Contributions}
NY and TO conducted resource estimation on classical computation. NY, YS and YK performed resource estimation on quantum algorithms. NY and WM conceived the idea. All authors contributed to the discussion and manuscript.

\section*{Competing Interests}
YS and WM own stock/options in QunaSys Inc. YS owns stock/options in Nippon Telegraph and Telephone.

\bibliography{bib.bib}




\let\addcontentsline\oldaddcontentsline
\onecolumngrid

\clearpage
\begin{center}
	\Large
	\textbf{Supplementary Notes for: Hunting for quantum-classical crossover in condensed matter problems}
\end{center}

\setcounter{section}{0}
\setcounter{equation}{0}
\setcounter{figure}{0}
\setcounter{table}{0}
\renewcommand{\thesection}{S\arabic{section}}
\renewcommand{\theequation}{S\arabic{equation}}
\renewcommand{\thefigure}{S\arabic{figure}}
\renewcommand{\thetable}{S\arabic{table}}

\setcounter{equation}{0}
\renewcommand{\theequation}{S\arabic{equation}}

\setcounter{section}{0}
\renewcommand{\thesection}{S\arabic{section}}
\setcounter{subsection}{0}
\renewcommand{\thesubsection}{\arabic{subsection}}
\setcounter{subsubsection}{0}
\renewcommand{\thesubsubsection}{\arabic{subsubsection}}
\setcounter{figure}{0}
\renewcommand{\thefigure}{S\arabic{figure}}
\setcounter{table}{0}
\renewcommand{\thetable}{S\arabic{table}}

\addtocontents{toc}{\protect\setcounter{tocdepth}{0}}


In this Supplemental Note, we present the details on the classical and quantum resource estimation on ground state problem in condensed matter physics.
In Sec.~\ref{sec:target_hamiltonian}, we provide the definitions and brief introduction to the nature of the representative lattice models considered in condensed matter physics.
In Sec.~\ref{sec:dmrg}, we describe how to estimate the classical runtime of tensor-network algorithms, namely the Density-Matrix Remormalization-Group (DMRG)  and the variational Projected Entangled-Pair State (PEPS) algorithms.
In Sec.~\ref{sec:state_preparation}, we introduce the adiabatic time evolution as the state preparation method and analyze its circuit complexity for 2d $J_1$-$J_2$ Heisenberg model.
In Sec.~\ref{sec:phase_estimation}, we provide a overview on the structure and cost of quantum algorithm to estimate ground state energy, namely the quantum phase estimation algorithm.
In Sec.~\ref{sec:post_trotter_oracles}, we describe the post-Trotter Hamiltonian simulation algorithms that are exponentially more efficient than the Trotter-based ones in terms of error scaling.
In Sec.~\ref{sec:cost_basic}, we summarize the gate complexity, or more concretely the $T$-count, to execute basic operations in fault-tolerant quantum computing.
In Sec.~\ref{sec:imperfect_prepare}, we describe how to modify an oracle in the qubitization algorithm such that the boundary conditions in the Hamiltonian does not heavily affect the gate complexity.
In Sec.~\ref{sec:distselect_analysis}, we provide a systematic framework to estimate the runtime of the qubitization-based quantum phase estimation algorithm assuming Clifford+$T$ formalism under surface code.
Finally, in Sec.~\ref{sec:total_resource}, we put all the analysis of quantum algorithm together and present the  estimation on total quantum resource including the runtime, physical qubits, and code distance of the surface code.

{
\hypersetup{linkcolor=blue}
\tableofcontents
}
\vspace{1cm}

\section{Definition of Target models}\label{sec:target_hamiltonian}
In this paper, we discuss the quantum-classical crossover in condensed matter problems where the Hamiltonian is defined on $d$-dimensional translationally symmetric lattice with at most $G$-local terms:
\begin{eqnarray}
    H = \sum_{\bp}\sum_{\{\bmu, \balpha\}}
    w_{\bmu, \balpha}
    \hat{\Lambda}_{\bp}^{(\alpha_1)} 
    \hat{\Lambda}_{\bp + \bmu_1}^{(\alpha_2)}\cdots 
    \hat{\Lambda}_{\bp + \bmu_{G-1}}^{(\alpha_G)},
\end{eqnarray}
where $\bp = (p_1, \dots, p_d)$ labels the lattice site (including the sublattice structure), $\bmu = (\bmu_1, ..., \bmu_{G-1})$ is a set of vectors that identifies the connection between interacting sites,
$\balpha$ discriminates the interaction on the sites,
$\hat{\Lambda}^{(\alpha)}_{\bp}$ is an operator for microscopic degrees of freedom (e.g. spin, fermion, boson) on site $\bp$,  and $w_{\bmu, \balpha}$ is the amplitude of the interaction.
In particular, we employ three representative models with long-standing problems; 2d $J_1$-$J_2$ Heisenberg model, 2d Fermi-Hubbard model, and spin-1 Heisenberg chain.
In the following, we provide the definition of the target models accompanied with brief descriptions.

\subsection{Spin-1/2 2d $J_1$-$J_2$ Heisenberg model}
The physical nature of the paradigmatic $J_1$-$J_2$ Heisenberg model on the square lattice is under debate for over decades, especially regarding the property of the ground state~\cite{zhang_valence_2003, jiang_spin_2012, hu_direct_2013, wang_critical_2018, nomura_dirac_2021}. 
The frustration, i.e., the competition between interactions that precludes satisfying all the energetic gain, is considered to suppress the formation of long-range order, and thus it amplifies the nontrivial effect of quantum fluctuation. Together with geometrically frustrated magnets defined on e.g., triangular, Kagom\'{e}, and pyrochlore lattices, the $J_1$-$J_2$ Heisenberg model on square lattice has been investigated intensively by classical algorithms.
Considering that the model is commonly recognized as one of the most challenging and physically intriguing quantum spin models, it is genuinely impactful to elucidate the quantum-classical crossover if exists at all.

The Hamiltonian of spin-1/2 $J_1$-$J_2$ Heisenberg model on the square lattice is defined as
\begin{eqnarray}
    H = J_1\sum_{\langle p, q\rangle} 
    \sum_{\alpha \in \{X, Y, Z\}} S_p^\alpha S_q^\alpha + 
    J_2 \sum_{\langle \langle p, q \rangle \rangle} 
    \sum_{\alpha \in \{X, Y, Z\}} S_p^\alpha S_q^\alpha,\label{eq:J1J2_definition}
\end{eqnarray}
where the summation of the first and second terms concern pairs of nearest-neighbor and next-nearest-neighbor sites, respectively, and $S_p^\alpha$ is the spin-1/2 operator on the $p$-th site that is related with the $\alpha$-th component of the Pauli matrices as $S_p^\alpha = \frac{1}{2} \sigma_p^\alpha$.

\subsection{2d Fermi-Hubbard model}
As the second representative model, we choose the Fermi-Hubbard model on the square lattice. The emergent phenomena in Fermi-Hubbard model is truly abundant, and provides a significant insights into electronic and magnetic properties of materials; the unconventional superfluidity or superconductivity, Mott insulating behaviour, and quantum magnetism.
However, the intricate nature of Fermi statistics has hindered us from performing scalable and quantitative description in the truly intriguing zero-temperature regime.

The Hamiltonian of Fermi-Hubbard model on the square lattice reads
\begin{eqnarray}
H = -t \sum_{\langle p, q \rangle, \sigma} (c_{p, \sigma}^\dag c_{q, \sigma}  + {\rm h.c.}) + U\sum_{p}n_{p, \uparrow} n_{p, \downarrow},\label{eq:FH-definition}
\end{eqnarray}
where $c_{p, \sigma}^{(\dag)}$ is the fermionic annihilation (creation) operator on site $p$ with spin $\sigma \in \{\uparrow, \downarrow\}$, $n_{p, \sigma} = c_{p, \sigma}^\dag c_{p, \sigma}$ is the corresponding number operator, $t$ is the hopping amplitude, and $U$ is the repulsive onsite potential. 

\subsection{Spin-1 Heisenberg chain}
The gap structure of translation and $U(1)$-invariant antiferromagnetic spin-$S$ chain has been argued to differ according to whether $S$ is half-integer or integer. Such a statement has been known as the Haldane conjecture~\cite{haldane_nonlinear_1983, haldane_continuum_1983}, which has attracted interest of condensed matter physicists for over decades. While the ground states for half-integer $S$ cases are considered to be gapless and hence there is no supporting argument to be simulated classically by polynomial time, it is reasonable to expect for integer $S$ cases to be computed efficiently using, e.g., the matrix product state (MPS), since the entanglement entropy shall obey area law if the system is gapped as is predicted in Haldane conjecture~\cite{Hastings_2007}. In this context, it is not clear if there is any quantum advantage at all for simulating the ground state or the gap structure in integer-$S$ Heisenberg chain.

In this paper, we consider the simplest gapped case in which quantum advantage is unlikely; the spin-1 antiferromagnetic Heisenberg chain. The Hamiltonian reads
\begin{eqnarray}
    H = \sum_p \sum_{\alpha \in \{X, Y, Z\}}S_p^\alpha S_{p+1}^\alpha,
    \label{eq:1dHeisenberg_definition}
\end{eqnarray}
where $S_p^\alpha$ is the spin-1 operator acting on the $p$-th site.

\section{Classical simulation of ground state using density-matrix renormalization group}\label{sec:dmrg}
\subsection{Overview of DMRG and PEPS algorithms}
The Density-Matrix Renormalization Group (DMRG) has been established as one of the most powerful numerical tools to investigate the strongly correlated one-dimensional (1d) quantum lattice models \cite{White1992}. DMRG can be understood as a variational calculation based on the Matrix Product State (MPS) \cite{OstlundR1995,DukelskyMNS1998}. Thanks to the area law of the entanglement entropy expected to hold for low-energy states of the 1d quantum lattice models \cite{EisertCP2010}, MPS is a quantitatively accurate variational ansatz for ground states of many quantum many-body systems \cite{Schollwock2011}. Although MPS cannot cover the area law scaling for two or higher dimensional systems, DMRG is still applicable for quasi 1d cylinders or finite-size 2d systems. Indeed, we can find recent applications of DMRG to the 2d models treated in this study \cite{LeBlanc2015,WangS2018}. 

We can also consider a 2d tensor network state, called Projected Entangled-Pair State (PEPS) or Tensor Product State (TPS), to calculate 2d lattice models \cite{NishinoHOMAG2001,VerstraeteC2004}. \black{Hereafter, we simply refer to such a state as PEPS.} Although it might be more suitable than DMRG for simulating 2d systems because PEPS can cover 2d area law of the entanglement entropy, it has been widely used for direct simulations of infinite 2d systems. \black{Furthermore, as we see later, for a small 2d systems treated in this study, the DMRG runs more efficiently.}

One problem with applying DMRG to 2d systems is that its computation cost may increase exponentially as we increase the system size $N_{\rm site}$ while keeping the same accuracy in the ground state energy. Because the computation cost of PEPS is expected to increase polynomially as we increase $N_{\rm site}$, it can be more efficient than DMRG for larger $N_{\rm site}$. However, as we see in the main text, the crossover between classical and quantum computations seems to occur at the relatively small systems for the models we treated. \black{For such small systems, the computation cost of the DMRG is indeed smaller than that of PEPS.} Thus, we believe DMRG gives proper estimations of computational resources to simulate 2d systems in classical computers. 

The computation time to obtain the ground state energy within a given target accuracy can depend on both the bond-dimension $D$ and the optimization steps. Although the error in a ground state energy calculated by DMRG or \black{variational PEPS} is expected to decrease exponentially as a function of $D$ asymptotically, its general relationship depends on the details of the target models. Similarly, it is not easy to estimate the necessary optimization steps {\it a priori}. Thus, to estimate the problem sizes where quantum computation will be superior to DMRG, we need to perform actual classical computations. In DMRG simulation, the computation time for the one optimization step, usually called sweep, scales as $O(N_{\textrm{site}}D^3)$, where $N_{\textrm{site}}$ is the number of lattice sites. 

\black{In the case of PEPS, two types of optimization algorithms have been widely used: imaginary time evolution \cite{VerstraeteC2004, JiangWX2008, JordanOVVC2008} and the variational optimization similar to DMRG algorithm \cite{Corboz2016, VanderstraetenHCV2016, LiaoLWX2019}. 
Usually, the yields more accurate simulation of ground states, although the environment tensor networks and contraction of them in the variational optimization becomes more complicated than the case of 1d tensor network. The recent development of the technique based on the automatic differentiation \cite{LiaoLWX2019} provides us with a more stable implementation of the variational optimization for PEPS. Thus, to estimate the computation time for PEPS, here we employ the variation optimization with the automatic differentiation. In this case, the computation time for the one optimization step scale as $O(N_{\textrm{site}}^\alpha D^\beta)$, where $\alpha = 1 \sim 1.5$ and $\beta=10 \sim 12$ depending on the approximate contraction algorithm. }

\subsection{Error analysis and computation time estimate for two-dimensional models by DMRG}
\begin{figure}[tbp]
    \centering
    \includegraphics[width=0.3\linewidth]{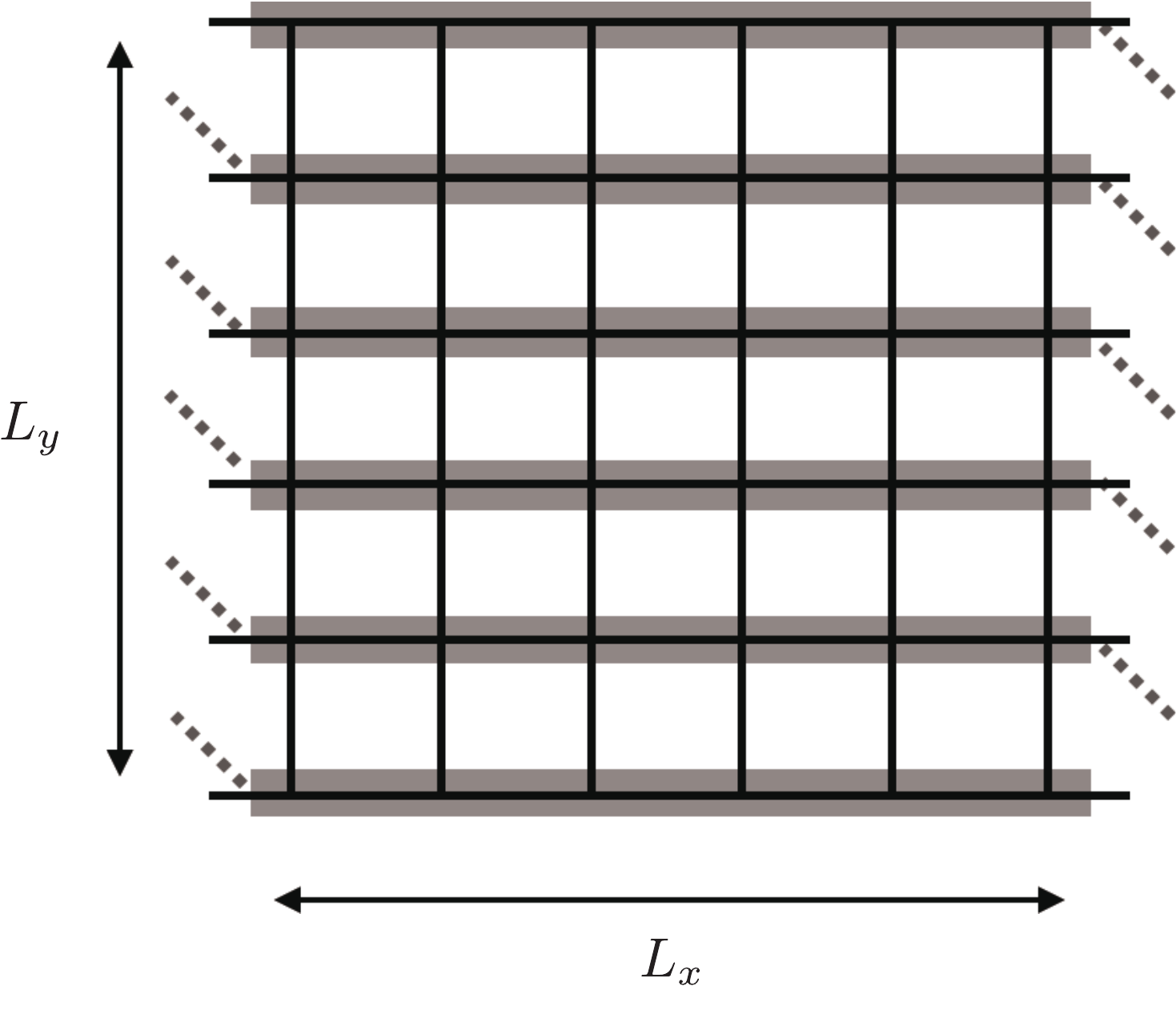}
    \caption{$L_x \times L_y$ square lattice used in the DMRG calculations. We consider the periodic boundary conditions along $y$ direction and the open boundary condition along $x$ direction. The gray lines indicate the assignment of MPS to the lattice sites.}
    \label{fig:MPS_Square}
\end{figure}

In this subsection, we explain our procedure to estimate the computation time for the ground state energy estimation in 2d systems.
For both the $J_1$-$J_2$ Heisenberg and the Hubbard models, we consider $L_x \times L_y$ square lattice with the periodic boundary condition along $x$ direction and the open boundary condition at $y$ direction so that the system forms a cylinder. In the DMRG calculation, the MPS string wraps around the cylinder as shown in Fig.~\ref{fig:MPS_Square}. 

To estimate the computation times in the ground state calculation by a classical computer, we performed DMRG simulation using ITensor \cite{itensor}. Typically, we set the maximum bond dimension as $D = D_{\mathrm{max}}$, and optimized the state starting from a random MPS with $D = 10$. We used $U(1)$ symmetric DMRG for the Heisenberg models, and similarly, we used particle number conservation for the Hubbard model. Most of the calculations have been done by a single CPU, AMD EPYC 7702, 2.0GHz, in the ISSP supercomputer center at the University of Tokyo. We performed multithreading parallelization using four cores. 

The estimation of computational time to reach desired total energy accuracy $\epsilon$ is performed by the following three steps.
Firstly, we optimize quantum states with various $D_{\mathrm{max}}$'s, and estimate the ground state energy by extrapolation. Then, we analyze the optimization dynamics of the energy errors. As we see in the main text, the dynamics of different $D_{\mathrm{max}}$ almost collapse into a single curve. Finally, we estimate the elapsed time to obtain a quantum state with the desired accuracy by extrapolating this scaling curve.
In the following, we elaborate on each step one by one.

\begin{figure}[tbp]
    \centering
    \includegraphics[width=0.9\linewidth]{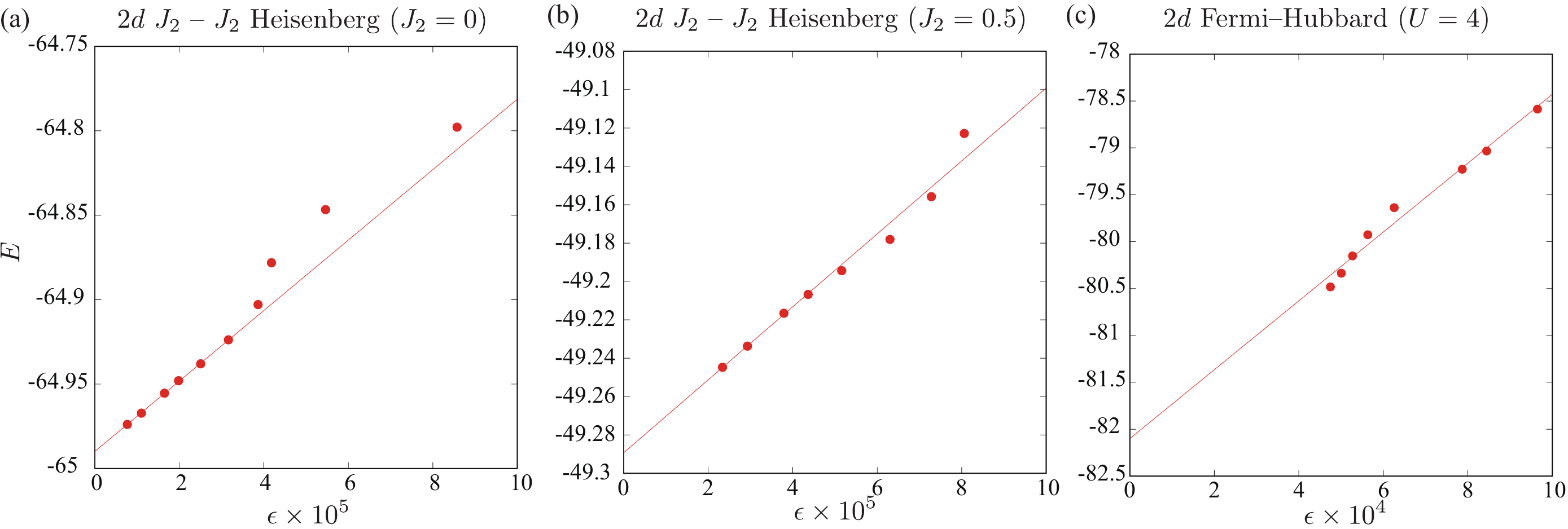}
    \caption{The extrapolations of the ground state energies for $10 \times 10$ square lattice models. Lines are the results of the linear fitting by Eq.~\eqref{eq:extrapolation_DMRG}(a) $J_1$-$J_2$ Heisenberg model with $J_2 = 0$. (b) $J_1$-$J_2$ Heisenberg model with $J_2/J_1 = 0.5$. (c) The Hubbard model with $U/t = 4$. }\label{fig:extrapolation_DMRG}
\end{figure}

As described above, firstly, we estimate the ground state energy by varying $D_{\mathrm{max}}$. It is known that the energy difference between the optimized and the exact energies is proportional to the so-called {\it truncation error}, which is a quantity obtained during the DMRG simulation. Namely, by iteratively updating the low-rank approximation by tensors for the MPS representation, one sums up the neglected singular values to compute the truncation error $\delta$. 
For sufficiently small truncation errors, the energy difference is empirically known to behave as 
\begin{equation}
 E_{\mathrm{DMRG}}(D_{\mathrm{max}}) - E_{\mathrm{exact}} \propto \delta (D_{\mathrm{max}}),
\label{eq:extrapolation_DMRG}
\end{equation}
where $\delta (D_{\mathrm{max}})$ is the truncation error for the optimized MPS state with $D_{\mathrm{max}}$ \cite{Schollwock2005,WhiteH1993, LegezaF1996}.
We extrapolated the obtained energies of target models to the zero truncation error using Eq.~\eqref{eq:extrapolation_DMRG}.  Fig.~\ref{fig:extrapolation_DMRG} shows typical extrapolation procedures for the $J_1$-$J_2$ Heisenberg model and the Fermi-Hubbard model on $10\times 10$ square lattice. When the truncation errors are sufficiently small, we observe the expected linear behaviors as in Figs.~\ref{fig:extrapolation_DMRG}(a) and (b). However, in the case of the Hubbard model (Fig.~\ref{fig:extrapolation_DMRG}(c)), the fitting by the linear function is not so precise. This is due to the limitation of the maximum bond dimension treated in our study. We also observed similar behavior in larger lattices in the $J_1$-$J_2$ model. Although the extrapolations seem to be not so accurate in such parameters, it is still sufficient to estimate the order of computation time, which is relevant in comparison with quantum computations.

\begin{figure}[tbp]
    \centering
    \includegraphics[width=0.95\linewidth]{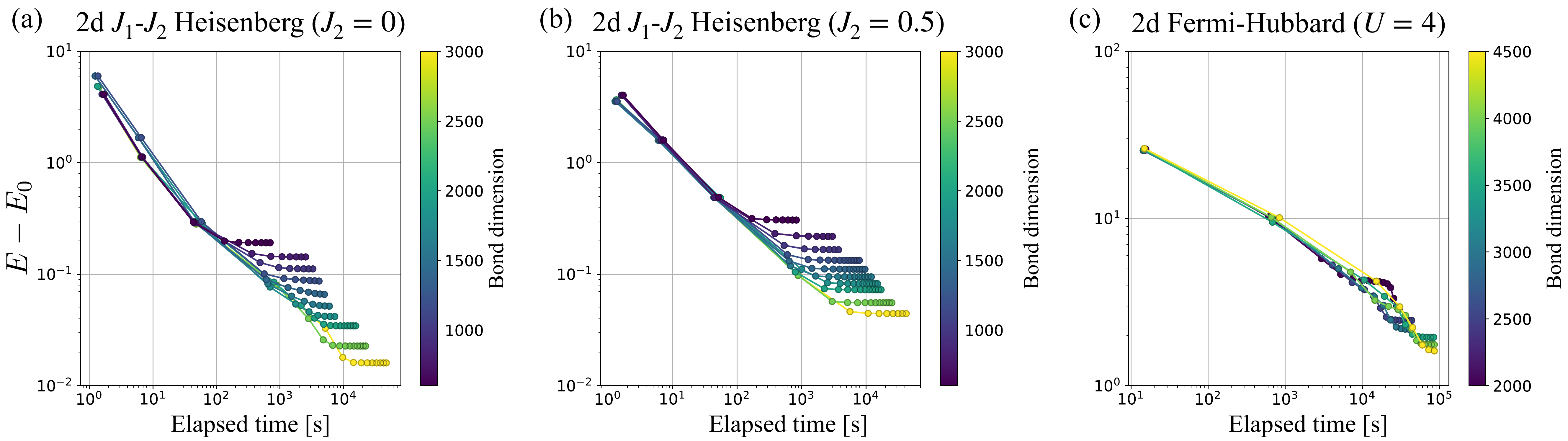}
    \caption{Dynamics of energy errors for (a) $J_1$-$J_2$ Heisenberg model with $J_2 = 0$. (b) $J_1$-$J_2$ Heisenberg model with $J_2/J_1 = 0.5$. (c) The Hubbard model with $U/t = 4$ in  $10 \times 10$ square lattice models with various $D_{\mathrm{max}}$. }\label{fig:dynamics_DMRG}
\end{figure}

Using the estimated value of ground state energy $E_0$, next we focus on the dynamics of the energy errors for each calculation. In Fig.~\ref{fig:dynamics_DMRG}, we show typical optimization dynamics as functions of the elapsed time. In the case of $J_1$-$J_2$ Heisenberg models with $J_2/J_1 =0, 0.5$, we see that curves corresponding to various $D_{\mathrm{max}}$ almost collapse to a universal optimization curve. Because the universal curve is likely a power function, we probably extrapolate optimization dynamics beyond the maximum value of $D_{\mathrm{max}}$ treated in the calculations properly. In the case of the Hubbard model with $U/t = 4$, we did not see a perfect collapse into a universal curve. In addition, the optimization curve seems to be deviated from the power law. Similar behaviors are also observed in other combinations of $L$ and $U/t$ for the Hubbard model. Although the extrapolation to larger $D_{\mathrm{max}}$ becomes inaccurate for such cases, to estimate the order of the computation time, we performed a power-law fitting of $E - E_0$ for the largest $D_{\mathrm{max}}$ before the saturation.

\begin{figure}[tbp]
    \centering
    \includegraphics[width=0.8\linewidth]{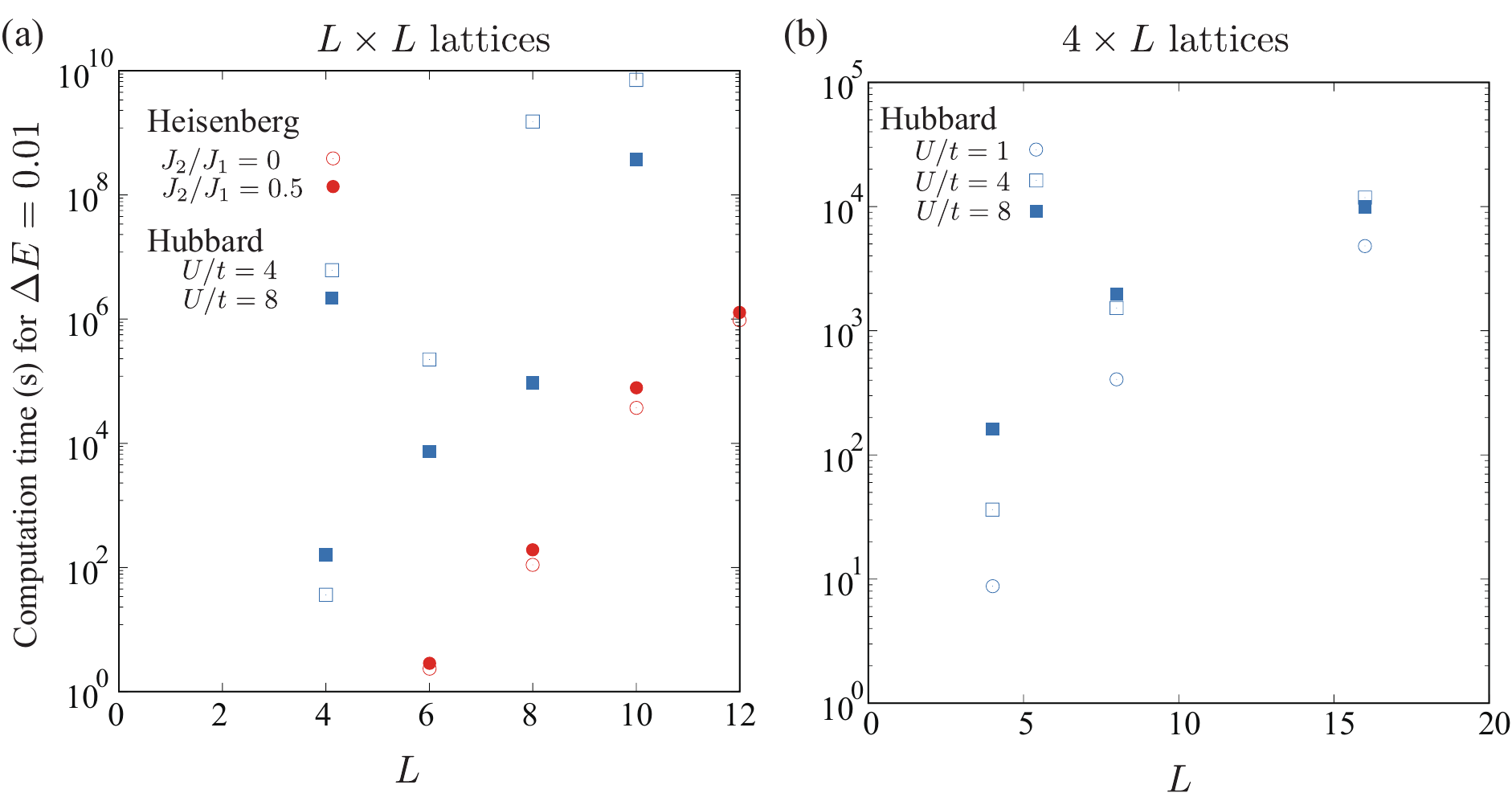}
    \caption{Estimated computation time to obtain the ground states with the total energy error $\Delta E \equiv E - E_{\mathrm{exact}} = 0.01$. (a) The estimated computation time for the $J_1$-$J_2$ Heisenberg and Hubbard models for $L \times L$ square lattices. (b) The estimated computation time for the Hubbard models on $4 \times L$ quasi-1d square lattices. }\label{fig:elapsedtime_DMRG}
\end{figure}

Based on such power-law fittings of the maximum $D_{\mathrm{max}}$ dynamics, finally, we estimate the computation time to obtain the ground state with the total energy error \black{$\epsilon:=E - E_{\rm exact} = 0.01$, where the unit of the energy is $J_1$ for the Heisenberg model and $t$ for the Hubbard model.
Since $\epsilon$ itself cannot be obtained in the simulation, we define the deviation from the extrapolated ground state energy as $\Delta E := E - E_0$, and assume $\Delta E \sim \epsilon$ such that $\Delta E$ is sufficiently accurate to predict the total runtime.
}
Fig.~\ref{fig:elapsedtime_DMRG} shows the estimated computation times for obtaining \black{$\Delta E = 0.01$} in classical DMRG simulations. The parameters for the DMRG calculations, together with the estimated computation times, are summarized in Tables \ref{table:param_Heisenberg_DMRG} and \ref{table:param_Hubbard_DMRG}. In the case of $L\times L$ geometry (Fig.~\ref{fig:elapsedtime_DMRG}(a)), we see an expected exponential increase of the elapsed time as we increase $L$. 
We find that the nearest neighbor Heisenberg model ($J_2/J_1 = 0$) needs a slightly shorter elapsed time than that of the frustrated model ($J_2/J_1 = 0.5$). 
Although this is related to the difficulty of the problem, the difference is not so large when we consider the orders of the elapsed times. 
We observe a rather larger difference between the Heisenberg and the Hubbard models, which is mainly attributed to the difference in the number of local degrees of freedom. In particular, the simulation in the Hubbard model is more time-consuming at $U/t = 4.0$ than that of the $U/t = 8.0$. This is probably explained by the larger energy gap in $U/t = 8.0$, which indicates lower entanglement in the ground state. 

It may be informative to mention that the runtime under the quasi-1d geometry $~(4 \times L)$ in the Hubbard model does not increase exponentially (See Fig.~\ref{fig:elapsedtime_DMRG}(b)). Indeed, in this geometry, we expect 1d area law of the entanglement entropy even when we increase $L$, and therefore, the required bond dimension $D$ to achieve a given target accuracy in the energy density becomes almost independent of $L$. Thus, we expect that the elapsed time to obtain $\epsilon = 0.01$ in the total energy increases in power law as a function of $L$.

\begin{table}[h]
 \centering
 \caption{Parameters for estimating the computation time in DMRG simulations of $J_1$-$J_2$ Heisenberg model.}\label{table:param_Heisenberg_DMRG}
 \begin{tabular}{c|c|c|c|c|c|c}
  $J_2/J_1$& $L_x$ & $L_y$ & largest $D_{\mathrm{max}}$& lowest energy & $E_0$ & Estimated computation time (s) for $\Delta E = 0.01$\\
  \hline
  \hline
  0& 6 & 6 & 800& -23.090068& -23.090117& $2.3$\\
  0& 8 & 8 & 3000& -41.348977& -41.349310& $1.1 \times 10^2$ \\
  0& 10 & 10 & 3000& -64.973863& -64.989896& $3.7 \times 10^4$ \\
  0& 12 & 12 & 2500& -93.847557& -93.951521& $9.6 \times 10^5$\\
  0.5& 6 & 6 &2000& -17.844007& -17.844015& $2.9$\\
  0.5& 8 & 8 & 1200& -31.538017& -31.544878& $1.9 \times 10^2$\\
  0.5& 10 & 10 & 3000& -49.244805& -49.289216& $7.9 \times 10^4$\\
  0.5& 12 & 12 & 2500& -70.699958& -70.782728& $1.3 \times 10^6$\\
  \hline
 \end{tabular}
\end{table}

\begin{table}[h]
 \centering
 \caption{Parameters for estimating the computation time in DMRG simulations of Hubbard  model.}\label{table:param_Hubbard_DMRG}
 \begin{tabular}{c|c|c|c|c|c|c}
  $U/t$& $L_x$ & $L_y$ & largest $D_{\mathrm{max}}$& lowest energy & $E_0$ & Estimated computation time (s) for $\Delta E = 0.01$\\
  \hline
  \hline
  1.0& 4 & 4 & 2500& -21.175239& -21.175265& $8.8$\\
  1.0& 4 & 8 & 3500& -43.486042& -43.487663& $4.1 \times 10^2$ \\
  1.0& 4 & 16 & 3000& -88.230867& -88.240000& $4.8 \times 10^3$ \\
  4.0& 4 & 4 &2500& -12.801348& -12.801536& $3.6 \times 10$\\
  4.0& 6 & 6 & 3500& -29.081849& -29.302504& $2.2 \times 10^5$\\
  4.0& 4 & 8 & 3500& -26.674309& -26.676826& $1.5 \times 10^3$\\
  4.0& 8 & 8 & 5000& -52.008298& -52.802798& $1.5 \times 10^9$\\
  4.0& 10 & 10 & 4500& -80.482124& -82.098947& $7.1 \times 10^9$\\
  4.0& 4 & 16 & 3500& -54.495584& -54.509963& $1.2 \times 10^4$\\
  8.0& 4 & 4 &2500& -7.660925& -7.660942& $1.6 \times 10^2$\\
  8.0& 6 & 6 & 3500& -17.737197& -17.767700& $7.4 \times 10^3$\\
  8.0& 4 & 8 & 3500& -16.079582& -16.079696& $2.0 \times 10^3$\\
  8.0& 8 & 8 & 5000& -31.806322& -31.993896& $9.4 \times 10^4$\\
  8.0& 10 & 10 & 5000& -49.287802& -51.699085& $3.7 \times 10^8$\\
  8.0& 4 & 16 & 3500& -32.922474& -32.923098& $9.9 \times 10^3$\\
  \hline
 \end{tabular}
\end{table}

In the above estimations, we used a single CPU. In practical calculations, we may expect a speed-up by GPU or parallelization using many CPUs. Indeed, in our simulation based on ITensor, we observed tens times faster execution by GPU when we did not impose the symmetries. Although ITensor simulation with the symmetry did not run with GPU, unfortunately, this observation indicates the estimated elapsed time can be shorter with a factor of $10^{-1}$. In the case of multi-node parallelization, we may consider the real-space decomposition of the DMRG algorithm \cite{StoudenmireW2013}. In this approach, each CPU optimizes different regions of the lattice (MPS), and communicates with each other to exchange information of the tensors. Because this approach needs to keep a sufficiently large block (set of tensors) for each CPU to obtain meaningful speed up, in the present system sizes, we divide the system into dozens at most. Thus, we estimate that the multi-node parallelization may give us a shorter elapsed time with a factor of $10^{-1}$.

We remark that here, we estimated the computation time to actually obtain a quantum state whose energy error is within $\epsilon = 0.01$ so that other observables can be extracted consistently. 
Meanwhile, the runtime becomes much shorter if we merely focus on estimating the energy within the error $\epsilon = 0.01$ since the zero-truncation extrapolation within error $\epsilon$ itself can be done with smaller $D$.  
Indeed, we have observed that extrapolations allow us to reach smaller errors than the energies obtained from the actual DMRG simulations.
This fact indicates that we might need shorter computation time to estimate the ground state energy within $\epsilon = 0.01$ in classical computers, although we point out that such a data analysis may also be available in quantum algorithms as well.

\black{\subsection{Error analysis and computation time estimate for two-dimensional models by PEPS}}
\black{In this subsection, we explain our procedure to estimate the computation time for the ground state energy estimation in 2d $J_1$-$J_2$ Heisenberg model. Different from the DMRG calculations, we consider $L_x \times L_y$ square lattice with the open boundary condition in both $x$ and $y$ directions due to the restriction of PEPS contraction efficiency. We represent the ground state of such a system by a simple PEPS where we put a local tensor on each vertex of the square lattice.}

\black{To estimate the computation time in the ground state calculation by PEPS, we performed variational optimization using QUIMB \cite{gray2018quimb}. Initially, we set random tensors with the bond dimension $D$\black{, which is fixed during the entire optimization process.
In order to compute expectation values and their derivatives to perform optimization, one must aim to compute the expectation values in an approximate sense, 
since the exact contraction of tensors in PEPS is a \#P-hard problem~\cite{schuch2007computational}. 
Namely, one must introduce environment tensors with bond dimension $\chi$ such that we can avoid superexponential computational cost but still maintain simulation accuracy.
In this regard, we emphasize that  simulation by PEPS is strictly different from DMRG; the calculation of the expectation value for a given PEPS contains errors that can be systematically controlled by using larger $\chi$. 
In our work, we have searched the appropriate value of $\chi \geq D^2$ so that the approximation error is sufficiently suppressed. 
}
Most of the calculations have been done by a single CPU, Intel Xeon Platinum 8280 2.7GHz (28core), in the ISSP supercomputer center at the University of Tokyo. We performed multithreading parallelization using 28 cores.}

\black{Similar to the DMRG calculations, the estimation of the computation time to reach desired total energy accuracy $\epsilon$ is performed by the following steps. Firstly, we calculate the exact ground state energy for the open boundary system by DMRG with the extrapolation used in the previous subsection. Then, we analyze the elapsed time scaling of energy errors for various $D$'s and $\chi$'s. As we will see later, the optimization curves of various $D$'s and $\chi$'s do not collapse into a single curve when we plot them as a function of the elapsed time. Thus, in the case of PEPS, we extrapolate the curve of $D = 4$ to $\Delta E = \epsilon$, and estimate the approximate lower bound of the computation time. 
\black{Considering that the bond dimension is expected to scale  as $D = O(\log N)$ in the asymptotic limit in our target models, we envision that the runtime scaling with $N$ becomes even worse than what we have observed in our work.}
}

\begin{figure}[tbp]
    \centering
    \includegraphics[width=0.95\linewidth]{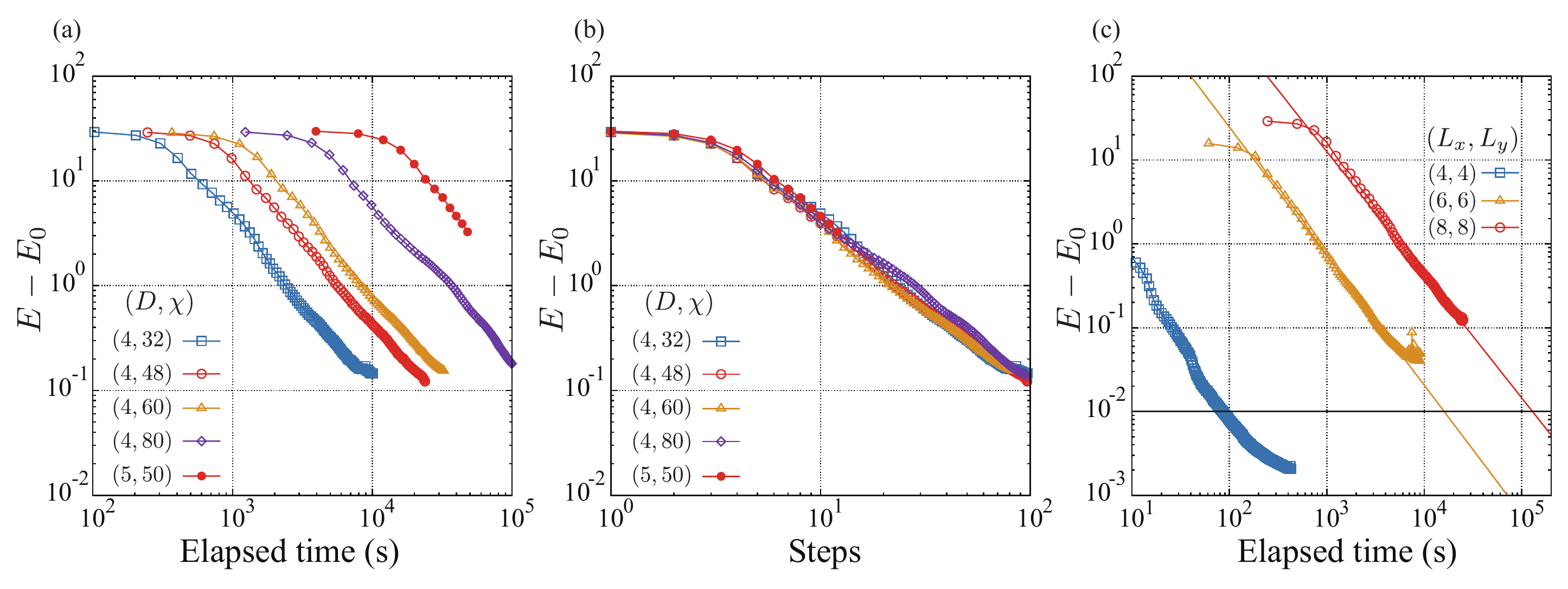}
    \caption{(a) \black{Optimization curves} of energy  for $J_1$-$J_2$ Heisenberg model with $J_2 = 0.5$ on $(L_x, L_y) = (8, 8)$ square lattice with various $D$ and $\chi$. (b) The energy error as a function of steps. (c) Typical optimization curves for $(L_x, L_y) = (4, 4)$, $(6,6)$, and $(8, 8)$ with $D=4$ and $\chi =48$.}\label{fig:2dJ1J2_PEPS}
\end{figure}

\black{In Fig.~\ref{fig:2dJ1J2_PEPS}(a) we show the \black{optimization curves} of energy for $J_1$-$J_2$ Heisenberg model with $J_2=0.5$ on $(L_x, L_y) = (8, 8)$ square lattice with various $D$ and $\chi$. Different from the case of the DMRG simulation, the optimization curves of PEPS method  do not show a collapse among different $D$ and $\chi$. 
One of the reasons for the difference is that we fix the bond dimensions $D$ during the entire optimization process; wed do not dynamically increase during the simulation in the case of the PEPS method. In contrast, in DMRG simulation, we \black{always start from a fixed value $D_0$ and dynamically increase $D$ according to a common scheduling until we reach $D_{\rm max}.$}
Thus, for sufficiently large $D_{\mathrm{max}}$, the bond dimension in the transient region is almost independent on $D_{\mathrm{max}}$ in DMRG method. Although we observe that the optimization curves in PEPS are dependent on $D$ and $\chi$, we still see a data collapse into a single curve when we plot the curves as a function of optimization steps, instead of the actual elapsed time, as shown in Fig.~\ref{fig:2dJ1J2_PEPS}(b). 
The obtained curve seems to be well described by a power-law scaling, which is in common with the case of DMRG method.}

\black{This almost bond-dimension independent power-law decay indicates that we can extrapolate the dynamics of energy errors toward our target $\Delta E = 0.01$ even if the actual simulation could not achieve this accuracy. Although we do not know the proper set of bond dimensions $D$ and $\chi$ to obtain $\Delta E = 0.01$, we can estimate the lower bound of the computation time by using smaller $D$ and $\chi$. Here we commonly use $D=4$ and $\chi =48$ for estimating such lower bounds for $(L_x, L_y) = (4, 4)$, $(6, 6)$, and $(8, 8)$. (Note that we clearly see the saturation of energy error for $D=4$ and $\chi =32$ in Figs.~\ref{fig:2dJ1J2_PEPS}(a). Thus we consider that $\chi =32$ is too small to obtain $\Delta E = 0.01$). The energy dynamics of these parameters are shown in Fig.~\ref{fig:2dJ1J2_PEPS} together with a fitting curve by the power law for $(6, 6)$ and $(8, 8)$. The extracted lower bound of the computation time to obtain the ground state with the total energy error $\Delta E = 0.01$ is shown in Table \ref{table:J1J2_PEPS}. Compared with the case of DMRG, the estimated computation times of PEPS are much longer for the present small sizes. Based on the present estimations, we can conclude that the computation time of DMRG is shorter than that of PEPS around the quantum-classical crossover, $L \simeq 10$, although the asymptotic scaling of the computation time of PEPS with respect to the system size $N_{\textrm{site}}$ is \black{expected to be } polynomial instead of exponential as in DMRG.}

\black{As in the case of DMRG, we may expect a speed-up of PEPS simulations by GPU or multi-node parallelization. Based on the similar consideration with DMRG, we estimated that these techniques may give us a shorter elapsed time with a factor of $10^{-1}$ for the present relatively small systems.}

\begin{table}[h]
 \centering
 \caption{Parameters for estimating the computation time in PEPS simulations of 2d $J_1$-$J_2$ Heisenberg model with $J_2 = 0.5$. The ground state energy $E_0$ was independently estimated from the extrapolation of the DMRG simulation.}\label{table:J1J2_PEPS}
 \begin{tabular}{c|c|c|c|c|c|c}
  $J_2/J_1$& $L_x$ & $L_y$ & $D$& $\chi$ & $E_0$ & Estimated computation time (s) for $\Delta E = 0.01$\\
  \hline
  \hline
  0.5& 4 & 4 &4& 48 & -7.505556& $93$\\
  0.5& 6 & 6 & 4& 48& -17.247352& $6.3 \times 10^4$\\
  0.5& 8 & 8 & 4& 48& -30.967671& $1.7 \times 10^5$\\
  \hline
 \end{tabular}
\end{table}

\begin{figure}[tbp]
    \centering
    \includegraphics[width=0.7\linewidth]{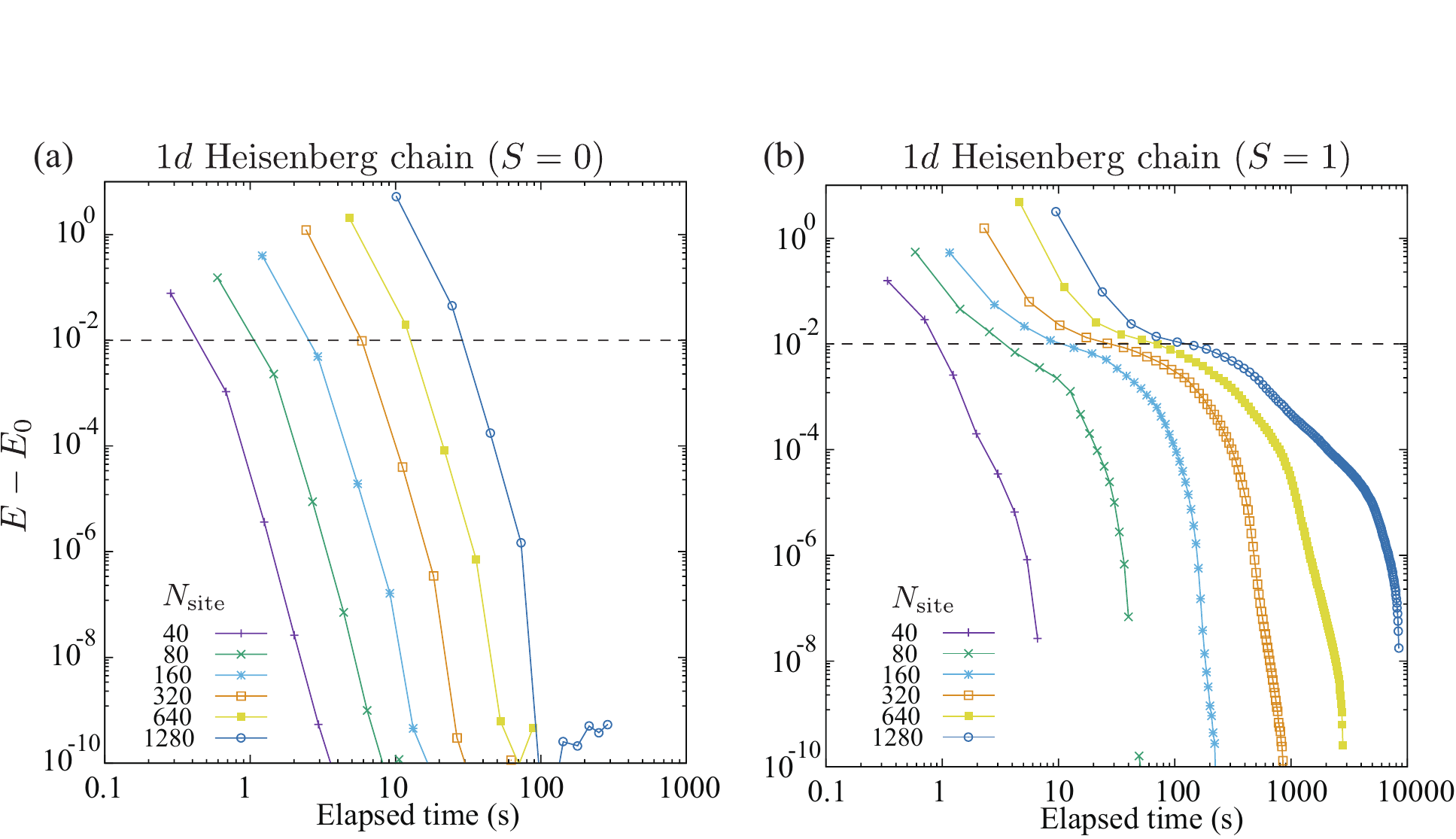}
    \caption{Dynamics of energy errors for $S=1$ Haldane chain for various $N_{\mathrm{site}}$. (a) $S=0$ sector (b) $S=1$ sector. Horizontal dashed line indicate $\epsilon = E -E_0 = 0.01$.}\label{fig:1DHeisenberg_DMRG}
\end{figure}

\subsection{Error analysis and computation time estimate for one-dimensional $S=1$ Heisenberg model}
In this subsection, we explain the runtime analysis of the 1d $S=1$ Heisenberg model. 
We consider the following Hamiltonian which is composed of $N_{\rm site}$ sites in total, with the edge sites being $S=1/2$ spins and the $N_{\rm site}-2$ bulk sites corresponding to $S=1$ spins as
\begin{equation}
 \mathcal{H} = \sum_{p=1}^{N_{\rm site}-3} 
 \bm{S}_p\cdot\bm{S}_{p+1} + 
 J_{\mathrm{end}}\left(\bm{s}_0\cdot \bm{S}_1 + \bm{S}_{N_{\rm site}-2}\cdot\bm{s}_{N_{\rm site}-1}
\right),
\label{eq:Haldane_DMRG}
\end{equation}
where $J_{\rm end}$ denotes the interaction amplitude of the $S=1$ and $S=1/2$ spins, \black{and also we have introduced dot product to denote the Heisenberg interaction as $\bm{S}_p \cdot \bm{S}_{q}:= \sum_{\alpha\in\{X, Y, Z\}} S_p^\alpha S_q^\alpha$.}
It has been pointed out that, thanks to the additional $S=1/2$ spins $\bm{s}_0$ and $\bm{s}_{N_{\rm site}-1}$, we avoid degeneracy between the total spin $S=0$ and $S=1$ sectors in the thermodynamic limit by taking $J_{\rm end}$ sufficiently large \cite{WhiteH1993, UedaK2011}. Here, we exclusively take $J_{\rm end}=0.7$ in the current work.
Since the entanglement entropy of the ground state is expected to obey 1d area law in this model, the DMRG algorithm is suitable to simulate larger system sizes. 
We also remark that, by extrapolating the energy gap between $S=0$ and $S=1$ sectors to $N_{\rm site} \to \infty$, we can estimate the famous Haldane gap \cite{Haldane1983}.

To estimate the runtime of the classical DMRG algorithm, we executed the optimization using ITensor library as in the previous sections \cite{itensor}. We used $U(1)$-symmetric DMRG to calculate the lowest energy states for $S_z=0$ and $S_z=1$ sectors, which we have confirmed to be the lowest energy states for $S=0$ and $S=1$, respectively.
In this model, we easily obtain sufficiently small truncation errors around $\delta(D_{\mathrm{max}}) \simeq 10^{-10}$ with $D_{\mathrm{max}} \lesssim 300$. Thus, we regard the lowest energy obtained in each sector along DMRG simulation as $E_0$ and see $\epsilon = E - E_0$. Most of the calculations have been done by a single CPU, 3.3GHz dual-core Intel Core i7 on MacBook Pro (13-inch, 2016). 

Fig.~\ref{fig:1DHeisenberg_DMRG} shows the dynamics of $\epsilon$ determined from the above procedure. 
In sharp contrast with the previous 2d models, we can easily reach $\epsilon \ll 10^{-2}$. In the case of the total $S=0$ sector, we expect a gapped ground state, 
and therefore the required $D_{\rm max}$ is almost independent of $N_{\mathrm{site}}$.
This nature is also reflected in the very rapid decay of $\Delta E$, as can be seen from Fig.~\ref{fig:1DHeisenberg_DMRG} (a). On the contrary, in the total $S=1$ sector, we expect the lowest energy state to be gapless. Thus, the entanglement entropy gradually increases as $\log N_{\mathrm{site}}$, and therefore we need to increase $D_{\mathrm{max}}$ as we increase $L$. This fact is related to relatively slower decays of $\epsilon$ in Fig.~\ref{fig:1DHeisenberg_DMRG}(b). 
We make another remark on the optimization dynamics in the total $S=1$ sector; here, we observed that $D_{\mathrm{max}}$ becomes the largest at an intermediate step of the optimization dynamics. Such a non-monotonic behavior in the bond dimensions might be possibly caused by local minima, which presumably arose by imposing symmetry on tensors of MPS ansatz: To escape from local minima, we may need a larger bond dimension. 

\begin{figure}[tbp]
    \centering
    \includegraphics[width=0.95\linewidth]{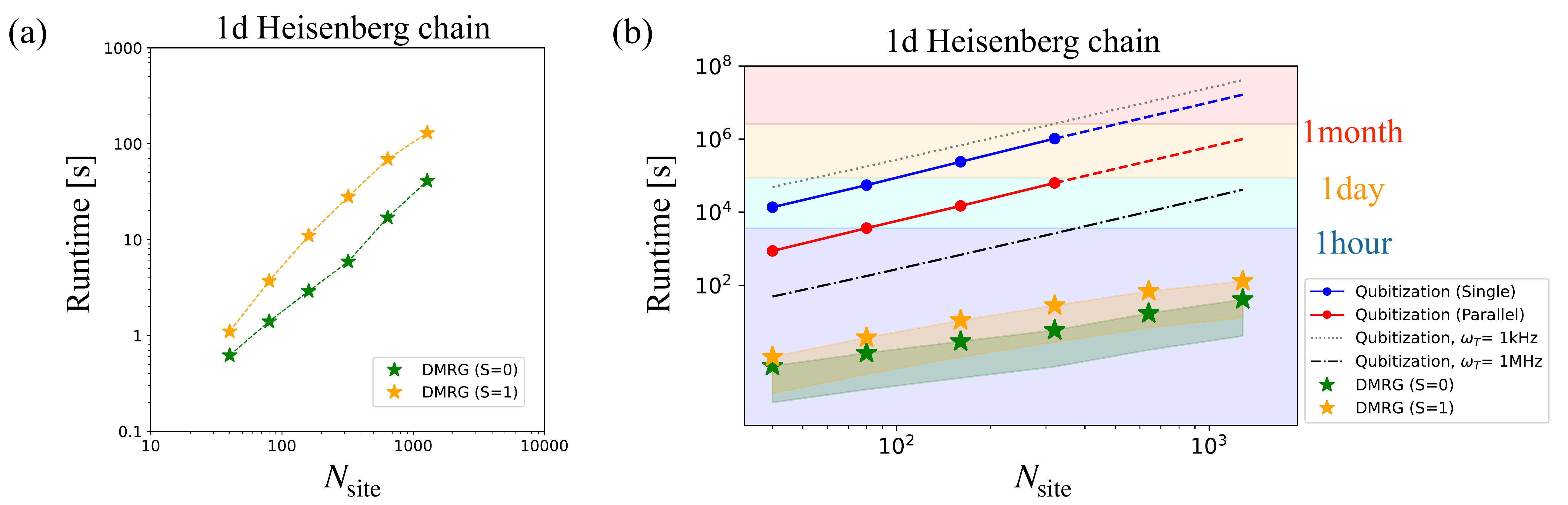}
    \caption{(a) Estimated computation time to obtain $\epsilon = 0.01$ in the $S=1$ Heisenberg chain defined in Eq.~\eqref{eq:Haldane_DMRG}. (b) Runtime comparison with qubitization-based quantum phase estimation. Notations follow those of Fig.~5 in the main text. }\label{fig:1DHeisenberg_time_DMRG}
\end{figure}

Finally, we estimate the computation time as the intersection of $\epsilon$ dynamics and $\epsilon = 0.01$. The obtained computation time is shown in Fig.~\ref{fig:1DHeisenberg_time_DMRG}(a), and the parameters in the simulation are summarized in Table \ref{table:param_Haldane_DMRG}. We see that the computation time is almost proportional to $N_{\mathrm{site}}$, which is expected from the almost constant $D_{\mathrm{max}}$ in the optimized state. We see that the $D_{\mathrm{max}}$ in the total $S_z=1$ sector slightly increases as we increase $N_{\mathrm{site}}$. This is probably related to the gapless nature of the lowest energy state within the total $S=1$ sector. Such weak $N_{\mathrm{site}}$ dependence also appears in the estimated computation time of the total $S_z=1$ sector, which is always longer than the computation time for the gapped ground state in the total $S = 0$ sector.

Let us remark that runtime analysis on the quantum algorithm reveals that the quantum advantage in this model is unlikely. In Fig.~\ref{fig:1DHeisenberg_time_DMRG}(b), we show the runtime in comparison with the qubitization-based quantum phase estimation~(whose detail is described in Sec.~\ref{sec:distselect_analysis} and~\ref{sec:total_resource}). 
Since the phase estimation algorithm aiming for the current $\epsilon$ requires runtime that scales at least quadratically with respect to the system size, the large separation already at the moderate size of $N_{\rm site} = O(10^2)$ implies that there is a very low chance of DMRG simulation being surpassed by the quantum algorithm.

\begin{table}[h]
 \centering
 \caption{Parameters for estimating the computation time in DMRG simulations of Haldane chain defined as Eq.~\eqref{eq:Haldane_DMRG} with $J_{\mathrm{end}} = 0.7$.}\label{table:param_Haldane_DMRG}
 \begin{tabular}{c|c|c|c|c}
  total $S_z$&$N_{\mathrm{site}}$& $D_{\mathrm{max}}$ in the optimized state& lowest energy &  Estimated computation time (s) for $\Delta E = 0.01$\\
  \hline
  \hline\
  0& 40 & 90& -53.293707633101& $0.62$\\
  0& 80 & 90& -109.35306916448&  $1.4$\\
  0& 160 & 90& -221.47179222696&  $2.9$\\
  0& 320 & 90& -445.70923835198& $5.9$\\
  0& 640 & 90& -894.18413060185& $17$\\
  0& 1280 & 90& -1791.13391510219& $41$\\
  1& 40 & 134& -52.861104764355&  $1.1$\\
  1& 80 & 156& -108.934890997408& $3.7$\\
  1& 160 & 169& -221.058987851845& $11$\\
  1& 320 & 175& -445.298116034289& $28$\\
  1& 640 & 177& -893.773481949644&  $69$\\
  1& 1280 & 179& -1790.7233920146&  $1.3 \times 10^2$\\
  \hline
 \end{tabular}
\end{table}

\section{Estimation of cost on ground state preparation}\label{sec:state_preparation}

\begin{figure}[thbp]
    \centering
    \includegraphics[width=0.45\linewidth]{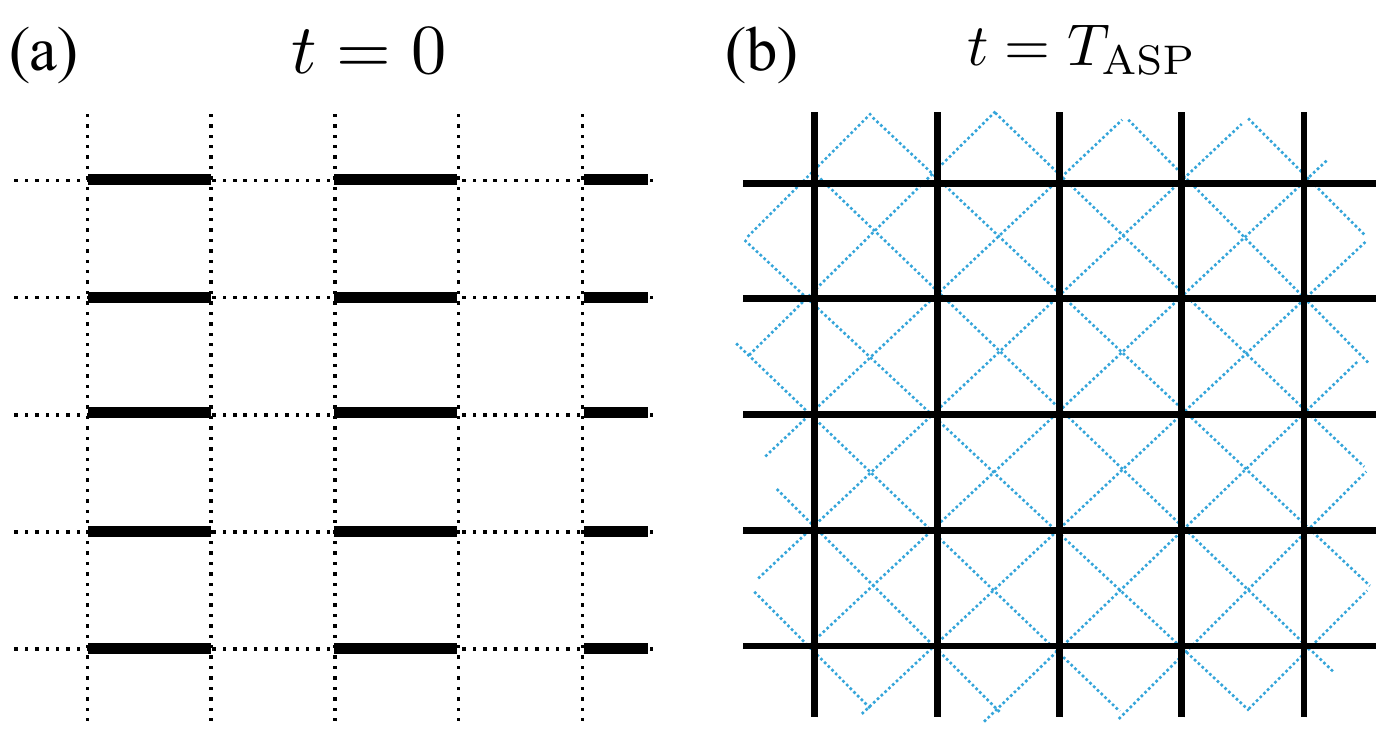}
    \caption{Schematic picture of the initial and target Hamiltonian of the ASP for 2d $J_1$-$J_2$ Heisenberg Hamiltonian on square lattice. (a) Interaction configuration $E_0$ of the initial Hamiltonian $H_0$.
    (b) Interaction configuration of the target Hamiltonian. The black lines denote the nearest-neighbor $J_1$ interaction, where the blue dashed lines represent the $J_2$ interaction.
    }\label{fig:asp_2dJ1J2_hamiltonian}
\end{figure}

\black{In this section, we provide details regarding the numerical simulation of Adiabatic State Preparation (ASP) of ground state.
As in the main text, we consider the ASP up to infidelity $\epsilon_f = 1 - |\braket{\psi_{\rm GS}| \psi(t_{\rm ASP})}|$ where $\ket{\psi_{\rm GS}}$ is the exact ground state and $\ket{\psi(t)}$ is a quantum state that is realized via the following time-dependent Schr\"{o}dinger equation:
\begin{eqnarray}
    i\frac{\partial }{\partial t} \ket{\psi(t)} = H(t)\ket{\psi(t)}.
\end{eqnarray}
Here, the time-dependent Hamiltonian is defined as $H(t)=H(s(t)) = sH_f + (1-s)H_0$ for the target Hamiltonian $H_f$ and the initial Hamiltonian $H_0$, and the smoothness of the dynamics is controlled by the interpolation function $s(t):\mathbb{R} \mapsto [0, 1]~(s(0)=0, s(t_{\rm ASP})=1)$.
For the simulation of 2d $J_1$-$J_2$ Heisenberg model of $J_2$=0.5, we have taken the initial state to denote a nearest-neighbor dimer covering $E_0$ as shown in Fig.~\ref{fig:asp_2dJ1J2_hamiltonian} as 
\begin{eqnarray}
    H_0 = J_1 \sum_{(p, q) \in E_0} \sum_{\alpha \in \{X, Y, Z\}} S_p^\alpha S_q^\alpha.
\end{eqnarray}
}

\black{It has been argued that the smoothness of the interpolating function $s(t)$ plays a crucial role to suppress nonadiabatic transition, such that we obtain logarithmic dependence $t_{\rm ASP} \propto \log(1/\epsilon_f)$ for carefully designed $s(t)$~\cite{ge2016rapid}. While the bounds are derived under assumptions that $s(t)$ belongs to the Gevrey class (whose derivative at $t=0, t_{\rm ASP}$ vanishes up to infinite order), it is practically sufficient to consider functions with vanishing derivative up to finite order (see Fig.~\ref{fig:2dJ1J2_ASP}(a) in the main text).
In this work, we have studied three types of interpolating functions as
\begin{eqnarray}
    s(t) = \begin{cases}
        t/t_{\rm ASP} \\
        \mathcal{S}_{\kappa}(t/t_{\rm ASP}) \\
        \mathcal{B}_{\kappa}(t/t_{\rm ASP})
    \end{cases}
\end{eqnarray}
where $\mathcal{S}_{\kappa}(x) := \sin^2(\frac{\pi}{2}\mathcal{S}^{\kappa-2}(x))$~($\mathcal{S}_{0}(x) = \frac{\pi}{2}x$). The last line is called the incomplete Beta functions that is defined as 
\begin{eqnarray}
 \mathcal{B}_\kappa(x) := \frac{B_{x}(1+\kappa, 1+\kappa)}{B_1(1+\kappa, 1+\kappa)}
\end{eqnarray}
with $B_\lambda(a, b):= \int_{0}^\lambda z^{a-1}(1-z)^{b-1}dz$.
The superscript $\kappa$ denotes that the derivatives of functions $\mathcal{S}_{\kappa}$ and $\mathcal{B}_{\kappa}$ vanishes up to $\kappa$-th order at $t=0, t_{\rm ASP}$.
Obviously we have $\kappa=0$ for the linear interpolation.
}

\black{
The numerical simulation of dynamics can be done by the using a standard technique to solve ordinary differential equation. We have concretely used the {\tt sesolve} function in the Library QuTiP~\cite{qutip}, which is based on the zvode routine provided by SciPy~\cite{scipy}.
On the other hand, the dynamics simulation of the MPS state have been done following the time-dependent variational principle (TDVP)~\cite{haegeman2011time} solver implemented in ITensor. 
In our simulation, the time evolution is sliced into a time step of $\delta t = 0.05$, and have kept the  maximum bond dimension to be $D=512$ since the DMRG simulations for current setups suggest $1 - |\braket{\psi_{D=512} | \psi_{D=3000}}|<10^{-3}$. This is sufficiently small compared to the target infidelity $\epsilon_f$ presented in Fig.~\ref{fig:2dJ1J2_ASP} in the main text, and therefore we safely take $D=512$ for the adiabatic time evolution.
}

\begin{figure}[thbp]
    \centering
    \includegraphics[width=0.85\linewidth]{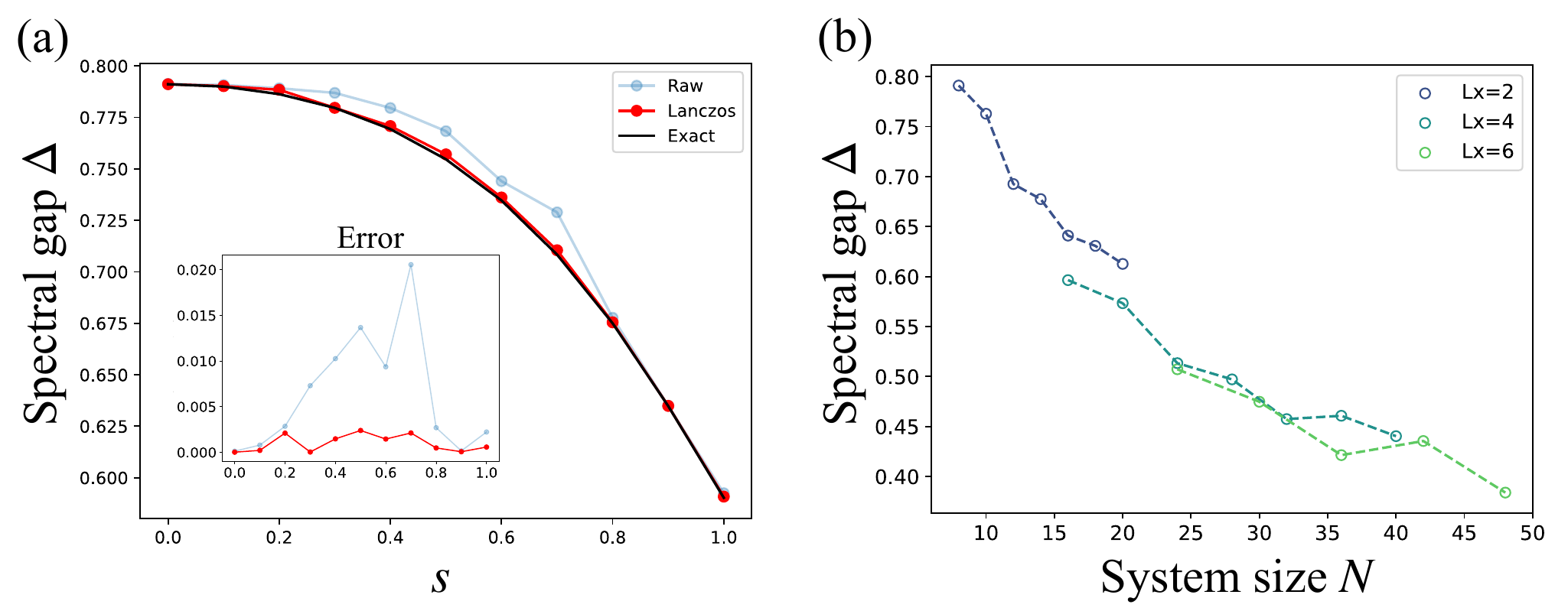}
    \caption{Estimation of spectral gap $\Delta$ between ground state and the first excited state in the symmetry sector $S=0$. (a) Improvement of accuracy along adiabatic path for 2d $J_1$-$J_2$ Heisenberg model on $4\times 4$ lattice with open boundary condition. The accuracy is improved by an order for all $s$. (b) Scaling of $\Delta$ with lattice size of $L_x \times L_y$ for $L_x=2, 4, 6$, where $L_y$ is taken as most $L_y \leq 10$.
    }\label{fig:asp_2dJ1J2_gap}
\end{figure}

\black{
Finally, we describe how to estimate the spectral gap using the MPS state.
Since the SU(2) symmetry is preserved along the entire ASP procedure, we focus on the spectral gap within the $S=0$ symmetry sector instead of other low-lying states such as the triplet excitations (i.e., the excited state with $S=1$).
We find that it is convenient to employ the penalty method to compute the first excited state in $S=0$ sector~\cite{Note3}.
Concretely, we modify the Hamiltonian $H$ as 
\begin{eqnarray}
    H_{\rm symm} = H + \upsilon S^2,
\end{eqnarray}
where $\lambda \in \mathbb{R}$ is the penalty coefficient for the total spin operator $S^2 = \sum_{\alpha \in \{X, Y, Z\}} (S_{\rm tot}^\alpha)^2 = \sum_{\alpha \in \{X, Y, Z\}} (\sum_p S_p^\alpha)^2$. 
We have exclusively used $\upsilon=5$ in our numerical simulation.
Once the $l$-th excited states of $H_{\rm symm}$ is simulated as $\ket{\phi_l},$ we perform a single Lanczos step so that 
\begin{eqnarray}
\ket{\Psi_{\rm ex}} = \sum_{l=1}^{l_{\rm max}} \eta_l \ket{\phi_l}    
\end{eqnarray}
denotes the first excited state with coefficients $\{\eta_l\}$.
For instance, a single Lanczos step with $l_{\rm max}=5$ improves the accuracy of $\Delta$ by an order for system size of $4\times 4$, and thus yields a reliable computational results (see Fig.~\ref{fig:asp_2dJ1J2_gap}).
The simulation is done with MPS of bond dimension $D \leq 3000$.
}

\section{Quantum algorithms for estimating energy spectra}\label{sec:phase_estimation}
\subsection{Overview on quantum phase estimation}
Quantum phase estimation (QPE) is a quantum algorithm designed to extract the eigenphase $\phi$ of a given unitary $U$ up to $m$ digits, using $m$ ancilla qubits that indirectly read out the complex phase of the target system.
More precise description of the algorithm originally proposed by Kitaev~\cite{kitaev1995quantum} is as follows: given an efficiently prepared trial state $\ket{\psi} = \ket{0}$ whose fidelity with the $k$-th eigenstate  $\ket{k}$ of the unitary is given as $f_k = \|\braket{k | \psi}\|^2$, a single run of QPE returns a random variable $\hat{\phi}$ which corresponds to a $m$-digit readout of $\phi_k$ with probability $f_k$. 
By plugging a time evolution operator $U=e^{-i H t}$ of Hamiltonian $H$ and take $\ket{\psi}$ to be efficiently-preparable approximation of the eigenstate, for instance, one can simulate the eigenspectrum of the target Hamiltonian $H$~\cite{Abrams99}.

In this work, we focus on the simulation of ground state and its eigenenergy, within some target accuracy, of a quantum many-body Hamiltonian considered in condensed matter physics.
Although the simulation of the ground state of a general Hamiltonian is a QMA-hard problem, we envision that quantum advantage can be realized in a practical sense, since the Hamiltonian simulation is BQP-complete and cannot be replaced be by any classical algorithm efficiently.
In this regard, our goal is to elucidate the quantum-classical crosspoint based on the actual runtime to prepare a many-body quantum state up to some fixed accuracy in terms of the energy.
The detailed procedure for classical algorithm has already been provided in Sec.~\ref{sec:dmrg}, and thus we detail on the quantum algorithm in the following.
In particular, we explain the abstract protocol of cost/runtime estimation strategy in Sec.~\ref{subsec:qpe_cost_general}, and then proceed to Sec.~\ref{subsec:hamiltonian_simulation_trotter},~\ref{subsec:hamiltonian_simulation_posttrotter} in order to provide different flavors of Hamiltonian simulation techniques, which constitute the main building block of the phase estimation algorithm.


\subsection{General cost estimation strategy for phase estimation} \label{subsec:qpe_cost_general}
As we have briefly introduced in the main text, the gate complexity for running the QPE can be abstractly given as
\begin{eqnarray}
    C_{\rm total} = C_{\rm SP} + C_{\rm HS} + C_{{\rm QFT}^\dag},\label{eqn:qpe_cost_full}
\end{eqnarray}
where the cost $C_{\rm SP}$ is for the state preparation of the approximate ground state, $C_{\rm HS} = 2^m C_{\rm CU}$ is controlled Hamiltonian simulation, and $C_{{\rm QFT}^\dag}$ for the inverse quantum Fourier transformation (QFT).
As we have explained in the main text, we exclusively focus on the cost from the second term of Eq.~\eqref{eqn:qpe_cost_full}, and discuss the runtime solely from it.

Estimation of the runtime of the phase estimation is performed via two steps. First, we perform a comprehensive comparison between multiple flavors of Hamiltonian simulation, which are roughly divided into Trotter-based and post-Trotter methods. Given the explicit quantum circuit structure in individual algorithms, we compute the $T$-count respectively to determine the best among the existing methods.
Note that such an estimation procedure is reasonable; in a situation when the logical Clifford operations can be efficiently implemented through stabilizer measurements and feedback, the runtime is dominated by the execution of $T$-gates, which is the most basic unit for non-Clifford operations under the surface code.
In the second step, which is detailed in Sec.~\ref{sec:distselect_analysis}, the runtime of the presumably best quantum algorithm is further analyzed by compiling the quantum circuit at the level of executable instructions. This yields a more precise runtime estimate with plausible quantum hardware setups.

\black{One shall keep in mind that there are multiple source of algorithmic error in phase estimation.
Our error analysis is based on findings in Ref.~\cite{Reiher201619152} that related the energy error $\epsilon$ with the error $\delta$ in the unitary circuit. For simplicity, in our work we have assumed $\delta = \epsilon/\lambda$ where $\lambda$ is the L1 norm of Pauli operator coefficients.
This total implementation error can be decomposed into several contributions as
\begin{eqnarray}
\delta =  \delta_{\rm PEA} + \delta_{\rm syn}
\end{eqnarray}
where $\delta_{\rm PEA}$ and $\delta_{\rm syn}$ denote the error from the phase estimation itself (due to finite readout digits) and the synthesis error, respectively.
Concretely, we have
\begin{eqnarray}
\delta_{\rm syn} = \begin{cases}
\delta_{\rm HS} + \delta_{\rm Rot}~(\text{qDRIFT})\\
\delta_{\rm HS} + \delta_{\rm Rot}~(\text{random Trotter})\\
\delta_{\rm HS} + \delta_{\tt PREP}~(\text{Taylorization})\\
\delta_{\tt PREP}~\ \ \ \ \ \ \ \ ~(\text{Qubitization})
\end{cases},
\end{eqnarray}
where $\delta_{\rm HS}$ is the error in approximating the Hamiltonian simulation unitary, $\delta_{\rm Rot}$ is the gate synthesis error for Pauli rotation, and $\delta_{\tt PREP}$ is the state synthesis error for {\tt PREPARE} oracle which is practically dominated by the rotation synthesis.
While previous works on Trotter-based method have revealed how to determine error budgets to minimize the error, there is no known closed-form expression that can analytically determine the ratio between error resources. We have performed numerical search on the error resource to find that we may practically fix as $\delta_{\rm PEA} = 0.9 \epsilon/\lambda$ and set other subdominant (i.e. logarithmically contributing) factors equally. Note that the scaling is inverse-linear with respect to $\delta_{\rm PEA}$, while for others it scales only polylogarithmically. This difference in the scaling already gives us sense that the dominant error resource shall be the finite-digit truncation in phase estimation.
}

In the remainder of this section, we introduce the Hamiltonian simulation techniques considered in our work, and also discuss their $T$-complexity and the effect of prefactors.

\subsection{Trotter-based Hamiltonian Simulation methods for phase estimation}\label{subsec:hamiltonian_simulation_trotter}
Among various flavors of Hamiltonian simulation algorithms for quantum computers, the ones based on the Suzuki-Trotter decomposition have been known from the earliest days.
Since universal quantum computers are mostly designed to execute quantum gates that operate only on constant number of qubits, they cannot directly implement the matrix exponentiation as $U = e^{- i H t}$ where we assume that the Hamiltonian is decomposed into sum of Pauli operators as $H=\sum_l w_l H_l$ (we tune the phase so that $w_l>0$).
The Suzuki-Trotter decomposition refers to the deterministic procedure that approximates the unitary evolution by using a sequence of Hermitian operators $\mathM = (..., M_j,...)$ to implement the product formula which is both spacially and temporally discretized as
\begin{eqnarray}
U := \exp\left( - i \sum_l w_l H_l t \right) \approx \left( \prod_j e^{-i M_j t/r}\right)^{r}.
\end{eqnarray}
For instance, in the case of the first-order product formula, the Hermitian operators are simply taken as $\mathM = (..., w_l H_l, ...)$ such that
\begin{eqnarray}
U_{{\rm Trotter}, 1} = \left( \prod_l e^{-i w_l H_l t/r}\right)^{r},
\end{eqnarray}
whose error from the exact implementation is bounded by the operator norm as 
\begin{eqnarray}
\left\|
U - U_{{\rm Trotter}, 1}
\right\| = \order\left(\frac{(L\Lambda t)^2}{r}\right).
\end{eqnarray}
Here, $L$ denotes the number of decomposition for Hamiltonian, $r$ is the number of repetitions, and $\Lambda = \max_l w_l$ is the largest coefficient. 
The higher order product formulae were also invented by Suzuki to improve the scaling with respect to the simulated time $t$~\cite{suzuki_general_1991}. 
For instance, the second order product formula is given by taking $\mathM = (w_1 H_1/2, ..., w_{L-1}H_{L-1}/2, w_L H_L, w_{L-1}H_{L-1}/2, ..., w_1 H_1/2)$ as 
\begin{eqnarray}
S_{2}(\tau) &:=& \prod_{l=1}^L \exp(w_l H_l \tau/2) \prod_{l=L}^1 \exp(w_l H_l \tau/2),\label{eq:second_trotter}
\end{eqnarray}
and further the $2k$-th order product formula is given by the following recursive procedure as
\begin{eqnarray}
S_{2k}(\tau) &:=& [S_{2k-2}(g_k \tau)]^2 S_{2k-2}((1-4g_k)\tau)[S_{2k-2}(g_k\tau)]^2,
\end{eqnarray}
where $ g_k = 1/(4-4^{1/(2k-1)})~(k>1)$.
By taking the time evolution unitary as $U_{{\rm Trotter}, 2k}=(S_{2k}(-it/r))^r$,
it follows that the error of the Hamiltonian simulation can be given as
\begin{eqnarray}
\left\| U - U_{{\rm Trotter},2k} \right\| = \order\left(\frac{(L\Lambda t)^{2k+1}}{r^{2k}} \right).
\end{eqnarray}

The product formula enables one to systematically improve the error with respect to, e.g., simulation time $t$. 
Meanwhile, when we count the gate complexity on quantum computers, the prefactor suffers from exponential increase with respect to the order $k$~\cite{campbell_random_2019}, and therefore is not preferred in practice.
An intriguing alternative is to introduce the randomization into the gate compilation~\cite{campbell_shorter_2017, hastings_turning_2017, childs_fasterquantum_2019}. 
For instance, it has been pointed out that, by randomly permuting the sequence of operators $\{H_l\}$ in Eq.~\eqref{eq:second_trotter} for the second-order product formula improves the error dependence on $L$~\cite{childs_fasterquantum_2019}. See Table~\ref{tab:trotter_scaling} for the comparison on the gate complexity and error scaling.

While the randomized Trotter methods have performed randomization concerning the order of the Pauli rotations, it was proposed by Campbell that we may stochastically choose the Pauli rotations according to their coefficients in the Hamiltonian to suppress the simulation cost for Hamiltonians that are composed of large numbers of Pauli with small coefficients~\cite{campbell_random_2019}.
To be concrete, according to the probability $p_l = w_l/\lambda$ where $\lambda:= \sum_l w_l$, we choose the Pauli operator $H_l$ and append the Pauli rotation $e^{-i H_l t/r}$ to the circuit, which is repeated for $r$ times.
It has been shown that the error scaling in such a technique, called the qDRIFT, is given rather by the L1 norm of the coefficients $\lambda$ rather than the number of Paulis $L$ [See Table~\ref{tab:trotter_scaling}.].

\renewcommand{\arraystretch}{2}
\begin{table}[tb]
    \centering
    \begin{tabular}{l|c|c}
    \hline
    Algorithm & Error & Gate complexity \\
    \hline
    Trotter, 1st order & $\order\left(\dfrac{(L \Lambda t)^2}{r}\right)$ & $\order\left(\dfrac{L^3(\Lambda t)^2}{\delta_{\rm HS}}\right)$ 
    \\[5mm]
    Trotter, $2k$-th order  & $
        \order\left(\dfrac{(L \Lambda t)^{2k+1}}{r^{2k}}\right)
        $
        & $\order\left(\dfrac{L^{2+1/2k}(\Lambda t)^{1 + 1/2k}}{\delta^{1/2k}_{\rm HS}}\right)$  \\[5mm]
    Random Trotter, $2k$-th order & 
    $\order\left(\dfrac{(L^{2k}(\Lambda t)^{2k+1})}{r^{2k}}\right)$& 
    $\order\left(\dfrac{L^2 (\Lambda t)^{1+1/2k}}{\delta^{1/2k}_{\rm HS}}\right)$\\[5mm]
    qDRIFT & $\order\left(\dfrac{2\lambda^2 t^2}{r}\right)$ & $\order\left(\dfrac{(\lambda t)^2}{\delta_{\rm HS}}\right)$ 
    \end{tabular}
    \caption{
        Gate complexity and error scaling of Trotter-based methods to perform Hamiltonian simulation for time $t$.
        }\label{tab:trotter_scaling}
\end{table} 
\renewcommand{\arraystretch}{1.0}

\renewcommand{\arraystretch}{2}
\begin{table}[tb]
    \centering
    \begin{tabular}{l|c|c|c}
    \hline
    Algorithm & $N_{\rm Rot}$ & $T$-count & $T$-complexity\\
    \hline
    Random, 2nd order Trotter & 
    $16\dfrac{\Lambda^3 L^2}{\epsilon^{3/2}}$ &
    $N_{\rm Rot} \times (\Gamma \log_2\delta_{\rm SS}^{-1} + \Xi)$ &  $\order\left(\dfrac{N^2}{\epsilon^2}\log (N/\epsilon^{3/4})\right)$ 
    \\[5mm]
    qDRIFT & 
    $35.5192\dfrac{\lambda^2}{\epsilon^2}~$ &
    $N_{\rm Rot} \times (\Gamma\log_2\delta_{\rm SS}^{-1} + \xi)$ & $\order\left(\dfrac{N^2}{\epsilon^2}\log (N/\epsilon)\right)$ 
    \end{tabular}
    \caption{
        $T$-count of Trotter-based methods to perform phase estimation up to energy accuracy $\epsilon$. Here we assume that the Pauli rotations are implemented with gate synthesis error \black{$\delta_{\rm SS} = \delta_{\rm syn}/2N_{\rm Rot}$ where $\delta_{\rm S}$ is the total gate synthesis error related with the total energy accuracy as $\delta_{\rm syn} = 0.1 \epsilon/\lambda$}.
        The number of controlled Pauli rotations are estimated under the assumption of using the Hodges-Lehmann estimator, from which the $T$-counts are calculated by multiplying the cost per controlled Pauli rotation (Also see Sec.~\ref{sec:cost_basic} for $T$-count of basic quantum operations).
        }\label{tab:trotter_tcount}
\end{table} 
\renewcommand{\arraystretch}{1.0}

\subsubsection{$T$-count in Trotter-based methods}
The error analysis on the Hamiltonian simulation eventually allows one to analyze the cost to perform the phase estimation up to a target energy accuracy $\epsilon.$ For instance, Campbell estimated the number of digits $m$ required to read out the energy with $\epsilon$ with success probability $1 - p_f$ as~\cite{campbell_random_2019}
\begin{eqnarray}
m = \log_2(\epsilon/2\lambda) + \log_2\left(\frac{1}{p_f} + 1\right)- 2.
\end{eqnarray}
By optimizing the gate synthesis accuracy for each control-$U^{2^{j}}$ such that the overall gate count is minimized under a fixed total error, one finds that the total number of rotation gates is given as
\begin{eqnarray}
N_{\rm Rot}\sim 133\frac{\lambda^2}{\epsilon^2 ((3/2)p_f)^3}.
\end{eqnarray}
While this analysis assumes to run the QPE for a single shot, Lee {\it et al.} introduced another layer of ``randomization"~\cite{lee_evenmore_2021}; by running the phase estimation multiple times, one may sample from the somewhat erroneous probability distribution, whose error can be efficiently suppressed by the Hodges-Lehman estimator to deal with the fat-tail errors. By assuming that the error in the output is symmetric, one obtains the improved number of rotations as 
\begin{eqnarray}
N_{\rm Rot} = 35.5192\frac{\lambda^2}{\epsilon^2}.
\end{eqnarray}
Given the number of rotations, one may estimate the $T$-count by multiplying with the gate synthesis cost for controlled rotations. See Table~\ref{tab:trotter_tcount} for the summary of the cost required to perform the phase estimation. Note that we also perform a similar analysis in the case of the 2nd order random Trotterization.

\subsection{Post-Trotter Hamiltonian simulation methods for quantum phase estimation}\label{subsec:hamiltonian_simulation_posttrotter}
One of the main bottlenecks of the Trotter-based Hamiltonian simulation techniques is that their gate complexity is polynomial regarding the dynamics simulation accuracy $\delta_{\rm HS}$.
Although one may in principle improve the gate-depth scaling as $O(\delta_{\rm HS}^{-1/2k})$ by employing $2k$-th order product formulas, one must pay the cost of significantly increased prefactor, which practically negates the advantage of the scaling in moderate-size problems.
This has been recognized as roadblock that prevents us from achieving quantum advantage in a practical sense.

A class of Hamiltonian simulation techniques that achieves exponential improvement with respect to the accuracy $\delta_{\rm HS}$ is called the ``post-Trotter methods."
While Trotter-based methods mainly consist of exponentiated Pauli rotations that act on the system qubits, in post-Trotter methods the class of available operations on the target system is vastly extended to include even non-unitary operations by making use of an extended Hilbert space. This technique is dubbed as {\it block-encoding}.
For instance, Ref.~\cite{berry_exponential_2014, berry_simulating_2015} proposed an algorithm that block-encodes the truncated Taylor series for the time evolution operator. The gate complexity to perform the phase estimation based on this so-called {\it Taylorization} method for lattice systems with size $N$ with energy accuracy $\epsilon$ is \black{$\frac{L \lambda}{\epsilon} \log(N/\delta_{\rm HS})/\log\log(N/\delta_{\rm HS})$}~\cite{babbush_exponentially_2016}.
A major innovation pointed out by Refs.~\cite{berry_improved_2018, poulin_quantum_2018} after the proposal of Taylorization is that one may completely eliminate the Trotter or Taylor errors in phase estimation by simulating $e^{i \arccos (H/\lambda)}$ instead of $e^{i Ht/\lambda}$. This idea has been combined with the concept of {\it qubitization}~\cite{low2016hamiltonian} to achieve even more efficient implementation of phase estimation algorithm~\cite{babbush_encoding_2018}.
As we later describe more in detail in Sec.~\ref{subsubsec:qubitization}, we construct oracles such that the operation of the block-encoding unitary can be ``qubitized" into a direct sum of two-dimensional subspace. Each of such ``single-qubit rotation" encodes the eigenvalue of the target operator, and therefore can be directly plugged in as a phase estimation subroutine.

In the following, we first introduce the concept of block encoding and its two core subroutines \prepare~and \select.
Then, we further introduce the qubitization and Taylorization algorithms, which are two representative post-Trotter methods to perform Hamiltonian simulation for phase estimation.

\subsubsection{Block encoding for linear combination of unitaries}
It is instructive to start from the definition of the term block encoding.
Let us define that block encoding of some bounded-norm target operator $A\in M_{d}(\mathbb{C})$~($d$: Hilbert space dimension) is realized when the action of an unitary operator in an extended Hilbert space $V \in M_{d+d_a}(\mathbb{C})$~($d_a$: auxiliary Hilbert space dimension) reproduces $A$ at specific subspace.
For concreteness, let us consider $n$-qubit target system with $a$ ancillary qubits. We say that
``$A$ is block-encoded into $V$ via signal state $\ket{\mathL}$" if the following holds:
\begin{eqnarray}
    (\bra{ \mathL} \otimes I^{\otimes n}) V (\ket{ \mathL} \otimes I^{\otimes n}) = A,~\label{eqn:block_encoding_operator}
\end{eqnarray}
where $\ket{\mathL}$ is a state in auxiliary $a$-qubit Hilbert space and $I$ denotes the identity.
Note that we have assumed that $\|A\| \leq 1$ without loss of generality, since we can rescale the operator as $A/\alpha~(\alpha\in\mathbb{R})$ if needed.
While we may also consider block encoding under multiple signal states as $\Pi V \Pi = A$ where $\Pi = \sum_{\mathL} \Pi_{\mathL} = \sum_{\mathL} \ket{\mathL}\bra{\mathL}$, here we focus on Eq.~\eqref{eqn:block_encoding_operator} for simplicity.
It is illustrative to take $\ket{\mathL} = \ket{0}^{\otimes a}$ and rewrite Eq.~\eqref{eqn:block_encoding_operator} into an equivalent expression that holds for any target state:
\begin{eqnarray}
    V &=& \left(\begin{array}{cc}
    A & *\\
    * & * \\
    \end{array}\right),\label{eqn:block_encoding}
\end{eqnarray}
Observe from Eq.~\eqref{eqn:block_encoding} why the technique is called ``block-encoding."
Since we can obtain the desired operation $A\ket{\psi}$ with finite probability by post-selecting the measured ancillary qubits to be $\ket{0}^{\otimes a}$, this is also referred to as ``probabilistic implementation of $A$."
In reality, we design the algorithm (e.g. use amplitude amplification) so that post selection is not explicitly required.

In phase estimation algorithm based on post-Trotter methods, one aims to extract information on the eigenspectra by making multiple queries to the oracular unitary which block-encodes the Hamiltonian.
The underlying assumption is that, there is a systematic and efficient way to encode the spectral information of the Hamiltonian into the oracular unitary $V$. 
As we later observe in the algorithms such as the qubitization and Taylorization, this question essentially falls into the problem of implementing two oracular subroutines named \prepare~and \select~efficiently.
Roughly speaking, \prepare~takes the ancillary quantum state to the properly block-encoded subspace, and \select~performs the block encoding of the Hamiltonian in the subspace given by the first oracle \prepare.

To explicitly write down the operation by two subroutines, 
first, let us note that any operator $A$ can be decomposed into a linear combination of unitaries as 
\begin{eqnarray}
    A &=& \sum_l w_l A_l~~(w_l > 0),\label{eq:LCU}
\end{eqnarray}
where $w_l$ is a real positive coefficient of the unitary operator $A_l$.
For $n$-qubit systems, $A_l$ can be thought to be a tensor product of multiple Pauli operators with arbitrary complex phase.
The first oracle \prepare~is often introduced to perform the following transformation that encodes all $L$ coefficients of the Hamiltonian into the ancillary qubits as
\begin{eqnarray}
    \ket{\mathL} :=\prepare \ket{0}^{\otimes a} &=& \sum_{l=1}^L \sqrt{\frac{w_l}{\lambda}} \ket{l},\label{eq:prepare_def}
\end{eqnarray}
where $\lambda = \sum_l w_l$ is the $L_1$ norm of the coefficients of Hamiltonian. 
Given the orthogonal bases $\{\ket{l}\}_l$ that encodes the information of the coefficients, next we construct the second oracle \select~so that the desired operator $A$ is realized in the subspace as 
\begin{eqnarray}\label{eq:select_block_encoding}
    (\bra{\mathL} \otimes I)\select (\ket{\mathL}\otimes I) = \frac{A}{\lambda}.
\end{eqnarray}
While the construction of \select~oracle is not unique, it is instructive to introduce an example as follows:
\begin{eqnarray}
    \select &=& \sum_l \ket{l}\bra{l} \otimes A_l + \sum_{\bar{l}}\ket{\bar{l}}\bra{\bar{l}} \otimes I,\label{eq:select_def}
\end{eqnarray}
where we have introduced $\bar{l}$ to denote computational bases of ancillary qubit system on which \select~acts trivially.

\begin{figure}[tbp]
    \centering
    \includegraphics[width=0.85\linewidth]{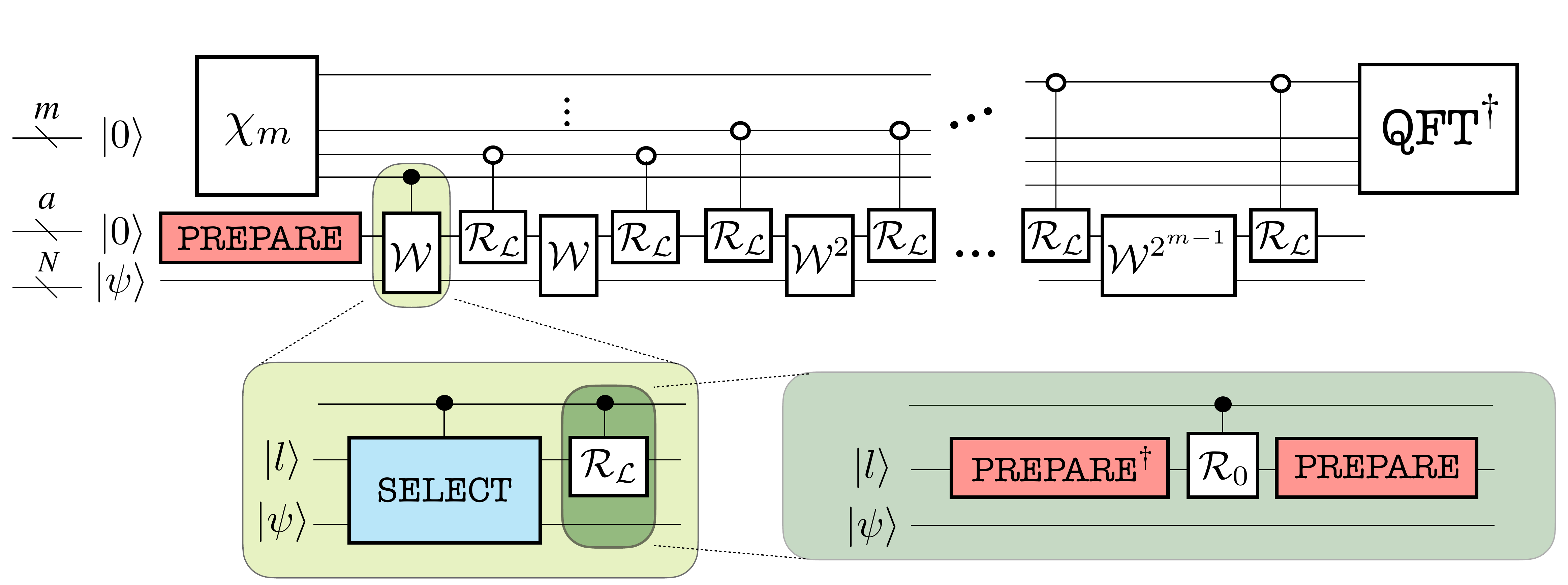}
    \caption{
    Quantum circuit structure of qubitization-based phase estimation algorithm. The definitions of \prepare~and \select~are given in Eqs.~\eqref{eq:prepare_def}, ~\eqref{eq:select_def}, where the target block-encoded operator $A$ is taken as the Hamiltonian here.
    }\label{fig:qubitization_circuit}
\end{figure}

\subsubsection{Qubitization}\label{subsubsec:qubitization}
Ref.~~\cite{low2016hamiltonian} pointed out that, by choosing the operator $A$ in Eq.~\eqref{eq:LCU} to be the Hamiltonian of the target system, one may construct a unitary that invokes  rotational operation for each eigenstates.
Since all the rotation is performed disjointly on two-dimensional subspace, such a specific form of block encoding is referred to as the {\it qubitization}.

Let $A=\sum_{\chi} \chi \ket{\chi}\bra{\chi}$ be a target Hermitian operator which we wish to block-encode into the \select~operator via the signal state~$\ket{\mathL} = \prepare \ket{0}$  as in Eq.~~\eqref{eq:prepare_def},~\eqref{eq:select_def},~\eqref{eq:select_block_encoding}.
To understand the ``qubitized" picture, we focus on a single eigenstate $\ket{\chi}$, and also introduce an orthogonal state $\ket{\perp_\chi} = \frac{1}{\sqrt{1-\chi^2}}(\select - \chi I)\ket{\mathL}\ket{\chi}$ so that 
\begin{eqnarray}
    \select\ket{\mathL}\ket{\chi} = \chi \ket{\mathL}\ket{\chi} + \sqrt{1 - \chi^2} \ket{\perp_\chi}.
\end{eqnarray}
We say that the oracles are qubitized if the action of the \select~is closed within the two-dimensional subspace $\mathH_\chi = {\rm Span}(\ket{\mathL}\ket{\chi}, \ket{\perp_\chi})$, which is known to be satisfied if and only if $\braket{\mathL | \select^2 | \mathL} = I$~\cite{low2016hamiltonian}. 
Observe that \select~defined as Eq.~\eqref{eq:select_def} satisfies this condition; $A$ is now Hermitian and hence can be decomposed as $A = \sum_l w_l A_l$ with $w_l > 0$ and $A_l^2 = I$. Meanwhile, this does not hold if $A$ is not Hermitian (as in the case of truncated Taylor series of unitary time evolution operator). 
We find that other transition amplitudes can be computed explicitly as 
\begin{eqnarray}
    \bra{\mathL}\bra{\chi} \select \ket{\perp_{\chi}} &=& \bra{\mathL}\bra{\chi} \select (\select - \chi I) \ket{\mathL}\ket{\chi}/\sqrt{1 - \chi^2}\\
    &=& \frac{1- \chi^2}{\sqrt{1 - \chi^2}} = \sqrt{1 - \chi^2}, \\
    \braket{\perp_\chi | \select | \perp_\chi} &=& \bra{\mathL}\bra{\chi} (\select - \chi I)  \select (\select - \chi I) \ket{\mathL}\ket{\chi} / (1 - \chi^2) \\
    &=& \bra{\mathL}\bra{\chi} (\select - 2\chi I + \chi^2 \select) \ket{\mathL}\ket{\chi} / (1 - \chi^2) \\
    &=& \frac{-\chi (1- \chi^2)}{1-1\chi^2} = -\chi,
\end{eqnarray}
and therefore the action within the effective Hilbert space $H_\chi$ is given as follows:
\begin{eqnarray}
    \select_{\chi} = \left( 
    \begin{array}{cc}
     \chi & \sqrt{1 - \chi^2} \\
     \sqrt{1 - \chi^2}& -\chi \\
    \end{array}
    \right).\label{eq:qubitized_oracle}
\end{eqnarray}
Observe that the action satisfies $\select_\chi \select_\chi^\dag = I$ and indeed closed within the subspace.

The framework of the qubitization enables one to perform a vast class of polynomial transformation on the eigenspectrum of $A$ as ${\rm Poly}(A)$. 
This implies that, one can build up a new block encoding of an operator from an existing block encoded operator, just as one develops a new algorithm by combining subroutines~\cite{martyn_2021}.
For instance, given a block encoding of $A$, one may construct a block encoding of polynomial approximation for $e^{iA}$ , from which one may also derive a block encoding of $A^{-1}$. 
Such a methodology to manipulate eigenvalues enable one to extract various information from the block encoded operator. 

Now we have confirmed that the oracles defined as in Eqs.~\eqref{eq:prepare_def}, ~\eqref{eq:select_def} are readily qubitized. 
Let us next proceed to construct a Szegedy-type walk opeartor as $\mathW = \mathRL\cdot \select$, which combines $\select$ with another reflection operator defined as $\mathRL = 2\ket{\mathL}\bra{\mathL} - I$.
It is straightforward to find that $\mathW$ is also qubitized in each two-dimensional subspace $\mathH_k$ which is defined from the $k$-th eigenstate of the Hamiltonian:
\begin{eqnarray}
    \mathW_k = \left( 
    \begin{array}{cc}
     E_k/\lambda & \sqrt{1 - (E_k/\lambda)^2} \\
     -\sqrt{1 - (E_k/\lambda)^2}& E_k/\lambda \\
    \end{array}
    \right) = e^{i \arccos (E_k/\lambda) Y},
\end{eqnarray}
where $E_k$ denotes the $k$-th eigenenergy.
Therefore, this walk operator not only block-encodes the Hamiltonian in the signal space as   $\braket{\mathL | \mathW | \mathL} = e^{i \arccos (H/\lambda)}$, but also provides efficient access to its entire eigenspectra.

\begin{figure}[tbp]
    \centering
    \includegraphics[width=0.9\linewidth]{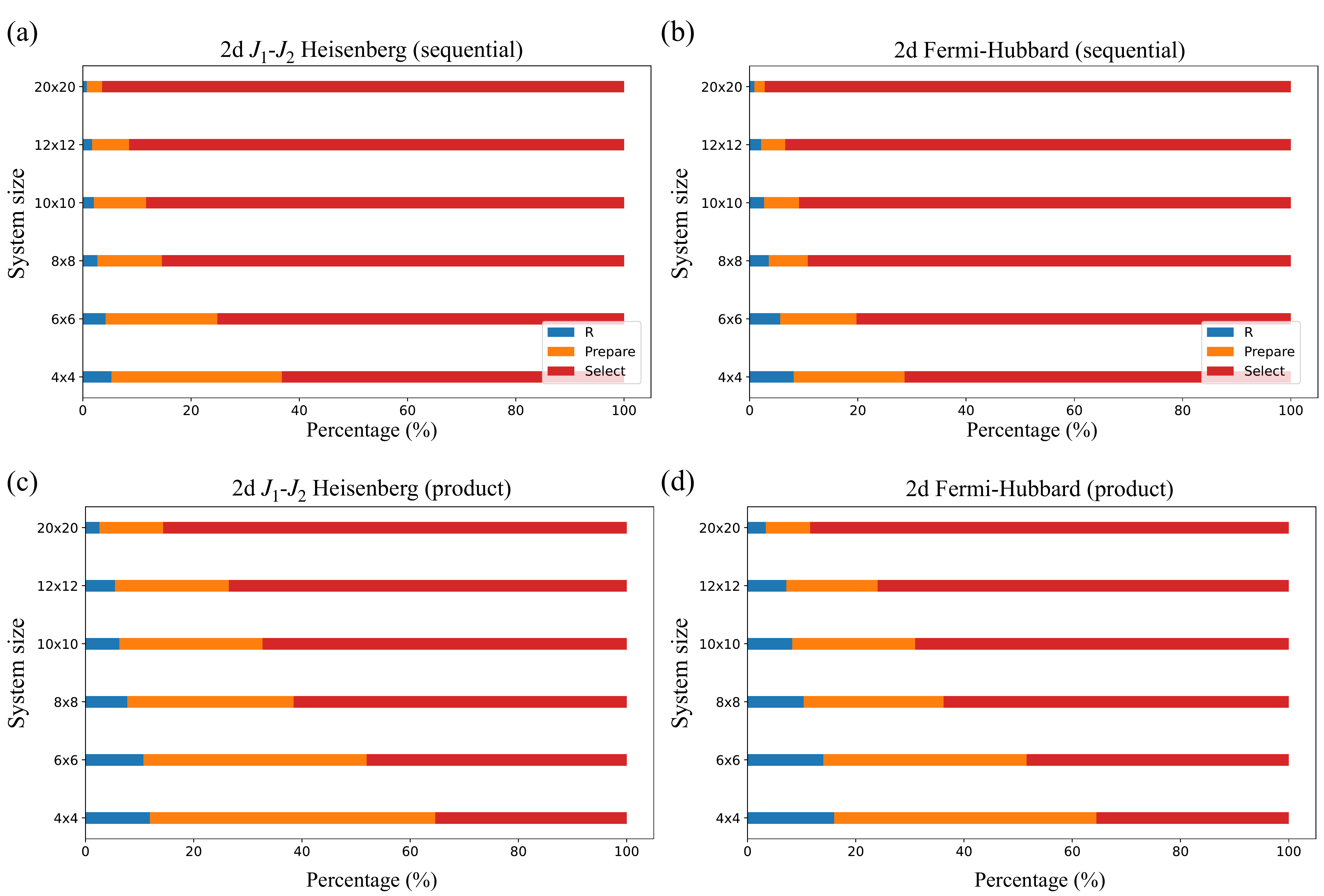}
    \caption{
    \black{
    The origin of $T$-count for qubitization algorithm with two flavors that sequentially block-encodes each Pauli one by one (sequential) and the one that exploits the translational symmetric local connectivity (product) for (a, c) 2d $J_1$-$J_2$ Heisenberg model ($J_2/J_1$=0.5) and (b, d) 2d Fermi-Hubbard model ($U/t$=4) on square lattice. The logarithmic contribution from the reflection and $\tt PREPARE$ oracles become negligible in the asymptotic limit.
    }
    }\label{fig:tcount_percentage}
\end{figure}

Using the oracles \prepare~and \select~introduced above, we construct the entire circuit as shown in Fig.~\ref{fig:qubitization_circuit}, where we have also introduced the preparation circuit for the readout qubits as~\cite{babbush_encoding_2018}
\begin{eqnarray}
\chi_m \ket{0}^m = \sqrt{\frac{2}{2^m+1}}\sum_{i=0}^{2^m-1}\sin\left( \frac{\pi(i+1)}{2^m+1}\right)\ket{i}.
\end{eqnarray}
This consumes $T$-gates of $\tilde{O}(m)$ and hence neglected in the resource analysis in the present work. It is crucial to the resource estimate that the number of readout digits $m$ is determined from the number of repetition $r$ regarding the the quantum walk operations as~\cite{gorecki_pi_2020, burg_quantum_2021, lee_evenmore_2021}
\begin{eqnarray}
r = \left\lceil \frac{\pi \lambda}{2 \epsilon} \right\rceil < 2^m.
\end{eqnarray}
This leads us to obtain the  relation between the cost (i.e., $T$-count or detailed runtime) of the global algorithm and the oracles as \black{
\begin{eqnarray}
C_{\rm HS} \sim r \times (C_{\rm P} + C_{{\rm P}^\dag} + C_{\rm S} + 2C_{\mathcal{R}_0}),
\end{eqnarray}
where $C_{{\rm P}^{(\dag)}}$ and $C_{\rm S}$ denote the cost required for the \prepare~and \select~oracles, respectively, and $C_{\mathcal{R}_0}$ is the cost for the reflection by $|0\rangle$.
By explicitly constructing the quantum circuits as shown in Sec.~\ref{sec:post_trotter_oracles}, we find that the cost for $\mathcal{R}_0$ and ${\tt PREPARE}$, which is only logarithmic with system size, indeed become asymptotically negligible (See Fig.~\ref{fig:tcount_percentage}). 
}

\begin{figure}[tbp]
    \centering
    \includegraphics[width=0.8\linewidth]{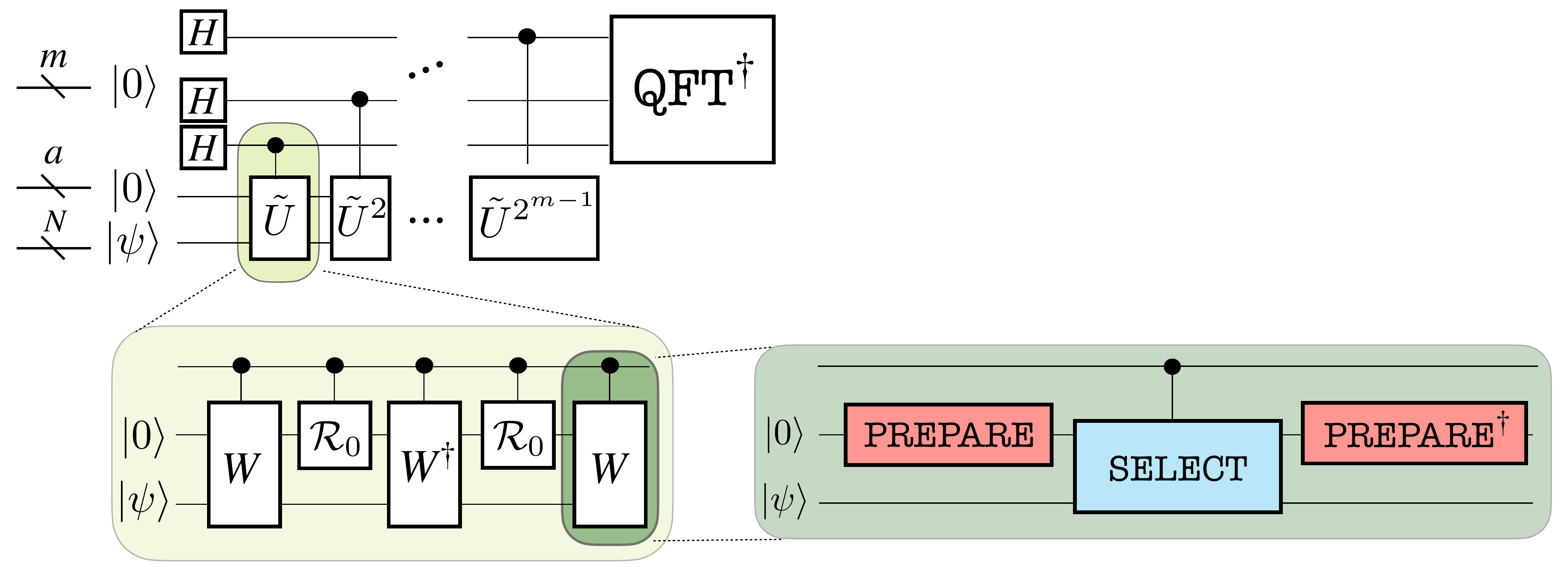}
    \caption{
    Quantum circuit structure of Taylorization-based phase estimation algorithm. The definitions of \prepare~and \select~are given in Eqs.~\eqref{eq:prepare_def}, ~\eqref{eq:select_def}, where the target operator is taken as the truncated Taylor series of the time evolution unitary.
    }\label{fig:taylorization_circuit}
\end{figure}
\subsubsection{Taylorization}\label{subsubsec:taylorization}
Taylorization is one of the earliest attempts to overcome the polynomial gate complexity ${\rm Poly}(1/\delta_{\rm HS})$ in Trotter-based Hamiltonian simulation methods~\cite{berry_improved_2018, poulin_quantum_2018}.
In essence, the algorithm uses the technique of block encoding to approximate the time evolution operator \black{$U = \exp(-iAt)$} with the repetitive operation of truncated Taylor series up to $K$-th order~(See Fig.~\ref{fig:taylorization_circuit}), which results in ${\rm polylog}(1/\delta_{\rm HS})$ gate complexity.
To be concrete, the time evolution is divided into $r$ steps as $U = (U_r)^r$ with $U_r = \exp(-iAt/r)$, where each step is approximated as 
\begin{eqnarray}
    \widetilde{U}_r &:=& \sum_{k=0}^K \frac{(-iAt/r)^k}{k!}H^k = \sum_{k=0}^K \sum_{l_1,...,l_k}w_{l_1}\cdots w_{l_K} A_{l_1}\cdots A_{l_K}.\label{eq:taylorization}
\end{eqnarray}
Hyperparameters such as $K$ and $r$ shall be chosen in order to predict the output energy with target precision $\epsilon$.
Following the discussion of the qubitization and also Ref.~\cite{berry_simulating_2015, Casares2022tfermionnon}, we take
\begin{eqnarray}
r = \frac{\pi}{2\delta_{PEA}}~\left(\sim\frac{\pi \lambda}{2\epsilon}\right),\ \ \ 
    K = \left\lceil -1 + \frac{2\log(2r/\delta_{\rm HS})}{\log \log (2r/\delta_{\rm HS})+1}\right\rceil,
\end{eqnarray}
where $\lambda=\sum_l w_l$ is the total sum of coefficients for $A = \sum_l w_l A_l$.
\black{Note that the scaling of $O(1/\epsilon)$ reflects the time-energy uncertainty relation in order to extract the eigenstate energy. 
}

Once we construct \prepare~and \select~for truncated Taylor series Eq.~\eqref{eq:taylorization}, we find that the block encoding is realized via a walk operator
\begin{eqnarray}
    W := (\prepare^\dag) (\select) (\prepare),
\end{eqnarray}
whose explicit operation is given using $s = \sum_{k=0}^K \sum_{l_1,...,l_K}w_{l_1}\cdots w_{l_K}$ as
\begin{eqnarray}
    W\ket{0}\ket{\psi} = \frac{1}{s}\ket{0}\widetilde{U}_r \ket{\psi} + \sqrt{1-\frac{1}{s^2}}\ket{\perp}.
\end{eqnarray}
Although the unwanted second term seem to deteriorate the simulation,
we may tune the amplitude of the target state via the oblivious amplitude estimation introduced in ~\cite{berry_exponential_2014}. Namely, we rescale the time evolution as $s=2$ so that the combination of the walk operator and a reflection operator on the ancillary qubits $\mathR_0 = I - 2\ket{0}\bra{0}$ gives~\cite{berry_simulating_2015}
\begin{eqnarray}
    W\mathR_0 W^\dag \mathR_0 W\ket{0}\ket{\psi} = - \ket{0}\widetilde{U}\ket{\psi}.
\end{eqnarray}
Finally, by plugging the controlled operation of $\widetilde{U}$ into the quantum circuit of phase estimation, we obtain the circuit structure of the entire algorithm as in Fig.~\ref{fig:taylorization_circuit}.
The overall cost is given as follows,
\begin{eqnarray}
C_{\rm HS} \sim r \times (3 (C_{\rm P} + C_{{\rm P}^\dag} + C_{\rm S}) + 2 C_{\mathR_0}),
\end{eqnarray}
where $C_{\rm P}$ and $C_{\rm S}$ is obtained from $K$-fold multiplication on cost for the qubitization algorithm, while $C_{\mathR_0}$ is simply the cost for multi-controlled Z-gate operated on ancillas.
\black{In practice, the number of repetitions $r$ is around tens of thousands (see also Sec.~\ref{sec:total_resource}) and $K$ is around 30 to 40 for the target models with system size of above hundreds. This indicates that each unit oracles are executed for $10^{5-6}$ times.
}

\section{Post-Trotter oracles for lattice models}\label{sec:post_trotter_oracles}
In Sec.~\ref{subsec:hamiltonian_simulation_posttrotter}, we have introduced the technique of block encoding, which allows us to perform Hamiltonian simulation with gate complexity that scales polylogarithmically with respect to the target accuracy $\delta_{\rm HS}$.
In this section, we further provide the explicit construction of quantum circuits that cover a broad range of lattice models considered in condensed matter physics. We introduce two methods to construct oracles, which we refer to as sequential and product-wise constructions. 

\black{The sequential construction is designed to block-encode each Pauli operators one by one. Given $L$ terms, we use $\lceil \log L \rceil$ ancillary qubits for $\prepare$~oracle, so that $l$-th computational basis is used to block-encode the $l$-th Pauli operator $P_l$ by the $\select$~oracle. On the other hand, the core idea of product-wise structure is utilize the translational symmetric structure of the connectivity; we encode the information on local connectivity and the interaction amplitude into {\tt PREPARE} oracle, while we keep {\tt SELECT} rather agnostic to the geometry. Such a construction turns out to be beneficial compared to the sequential one when the connectivity is dense; for instance, when we have all-to-all $G$-local interaction, the gate complexity drops from $O(N^G)$ to $O(GN)$~\cite{koizumi2024}. 
}

In the following, abstract descriptions are provided each for \prepare~and \select~oracles in Secs.~\ref{subsec:prepare_abstract} and \ref{subsubsec:select_abstract}, respectively.
Subsequently we give constructions for quantum spin models and fermionic models whose interaction is geometrically local and translationally invariant. 
Then, we proceed to show examples in concrete models: spin-1/2 $J_1$-$J_2$ Heisenberg model and Fermi-Hubbard model on the square lattice, and also the spin-1 Heisenberg chain.
From the results of $T$-count analysis combined with that for Trotter-based methods presented in Sec.~\ref{subsec:hamiltonian_simulation_trotter}, we compare the order of computational cost between various phase estimation subroutines to conclude that the qubitization algorithm gives the best practice among the Hamiltonian simulation subroutines. 
We further guide readers to Sec.~\ref{sec:distselect_analysis} for more detailed runtime analysis that compiles quantum circuit of the qubitization algorithm into executable instructions.
We importantly remark that we have chosen the product-wise construction to the compare $T$-counts between Hamiltonian simulation subroutines, while the for the runtime analysis we focus on the sequential type of \select~oracle for the sake of simplicity.

\subsection{Abstract structure of \prepare}\label{subsec:prepare_abstract}
\label{subsubsec:prepare_abstract}
\begin{figure}[h]
    \centering
    \includegraphics[width=0.8\linewidth]{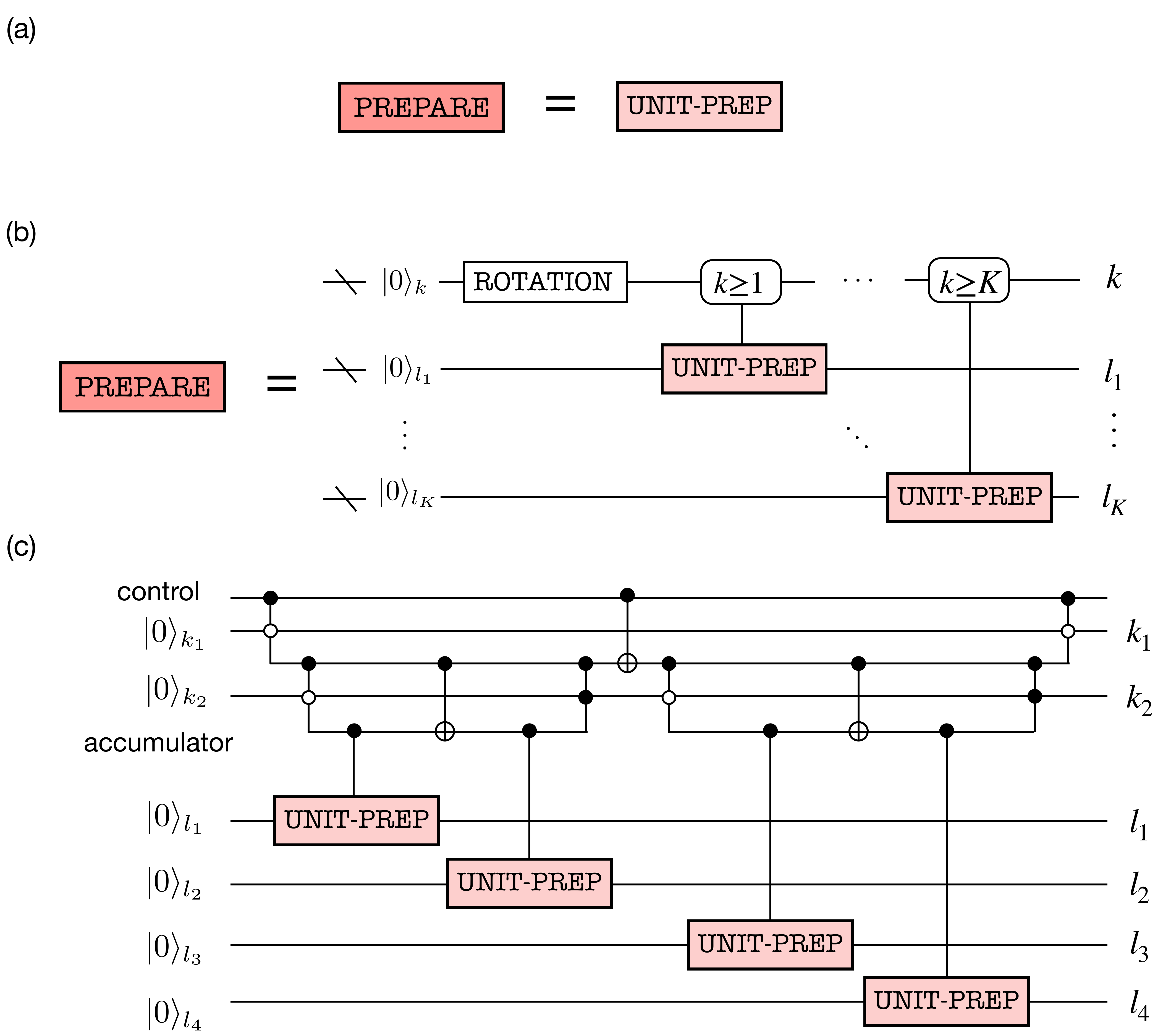}
    \caption{
    \prepare~structure of (a) qubitization and (b) Taylorization, whose detailed circuit structure for $K=4$ case is described in (c). 
    }
    
    \label{fig:prepare_qubitization_taylorization}
\end{figure}

Here, we present the abstract structure of \prepare~circuit that achieves polylogamithmic cost, in terms of both $T$-count and ancillary qubit count, for translationally-invariant $G$-local Hamiltonians introduced in Sec.~\ref{sec:target_hamiltonian} as
\begin{eqnarray}
    H = \sum_{\bp}\sum_{\{\bmu, \balpha\}}
    w_{\bmu, \balpha}
    \hat{\Lambda}_{\bp}^{(\balpha)} 
    \hat{\Lambda}_{\bp + \bmu_1}^{(\balpha)}\cdots 
    \hat{\Lambda}_{\bp + \bmu_{G-1}}^{(\balpha)},
\end{eqnarray}
where $\bp = (p_1, \dots, p_d)$ labels the site (including the sublattice structure), $\bmu = (\bmu_1, ..., \bmu_{G-1})$ is a set of vectors that identifies the connection between interacting sites, $\hat{\Lambda}^{\balpha}_{\bp}$ is an operator for microscopic degrees of freedom (e.g. spin, fermion, boson) on site $\bp$,  and $w_{\bmu, \balpha}$ is the amplitude of the interaction.
As we display in Fig.~\ref{fig:prepare_qubitization_taylorization}, both the \prepare~for qubitization and Taylorization consist of the identical subroutine which we call in this paper as \unitprep.
Therefore, let us first focus on the case of qubitization and later mention the case of Taylorization.

\begin{figure}[h]
    \centering
    \includegraphics[width=1.05\linewidth]{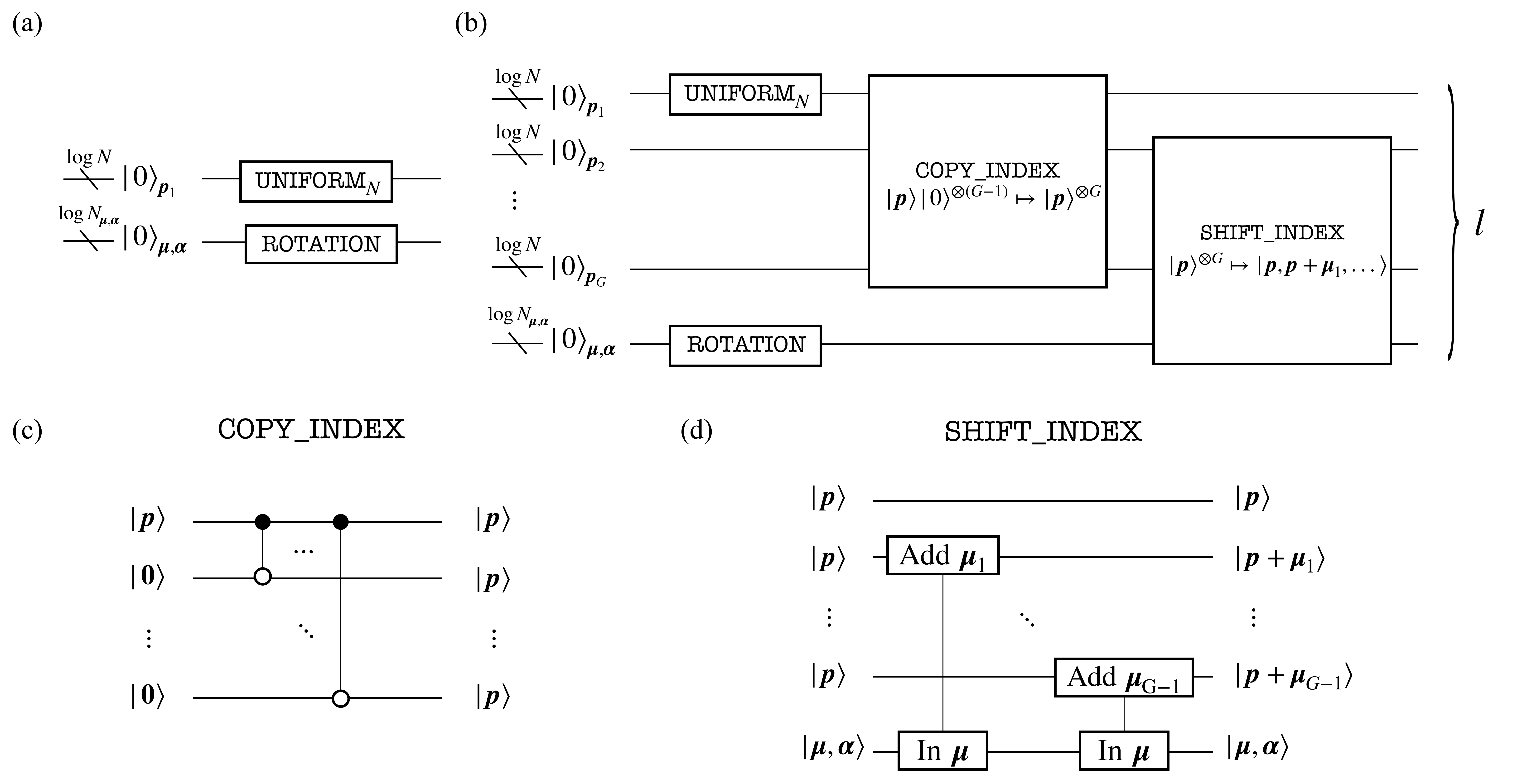}
    \caption{
    General structure of \prepare~for lattice systems with $G$-local translationally invariant interactions. \black{(a) {\tt PREPARE} oracle for sequential-type construction. (b) {\tt PREPARE} oracle for product-wise construction. The quantum circuit structure are provided for (c) ${\tt COPY\_INDEX}$ and (d) {\tt SHIFT\_INDEX}. For detailed information on constant adders, see Sec.~\ref{sec:cost_basic}.}
    }\label{fig:prepare_general}
\end{figure}

First we introduce the sequential construction, which straightforwardly encodes all the information of coefficients as 
\begin{eqnarray}
{\tt UNIT}\text{-}{\tt PREP}:\ket{0} \mapsto 
    \ket{\mathL} = 
    \sum_{\bp}\sum_{\{\bmu, \balpha\}} \sqrt{\frac{w_{\bmu, \balpha}}{\lambda}} 
    \ket{\bp} \ket{\bmu, \balpha}.
\end{eqnarray}
Thanks to the translational invariance of the Hamiltonian, we may split the ancilla for geometrical position $\bp$ and interaction configurations $\bmu, \balpha$. 
Note that in the graphical description in Fig.~\ref{fig:prepare_general}(a) we have introduced ${\tt UNIFORM}_N \ket{0} = \frac{1}{\sqrt{N}}\sum_{n=1}^N\ket{n}$.

One evident difference between the product-wise construction from the sequential one is the $G$-fold increase in use of ancilla qubits, from which we benefit in terms of $T$-count in \select~oracle (See.~\ref{subsubsec:select_abstract}). The general form of product-wise  \unitprep~encodes the interaction amplitudes into ancillary qubits as follows:
\begin{eqnarray}
    {\tt UNIT}\text{-}{\tt PREP}:\ket{0} \mapsto 
    \ket{\mathL} = 
    \sum_{\bp}\sum_{\{\bmu, \balpha\}} \sqrt{\frac{w_{\bmu, \balpha}}{\lambda}} 
    \ket{\bp, \bp + \bmu_1, ..., \bp + \bmu_{G-1}} \ket{\bmu, \balpha}.
\end{eqnarray}
As shown in Fig.~\ref{fig:prepare_general}(b), this can roughly be decomposed into three steps as
\begin{eqnarray}
\begin{split}
    \ket{0} &\xrightarrow{\rm Rotations} 
    \left(\sum_{\bp} \ket{\bp} \ket{0}^{(G-1)\log N} \right) \otimes \left( \sum_{\{\bmu, \balpha\}} 
    \sqrt{w_{\bmu, \balpha}}\ket{\bmu, \balpha}\right)
    = \sum_{\bp}\sum_{\{\bmu, \balpha\}}\sqrt{w_{\bmu, \balpha}}\ket{\bp}\ket{0}^{(G-1)\log N} \ket{\bmu, \balpha}
    \\
    &\xrightarrow{{\tt COPY\_INDEX}} 
    \sum_{\bp}\sum_{\{\bmu, \balpha\}}\sqrt{w_{\bmu, \balpha}}\ket{\bp, \bp, ..., \bp} \ket{\bmu, \balpha}\\
    &\xrightarrow{{\tt SHIFT\_INDEX}}
    \sum_{\bp}\sum_{\{\bmu, \balpha\}}
    \sqrt{\frac{w_{\bmu, \balpha}}{\lambda}}
    \ket{\bp, \bp + \bmu_1, ..., \bp + \bmu_{G-1}} \ket{\bmu, \balpha} (= \ket{\mathL}),
\end{split}    
\end{eqnarray}
where $N$ is the total system size. Note that we have abbreviated overall coefficients in the first and second lines for the sake of simplicity.

The main resource of $T$-count is the  synthesis of the rotation gates in the first step, and the adder circuit for {\tt SHIFT\_INDEX} \black{(note that {\tt COPY\_INDEX} can be performed solely by CNOT gates)}. Regarding the dynamics simulation error $\delta_{\rm HS}$, the order of the $T$-count can be obtained as $8 \log N + N_{{\bm \mu}, {\bm \alpha}} \log(1/\delta_{\rm ss})$ where $\delta_{\rm ss} = \delta_{\rm syn}/r$
where the required number of ancillary qubits can be estimated as $(G+1)\log N + \log N_{\balpha, \bmu} + \order(1)$, with $N_{\balpha, \bmu} = N_{\balpha}N_{\bmu}$ denoting the total number of combinations of $(\balpha, \bmu)$. 
Here, we assume to use the efficient implementation of logical adder that use auxiliary qubits to store the information of ``carry-on"~\cite{gidney_halving_2018}, resulting in  additional use of ancillary qubits of $\log N_{\balpha, \bmu}$.
Note that, for the \prepare~circuit for Taylorization, we prepare $N_{\balpha, \bmu}\log L + \log N_{\balpha, \bmu}$ ancilla qubits to encode information on higher order contribution from Taylor expansion, as can be seen from Fig.~\ref{fig:prepare_qubitization_taylorization}.

Last but not least, let us mention how to deal with the boundary condition. 
Be aware that the above-mentioned construction assumes translationally invariant Hamiltonian defined under the periodic boundary condition.
However, of course we are not limited to this situation: open, cylinder, or twisted boundary conditions can also be readily treated as well. One straightforward way is to encode the boundary terms individually, which can be done by sparing $T$-count of $\log(N^{d-1/d})$. However, in this work we consider an alternative method to simulate Hamiltonians defined on cylindrical boundary condition which solely relies on simple postprocessing and rescaling of the Hamiltonian. The details are explained in Sec.~\ref{sec:imperfect_prepare}.

\begin{table}[]
\begin{tabular}{|c|c|ccc|} \hline
                               & Sequential                                       & \multicolumn{3}{c|}{Product}                                                                                                                                            \\ \hline \hline
\multirow{4}{*}{$T$-count}     & \multirow{4}{*}{$4L-4$}                          & \multicolumn{1}{c|}{Rotations}                                                & \multicolumn{1}{c|}{\tt COPY\_INDEX} & {\tt SHIFT\_INDEX}                                         \\ \cline{3-5} 
                               &                                                  & \multicolumn{1}{c|}{$8 \log N + (N_{\bmu, \balpha}+2) \log(1/\delta_{\rm ss})$} & \multicolumn{1}{c|}{0}           & $8(G-1) \log N + 4 (G-1)\log N_{\bmu, \balpha} + O(1)$ \\ \cline{3-5} 
                               &                                                  & \multicolumn{3}{c|}{Total}                                                                                                                                              \\ \cline{3-5} 
                               &                                                  & \multicolumn{3}{c|}{$8G \log N + 4(G-1) \log N_{\mu, \alpha} + (N_{\bmu, \balpha} + 2)\log (1/\delta_{\rm SS})$}                                                          \\ \hline \hline
\multirow{4}{*}{Ancilla count} & \multirow{4}{*}{$\log N + \log N_{\bmu, \balpha}$} & \multicolumn{1}{c|}{Rotations}                                                & \multicolumn{1}{c|}{\tt COPY\_INDEX} & {\tt SHIFT\_INDEX}                                         \\ \cline{3-5} 
                               &                                                  & \multicolumn{1}{c|}{$\log N + \log N_{\bmu, \balpha}$}                          & \multicolumn{1}{c|}{$G \log N$}  & $(G+1)\log N + \log N_{\bmu, \balpha}$                 \\ \cline{3-5} 
                               &                                                  & \multicolumn{3}{c|}{Total}                                                                                                                                              \\ \cline{3-5} 
                               &                                                  & \multicolumn{3}{c|}{$(G+1)\log N + \log N_{\bmu, \balpha}$} \\ \hline 
\end{tabular}
\caption{
\black{
Quantum resource required to implement ${\tt PREPARE}$ oracle of sequential and product-wise construction of qubitization method. Note that the ancilla qubit count for the product-wise construction denotes the number of involved qubits, and thus not simply the summation over the ancilla qubit counts.  Also, we have neglected the $O(1)$ contributions here for simplicity.}
}
\end{table}                                                   

\subsection{Abstract structure of \select}\label{subsubsec:select_abstract}
\begin{figure}[h]
    \centering
    \includegraphics[width=0.9\linewidth]{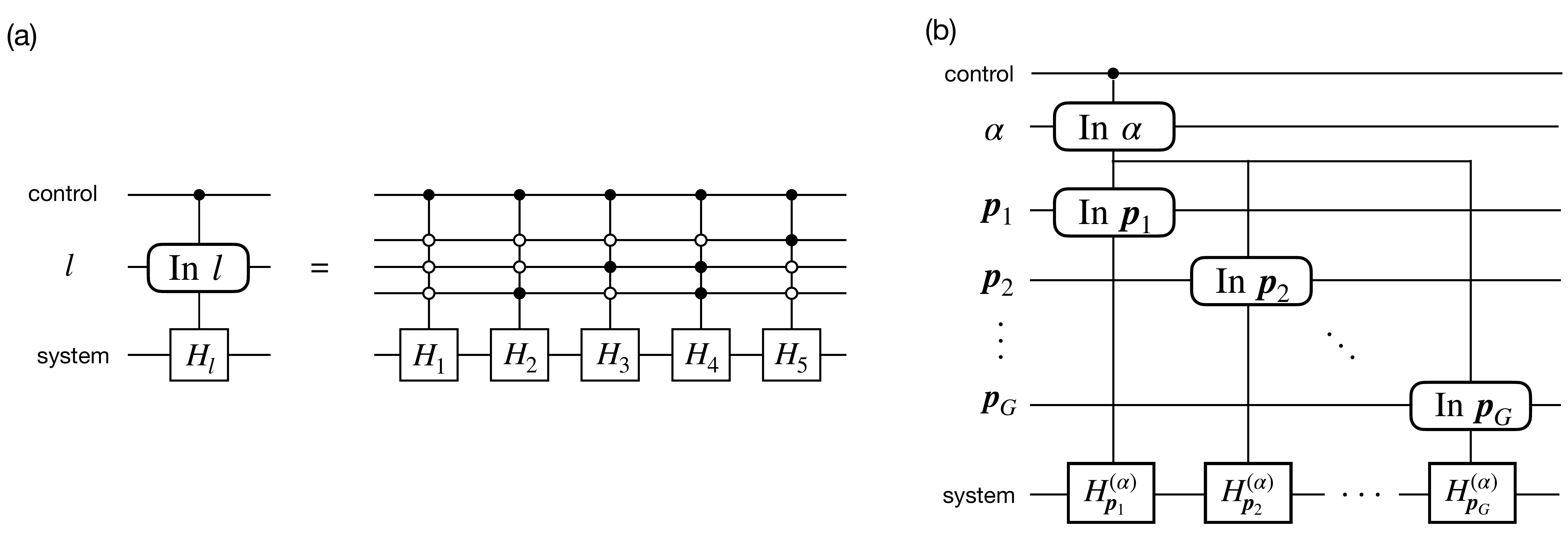}
    \caption{
    General quantum circuit structure of \select~oracle.
    (a) Sequential structure based on the unary encoding of efficiently-implementable unitaries. Here we assume that the Hamiltonian is a sum over $L=5$ Pauli terms.
    (b) Product-wise structure exploiting the $G$-local translational symmetric structure of Hamiltonian.
    \black{Note that, instead of sharing the carry-on register for $\balpha$, we may also consider $G$ consecutive structure by uncomputing $\balpha$ every time, which can be done without increasing the $T$-count. Meanwhile, in practice the proposed approach shall reduce the $T$-{\it depth}, since it allows us to perform the gates in parallel, as discussed in Sec.~\ref{sec:distselect_analysis}.}
    }\label{fig:select_general}
\end{figure}

Next, we present the sequential and product-wise structure of \select~oracles that are used to block-encode an operator $H=\sum_{l=1}^L w_l H_l$.
In the sequential construction, we perform one-hot encoding of individual unitary $H_l$ as defined in Eq.~\eqref{eq:select_def} as
\begin{eqnarray}
\select = \sum_l \ket{l}\bra{l} \otimes H_l + \sum_{\bar{l}} \ket{\bar{l}}\bra{\bar{l}} \otimes I.~\label{eq:select_def_again}
\end{eqnarray}
In particular, we assume that every $H_l$ is a tensor product of Pauli operators (with global phase $\pm i$ or $\pm 1$) so that we can implement $H_l$ and controlled $H_l$ with Clifford operations. 
Babbush {\it et al.} has proposed a quantum circuit that realizes Eq.~\eqref{eq:select_def_again} with $T$-count of  $4L-4$~\cite{babbush_encoding_2018}, whose structure is explicitly shown in Fig.~\ref{fig:select_general}(a).

\black{
While Fig.~\ref{fig:select_general}(a) is applicable to general form of Hamiltonian, we find that in many cases the translational symmetry allows even more efficient implementation via product-wise construction.
The product-wise qubitization is designed so that the (hyper-)connectivity of local terms are encoded solely in {\tt PREPARE}, while the {\tt SELECT} oracle is agnostic to the geometric information.
Assume that we have {\tt PREPARE} oracle that generates a signal state 
\begin{eqnarray}
{\tt PREPARE} |0\rangle = |\mathcal{L}\rangle = \sum_{{\bm \alpha}, {\bm \mu}} \sum_{{\bm p}} \sqrt{\frac{w_{\bm{\alpha}, \bm{\mu}}}{\lambda}} |{\bm p}_1 , {\bm p}_1+{\bm \mu} _1, ..., {\bm p}_1 + {\bm \mu}_{G-1}\rangle,
\end{eqnarray}
using the construction method discussed in the previous subsection. Then, it suffices to construct the {\tt SELECT} oracle without encoding the information of ${\bmu}$ to block-encode the entire Hamiltonian. Concretely, consider the following operation:
\begin{eqnarray}
{\tt SELECT} = \sum_{{\bm \alpha}, {\bm p}_1, ..., {\bm p}_G} |{\bm \alpha}\rangle \langle {\bm \alpha}| \otimes |{\bm p}_1, ..., {\bm p}_G\rangle \langle {\bm p}_1 , ..., {\bm p}_G| \otimes H_{{\bm p}_1}^{(\balpha)}H_{{\bm p}_2}^{(\balpha)}\cdots H_{{\bm p}_G}^{(\balpha)}.\label{eq:select_product}
\end{eqnarray}
Observe that this satisfies ${\tt SELECT}^2 = I$ and  the following:
\begin{eqnarray}
	\langle \mathcal{L}| {\tt SELECT} | \mathcal{L}\rangle = \sum_{{\bm \alpha}, {\bm \mu}, {\bm p}} \frac{w_{{\bm \alpha}, {\bm \mu}}}{\lambda} H_{{\bm p}}^{(\balpha)} H_{{\bm p} + {\bm \mu}_2}^{(\balpha)} \cdots H_{{\bm p} + {\bm \mu}_{G-1}}^{(\balpha)} = H/\lambda.
\end{eqnarray}
This allows us to block-encode the Hamiltonian with simpler quantum circuit structure.
In fact, by observing that Eq.~\eqref{eq:select_product} can be written as tensor product as 
\begin{eqnarray}
    {\tt SELECT} = \sum_{\balpha} |\balpha\rangle\langle  \balpha| \otimes \left[ 
    \left(\sum_{\bp} |\bp_1\rangle\langle \bp_1|\otimes H_{\bp_1}^{\balpha}\right) \otimes
    \left(\sum_{\bp} |\bp_2\rangle\langle \bp_2|\otimes H_{\bp_2}^{\balpha}\right) 
    \otimes \cdots \otimes
    \left(\sum_{\bp} |\bp_G\rangle\langle \bp_G|\otimes H_{\bp_G}^{\balpha}\right)
    \right],
\end{eqnarray}
it is natural to employ the structure as shown in Fig.~\ref{fig:select_general}(b).
We find that each ``thread" is used to block-encode $2SN_{\balpha} N_{\rm site}$ Pauli operators, respectively.
Namely, we have $G \times 2SN_{\balpha} N_{\rm site}$ terms in total, and thus the $T$-count per $\select$~oracle is $4\times G\times 2SN_{\balpha} N_{\rm site} -4$ (see Table~\ref{tab:tcount_select_heisenberg}.)
For concrete circuit structures of 2d $J_1$-$J_2$ Heisenberg model, 2d Fermi-Hubbard model, and spin-$S$ Heisenberg chain, see Sec.~\ref{subsubsec:oracle_J1J2},~\ref{subsubsec:oracle_2dFH}, and ~\ref{subsubsec:oracles_haldane}, respectively.
}


\black{The connectivity-agnostic construction of {\tt SELECT} oracle turns out to be beneficial in particular when we consider a densely coupled system due to, e.g., high dimensionality or long-range interaction.}
For instance, let us consider a spin-$S$ Heisenberg model on $d$-dimensional hypercube with $r$-nearest-neighbor interaction, each site of which is coupled to $\gamma_{r, d}$ sites.
Under such a procedure, the number of Pauli operator is given as 
\begin{eqnarray}
L=3 \times(2S)^2 \times \frac{\gamma_{r, d}}{2}\times N_{\rm site}.
\end{eqnarray}
\black{Note the factor $2S$ due to encoding of high-spin operator as a sum of Pauli operator as $S_p^a = \frac{1}{2}\sum_{\nu=1}^{2S} \sigma_{p, \nu}^a$.}
Meanwhile, 
as we show in Table~\ref{tab:tcount_select_heisenberg}, the product-wise construction of \select~is 
not affected by the connectivity of the interaction profile; this gives improvement for arbitrary interaction profile except for the case of spin-1/2 1d nearest-neighbor interaction.
\black{Furthermore, when we consider a quantum spin-$S$ model with all-to-all $G$-local interactions, the reduction on the $T$-count is even more significant; each sequential-type {\tt SELECT} oracle requires $O(S^G N^G)$ while the product-wise type can be implemented by $O(GSN)$, in exchange for additional $O(G\log N)$ ancilla qubits. 
Systematic  discussion on such advantage is discussed elsewhere~\cite{koizumi2024}.
}

Let us remark briefly on the estimation of the actual runtime. While we may use $T$-counts as the baseline to compare the performance between different Hamiltonian simulation algorithms, we further discuss in Sec.~\ref{sec:distselect_analysis} that the actual runtime (which is more related with the $T$-depth) can be further improved by distributed implementation of the controlled $H_l$ operation.

\begin{table}[]
    \centering
    \begin{tabular}{c|c||c|c|c|c|c|}
         \select~type 
         & 
         General $T$-count
         & 
         
            Heisenberg chain
         &
            2d $J_1$-$J_2$ Heisenberg
         &
         3d $J_1$-$J_2$-$J_3$ Heisenberg 
         &
         \begin{tabular}{c}
            General lattice Heisenberg\\
             $(r, d, S)$
         \end{tabular}         
         \\ 
         \hline \hline
         Sequential & $4L-4$ & $48 S^2 N_{\rm site} - 4$ & $192 S^2 N_{\rm site} - 4$ & $624 S^2 N_{\rm site} - 4$ & $24S^2 \gamma_{r, d}N_{\rm site} - 4$\\
         Product-wise & $8S G N_{\balpha}N_{\rm site} - 4$ & $48 S N_{\rm site} - 4 $ & $48 S N_{\rm site} - 4$ & $48 S N_{\rm site}-4$ & $48 S N_{\rm site} - 4$
    \end{tabular}
    \caption{$T$-count of \select~for spin-$S$ Heisenberg model on $d$-dimensional hypercube with $r$-nearest-neighbor interactions. Here, we denote the number of edges per site as $\gamma_{r, d}$ and the total number of sites as $N_{\rm site}$. To be concrete, Heisenberg chain satisfies $(r, d) = (1, 1)$ with $\gamma_{r, d} = 2$, 2d $J_1$-$J_2$ Heisenberg corresponds to $(r, d) = (2,2)$ with $\gamma_{r, d} = 8$, where 3d $J_1$-$J_2$-$J_3$ Heisenberg model is described by $(r, d) = (3, 3)$ with $\gamma_{r, d}=26$.
    }
    \label{tab:tcount_select_heisenberg}
\end{table}

\subsection{Explicit construction of oracles}\label{subsec:oracles_model}
We are now ready to dive into the explicit construction of quantum circuits for executing post-Trotter methods.
Since it is straightforward to modify the circuits generated for qubitization into those for Taylorization, we exclusively discuss the oracles for qubitization. 

\subsubsection{2d $J_1$-$J_2$ Heisenberg model}\label{subsubsec:oracle_J1J2}
The Hamiltonian of spin-1/2 $J_1$-$J_2$ Heisenberg model on the square lattice is defined as in Eq.~\eqref{eq:J1J2_definition} as
\begin{eqnarray}
    H = J_1\sum_{\langle p, q\rangle} 
    \sum_{\alpha \in \{X, Y, Z\}} S_p^\alpha S_q^\alpha + 
    J_2 \sum_{\langle \langle p, q \rangle \rangle} 
    \sum_{\alpha \in \{X, Y, Z\}} S_p^\alpha S_q^\alpha,
\end{eqnarray}
where the summation of the first and second terms concerns nearest-neighbor and next-nearest-neighbor sites, respectively, and $S_p^\alpha$ is now the spin-1/2 operator.
Here we exclusively consider a square lattice of $N=M\times M$ sites, and focus on the model with
 $J_1 = 1, J_2 = 0.5$ where quantum spin liquid phase is expected to show up.
 
\begin{figure}[tbp]
    \centering
    \includegraphics[width=0.9\linewidth]{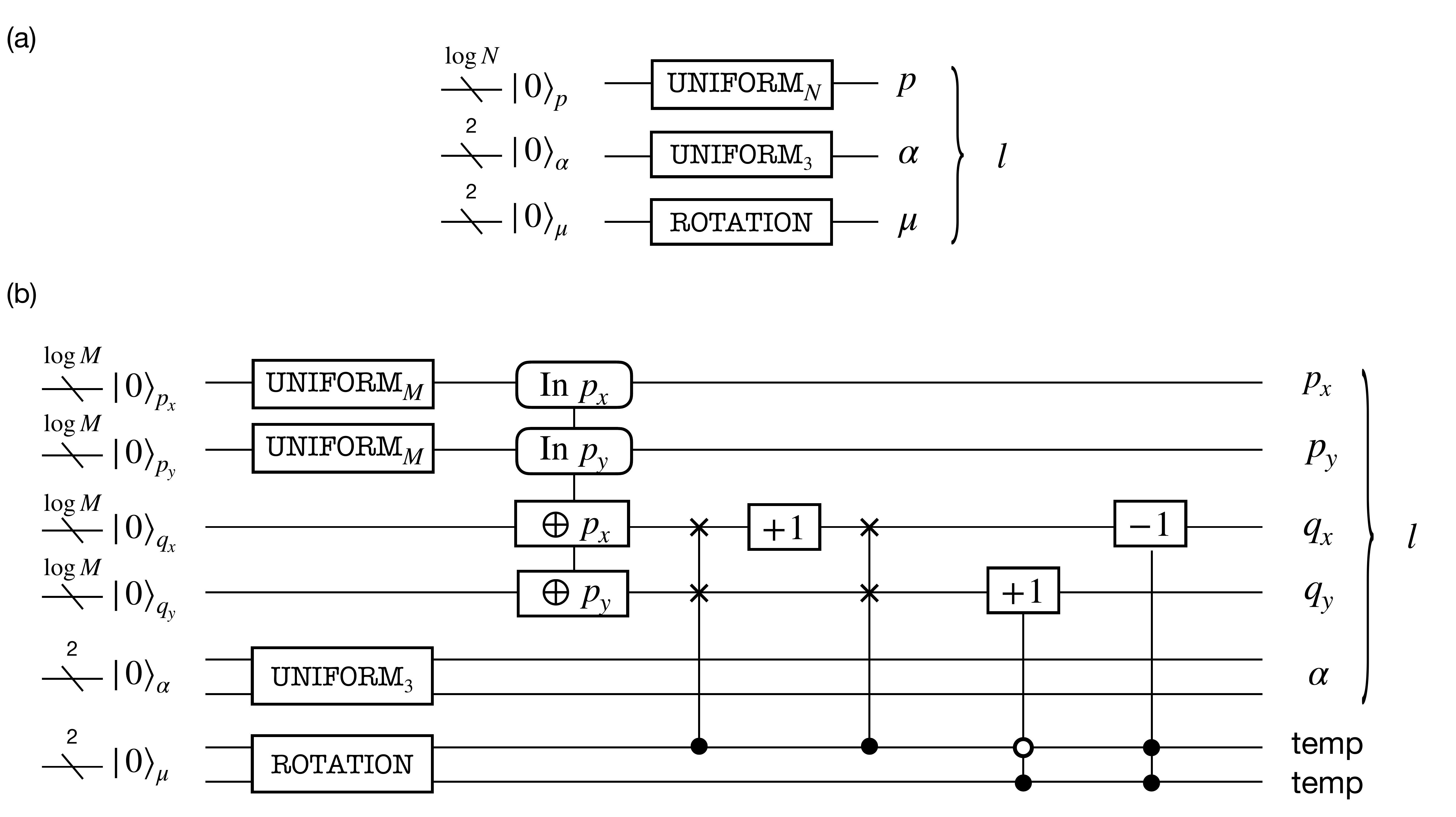}
    \caption{
    Circuit structure of \unitprep~for spin-1/2 $J_1$-$J_2$ Heisenberg model on square lattice. (a) The sequential construction requires $T$-count of $8\lceil\log N\rceil + 4\Gamma\log(1/\delta_{\rm SS})$ and  $\log N + 4$ ancilla qubits, where (b) product-wise construction consumes $T$-count of \black{$14 \log N + 7\Gamma \log(1/\delta_{\rm SS})$} and spans $3\log N + 4$ ancillary qubits.
    Note that we have introduced $M^2 = N$ in the figure, and denoted the synthesis-dependent constant as $\Gamma$ which are known to be $\Gamma\sim 1$~\cite{Kliuchnikov2023shorterquantum}.
    }
    \label{fig:prepare_2dJ1J2}
\end{figure}

First, let us briefly introduce the sequential construction of \unitprep~oracle. The circuit structure shown in Fig.~\ref{fig:prepare_2dJ1J2}(a) realizes the following action:
\begin{eqnarray}
\begin{split}
    {\tt UNIT}\text{-}{\tt PREP} \ket{0} \mapsto \sum_{p} \sum_{\alpha \in \{X, Y, Z\}} \sum_{\mu=1,2}
    \sqrt{\frac{J_1}{\lambda}} \ket{p} \ket{\alpha}\ket{\mu} 
    +  \sum_{\mu=3, 4}\sqrt{\frac{J_2}{\lambda}} \ket{p} \ket{\alpha} \ket{\mu},
\end{split}    
\end{eqnarray}
where $\mu\in\{1, 2, 3, 4\}$ correspond to the coupling between $(p_x, p_y)$ and $(p_x+1, p_y), (p_x, p_y+1), (p_x+1, p_y+1), (p_x-1, p_y+1)$, respectively.

On the other hand, for product-wise construction, \unitprep~circuit is designed  to generate the following state:
\begin{eqnarray}
\begin{split}
    {\tt UNIT}\text{-}{\tt PREP} \ket{0} \mapsto \sum_{p_x, p_y} \sum_{\alpha \in \{X, Y, Z\}} 
    &\sqrt{\frac{J_1}{\lambda}} \ket{p_x, p_y} \ket{p_x + 1, p_y} \ket{\alpha} \\
    + &\sqrt{\frac{J_1}{\lambda}} \ket{p_x, p_y} \ket{p_x, p_y+1}\ket{\alpha} \\
    + & \sqrt{\frac{J_2}{\lambda}} \ket{p_x, p_y} \ket{p_x + 1, p_y+1} \ket{\alpha}\\
    + & \sqrt{\frac{J_2}{\lambda}} \ket{p_x, p_y} \ket{p_x - 1, p_y+1}\ket{\alpha},
\end{split}    
\end{eqnarray}
whose graphical description is shown in Fig.~\ref{fig:prepare_2dJ1J2}(b). Note that, the required number of ancillary qubits is $3\log N + \order(1)$ rather than $2 \log N + \order(1)$ that is obtained by simply adding up the numbers in Fig.~\ref{fig:prepare_2dJ1J2}(b). This is because the most efficient implementation of adders (in terms of $T$-count) require as many quantum registers as the number of qubits to be simulated~\cite{gidney_halving_2018}.
Furthermore, by counting the number of required $T$-gates using the table presented in Sec.~\ref{sec:cost_basic}, we find that the \unitprep~can be implemented with $T$-count of $20 \log N + 48 \log(1/\delta_{\rm SS})$.

As for the \select~oracle, we follow the framework presented in Sec.~\ref{subsubsec:select_abstract} for sequential construction. 
Meanwhile, if we use the product-wise \select~oracle, the operation on computational bases that are not included in the signal state also becomes nontrivial, while this does not affect the dynamics encoded in the signal state.
In this case, the $T$-count becomes $24N -4$, which is approximately halved compared to the vanilla sequential type of \select~oracle.

\subsubsection{2d Fermi-Hubbard model}\label{subsubsec:oracle_2dFH}
The Hamiltonian of Fermi-Hubbard model on the square lattice reads as defined in Sec.~\ref{sec:target_hamiltonian} as
\begin{eqnarray}
H = -t \sum_{\langle p, q \rangle, s} (c_{p, s}^\dag c_{q, s}  + {\rm h.c.}) + U\sum_{p}n_{p, \uparrow} n_{p, \downarrow},
\end{eqnarray}
where $c_{p, \sigma}^{(\dag)}$ is the fermionic annihilation (creation) operator on site $p$ with spin $\sigma \in \{\uparrow, \downarrow\}$, $n_{p, \sigma}=c_{p, \sigma}^\dag c_{p, \sigma}$ is the corresponding number operator, $t$ is the hopping amplitude, and $U$ is the repulsive onsite potential. As in the case for spin-1/2 $J_1$-$J_2$ Heisenberg model, we mainly consider the lattice of $M\times M$ sites, and thus the total number of qubits required to simulate the system is $N=2M^2$. 
While we take $t$ to be unity in this work, we explicitly keep it among expressions when it is more informative.

For concreteness, we assume to employ the Jordan-Wigner transformation to encode fermionic degrees of freedom into quantum computer:
\begin{eqnarray}
c_{p,s}^{(\dag)} = {\rm sgn}(s) \prod_{r<p}\prod_{s=\uparrow, \downarrow}\sigma_{r,s}^{+(-)}
\end{eqnarray} 
where $p = p_x+p_yM$ and ${\rm sgn}(s)$ takes $\pm 1$ according to whether the spin is up or down.
This leads us to the expression of Hamiltonian as follows,
\begin{eqnarray}
H=-\frac{t}{2}\sum_{\langle p, q\rangle, s}
(
\sigma_{p, s}^X \sigma^{\vec{Z}} \sigma_{q, s}^X + 
\sigma_{p, s}^Y \sigma^{\vec{Z}} \sigma_{q, s}^Y)
+ \frac{U}{4}\sum_{p} \sigma^Z_{p,\uparrow} \sigma^Z_{p, \downarrow} - \frac{U}{4} \sum_{p, s} \sigma_{p, s}^Z + \frac{UN}{8}I,
\end{eqnarray}
where we have introduced $\sigma^{\vec{Z}}$ to represent the Z-string generated by Jordan-Wigner transformation.
\black{While one may use this definition, we may also employ an alternative definition with shift in the potential term as $\sum_p n_{p, \uparrow} n_{p, \downarrow} \rightarrow \sum_{p} n_{p, \uparrow} n_{p, \downarrow}- \frac{U}{2}\sum_{p, s} n_{p, s} + \frac{U}{4}I$, which results in the simplified form as
\begin{eqnarray}
    H = -\frac{t}{2}\sum_{\langle p, q\rangle} 
    (
\sigma_{p, s}^X \sigma^{\vec{Z}} \sigma_{q, s}^X + 
\sigma_{p, s}^Y \sigma^{\vec{Z}} \sigma_{q, s}^Y)
 + \frac{U}{4} \sum_p \sigma_{p, \uparrow}^Z \sigma_{p, \downarrow}^Z.
\end{eqnarray}
While this shift does not significantly change the $T$-count of oracles, it slightly modifies the L1 norm of the Hamiltonian which affects the total gate count, as discussed below.
}

\begin{figure}[tbp]
    \centering
    \includegraphics[width=0.9\linewidth]{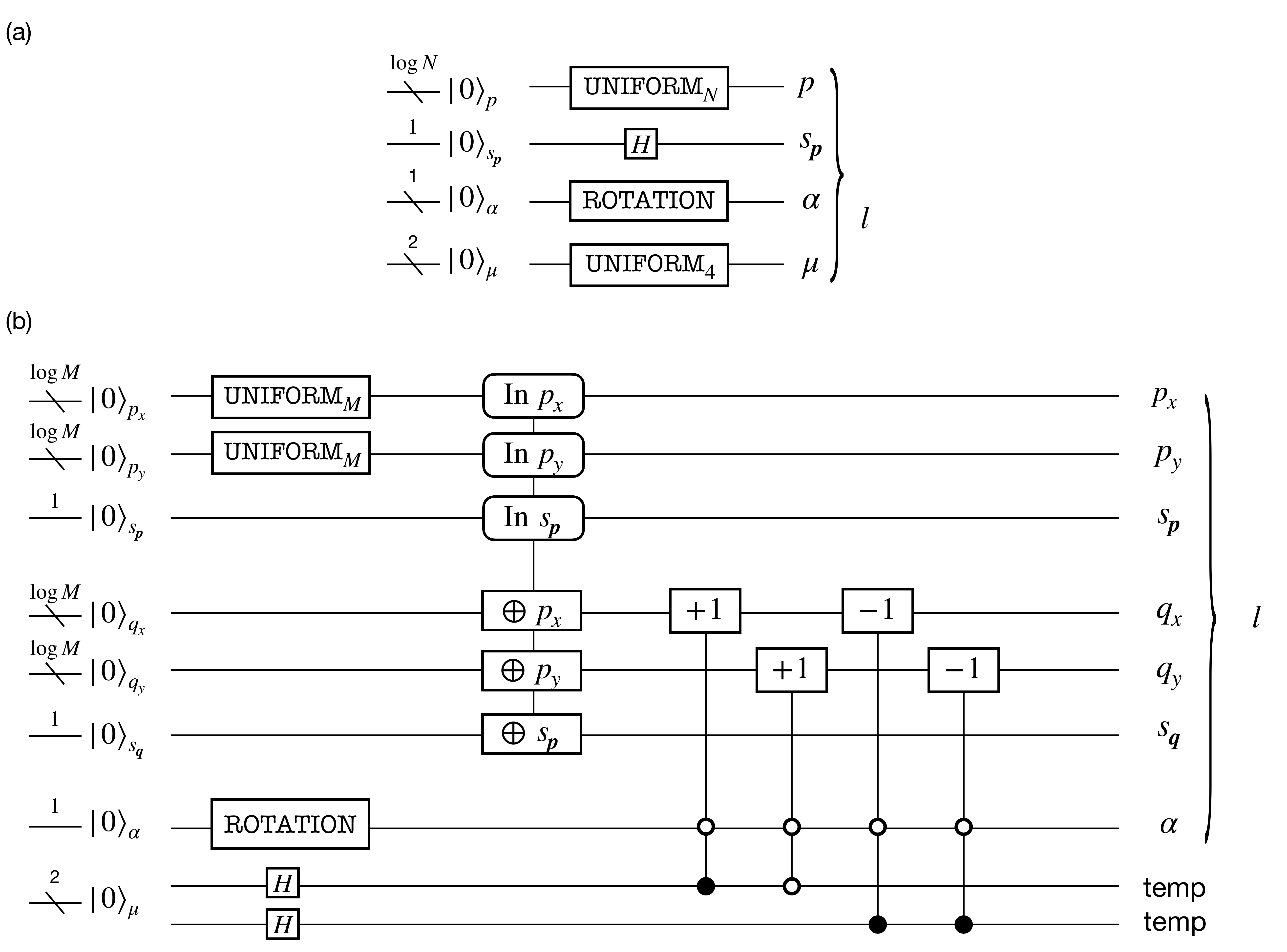}
    \caption{
    Circuit structure of \unitprep~for 2d Fermion Hubbard model on square lattice. (a) Sequential construction requires $T$-count of $8\log N + 4 \log (1/\delta_{\rm SS}) + O(1)$ with $\log N + O(1)$ ancilla qubits, while (b) product-wise construction  consumes $T$-count of $16\log N + 6\Gamma \log (1/\delta_{SS}) + O(1)$ and $3\log N + O(1)$ ancilla qubits. 
    The construction of {\tt SHIFT}\_{\tt INDEX} is slightly more efficient than the one presented in  Ref.~\cite{babbush_encoding_2018}; the required number of $T$-gates is reduced from $32 \log N$ to \black{$16\log N$}.
    However, the dominant factor is the \select~circuit that consumes $T$-count of $\order(N)$ and hence the difference is negligible in large-scale limit.
    }\label{fig:prepare_2dFH}
\end{figure}

In similar to the case with 2d $J_1$-$J_2$ Heisenberg model, we first introduce the sequential construction of \unitprep. Our construction simply aims to generate
\begin{eqnarray}
\begin{split}
    {\tt UNIT}\text{-}{\tt PREP} \ket{0} \mapsto &\sum_{p_x, p_y}
    \Bigg[ \sqrt{\frac{U}{4\lambda}} \ket{p, \uparrow}
    \ket{\text{onsite interaction}}_{\alpha, \mu}  \\
    + & \sum_{s = \uparrow, \downarrow}\sum_{\mu=e_x, e_y, e_{-x}, e_{-y}} \sqrt{\frac{t}{2\lambda}} 
    \ket{p, s} \ket{\text{hopping towards }\mu}_{\alpha, \mu}  \Bigg],\label{eq:unitprep_2dFH}
\end{split}    
\end{eqnarray}
which is realized by circuit shown in Fig.~\ref{fig:prepare_2dFH}(a).
Note that the ancillary quantum states such as $\ket{\text{onsite~interaction}}$ functions as ``flag" to tell the \select~oracle which interaction term must be operated.
\black{Here, the L1 norm is explicitly given as $\lambda = \lambda_t + \lambda_U$ where the contribution from the hopping term is $\lambda_t = t\cdot {\rm (\#edges)}$ and that from the interaction term is \black{$\lambda_U = \frac{1}{8}UN$}. We comment that this value is slightly different from $\lambda_U = \frac{1}{2}UN$ as employed in Ref.~\cite{babbush_encoding_2018}, which takes the constant term into account. However, the constant term can be eliminated from the QPE algorithm since we do not need to estimate its value. This results in a slightly smaller $\lambda$ so that the repetition count of \select~oracle is suppressed.}

By carefully examining the coefficients and the commutation relation, we find out that \unitprep~presented in Fig.~\ref{fig:prepare_2dFH}(b) provides the the product-wise construction via the following operation:
\begin{eqnarray}
\begin{split}
    {\tt UNIT}\text{-}{\tt PREP} \ket{0} \mapsto &\sum_{p_x, p_y}
    \Bigg[ \sqrt{\frac{U}{4\lambda}} \ket{p_x, p_y, \uparrow}\ket{p_x, p_y, \downarrow} \ket{1}_{\alpha}\\
    + & \sum_{s = \uparrow, \downarrow} \sqrt{\frac{t}{2\lambda}} \Bigl( 
    \ket{p_x, p_y, s}\ket{p_x + 1, p_y, s} \ket{0}_\alpha+ \ket{p_x, p_y, s}\ket{p_x, p_y + 1, s}\ket{0}_\alpha
    \Bigr) \\
    + & \sum_{s = \uparrow, \downarrow}\sqrt{\frac{t}{2\lambda}} \Bigl( 
    \ket{p_x, p_y, s}\ket{p_x-1, p_y, s}\ket{0}_\alpha+ \ket{p_x, p_y, s}\ket{p_x, p_y-1, s}\ket{0}_\alpha \Bigr) \Bigg],\label{eq:unitprep_productwise_2dFH}
\end{split}    
\end{eqnarray}
where $\alpha$ specifies the type of interaction term.

\begin{figure}[tbp]
    \centering
    \includegraphics[width=0.8\linewidth]{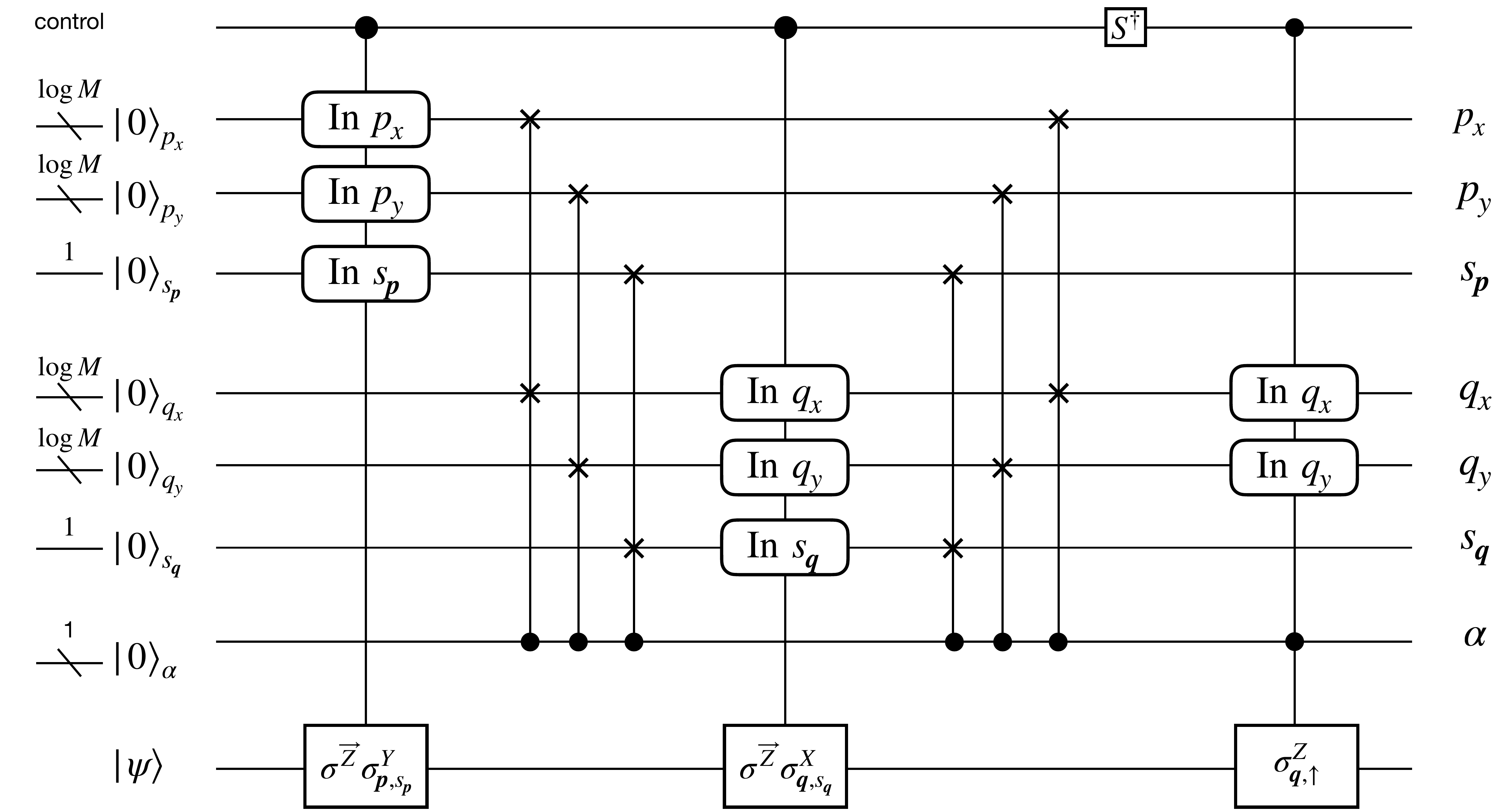}
    \caption{
    Product-wise construction of \select~for 2d Fermion Hubbard model on square lattice proposed in Ref.~\cite{babbush_encoding_2018}.
    }\label{fig:select_2dFH}
\end{figure}

Now we proceed to the operation of \select. We remark that the sequential construction is straightforward, and can be done by following the procedure explained as in Fig.~\ref{fig:select_general}(a).
In contrast, the product-wise construction of \select~is rather nontrivial; given a signal state as presented in Eq.~\eqref{eq:unitprep_productwise_2dFH}, the operation of \select~shall satisfy
\begin{eqnarray}
&&\select \ket{p_x, p_y, s_{\bf p}, q_x, q_y, s_{\bf q}, \alpha} \ket{\psi} \\
&=& \ket{p_x, p_y, s_{\bf p}, q_x, q_y, s_{\bf q}, \alpha} \otimes
\begin{cases}
\sigma_{p, \uparrow}^Z\sigma_{q, \downarrow}^Z\ket{\psi} & \text{if~}p = q,  (s_p, s_q) = (\uparrow, \downarrow), \alpha  = 1\\
\sigma_{p, s_p}^X(\prod_{p < r < q, s_r}\sigma_{r, s_r}^Z) \sigma_{q, s_q}^X\ket{\psi} & \text{if~}p < q, s_p = s_q, |{\bf p}- {\bf q}| = 1, \alpha=0\\
\sigma_{q, s_q}^Y(\prod_{q < r < p, s_r}\sigma_{r, s_r}^Z) \sigma_{p, s_p}^Y\ket{\psi} & \text{if~}q < p, s_p = s_q, |{\bf p}- {\bf q}| = 1, \alpha=0
\end{cases}
\end{eqnarray}
where the operation on subspace outside the signal state is unrestricted as long as $\select^2 = I$ is satisfied. Note that comparison between $p$ and $q$ is based on the integer $p=p_x+p_yM$ and $q=q_x+q_yM$ while $\|{\bf p}- {\bf q}\|$ denotes the Manhattan distance on the square lattice.
Product-wise construction of \select~oracle for the Fermi-Hubbard model has been suggested by Babbush et al., which pointed out that
the construction can be done by structure shown in Fig.~\ref{fig:select_2dFH} with $T$-count of $10 N$~\cite{babbush_encoding_2018}, as opposed to the one by the sequential type of \black{$18N$}. \black{Note that, the $T$-count of sequential type without shifting the potential term is $22N$.}


\subsubsection{Spin-1 antiferromagnetic Heisenberg chain}\label{subsubsec:oracles_haldane}
Recall that the Hamiltonian of spin-1 antiferromagnetic Heisenberg chain reads
\begin{eqnarray}
    H = \sum_{p=1}^{N_{\rm site}} \sum_{\alpha \in \{X, Y, Z\}}S_p^\alpha S_{p+1}^\alpha,
\end{eqnarray}
where $S_p^\alpha$ is the spin-1 operator acting on the $p$-th site and $N_{\rm site}$ denotes the number of sites in the system.
To render the problem compatible with quantum computers that take qubits as a elementary unit of quantum registers, we shall encode the system into enlarged Hilbert space spanned solely by qubits.
In this regard, we represent each spin operator $S_p^\alpha$ as a sum of  Pauli operators as $S_p^\alpha = \frac{1}{2}\sum_{\nu=1}^{2S}\sigma_{p, \nu}^\alpha$ and rewrite the Hamiltonian as
\begin{eqnarray}
H = \frac{1}{4}\sum_{p=1}^{N_{\rm site}} \sum_{\nu,\tau=1}^{2S} \sum_{\alpha \in \{X, Y, Z\}} \sigma_{p,\nu}^\alpha \sigma_{p+1, \tau}^\alpha.
\end{eqnarray}
Note that we also assume that the trial state for quantum phase estimation is also represented using the qubit encoding.

\begin{figure}[tbp]
    \centering
    \includegraphics[width=0.99\linewidth]{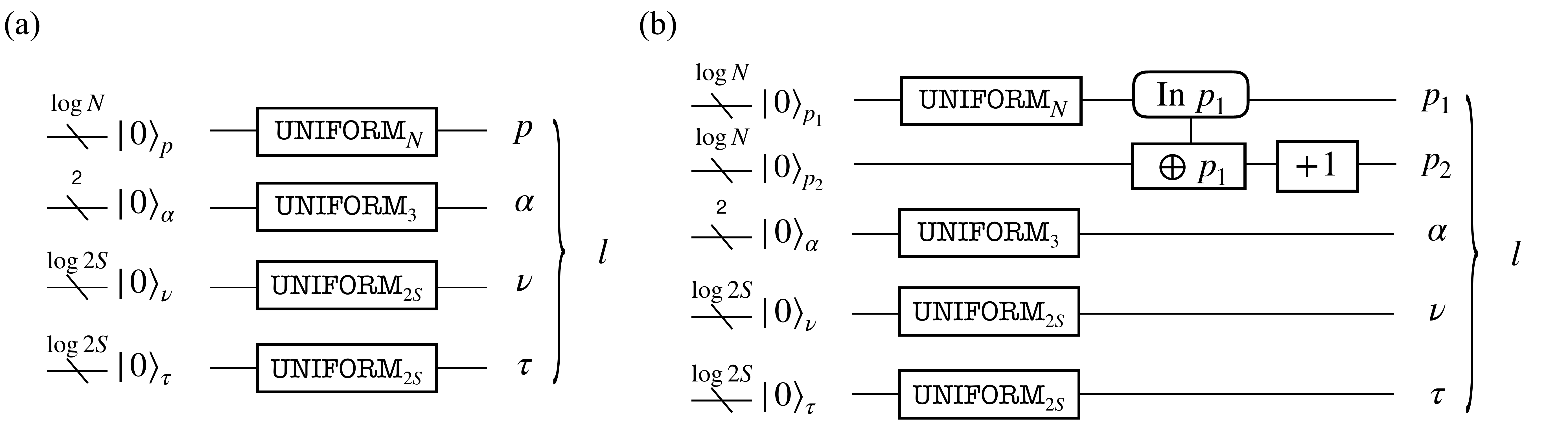}
    \caption{
    Circuit structure of \unitprep~for spin-$S$ Heisenberg chain. We display both (a) sequential construction that consumes $T$-counts of  $8\log N + 8\Gamma \log(1/\delta_{\rm SS})$ and $\log N + 2\log S + O(1)$ ancillary qubits and (b) product-wise construction that consumes $T$-counts of $12\log N + 8\log(1/\delta_{\rm SS})$ and $3\log N + 2\log S + O(1)$ ancillary qubits.
    Note that it is redundant to include {\tt COPY\_INDEX} and {\tt SHIFT\_INDEX} for (b), since the site index is unambiguously related with the interaction terms. 
    }\label{fig:prepare_1dAFH}
    \end{figure}

From the expression of Hamiltonian, we deduce that 
the \unitprep~circuit under periodic boundary condition can be composed as shown in Fig.~\ref{fig:prepare_1dAFH} to generate the following state:
\begin{eqnarray}
{\tt UNIT}\text{-}{\tt PREP}\ket{0}^{\log N_{\rm site} + 2} \mapsto \sum_{p, \alpha\in\{X, Y, Z\}} \sqrt{\frac{1}{\lambda}}\ket{p}\ket{\alpha}.
\end{eqnarray}
Correspondingly, the \select~in the case of qubitization is constructed following the one presented in Fig.~\ref{fig:select_general}(a) such that
\begin{eqnarray}
\select\ket{p,\alpha, \nu,\tau}\ket{\psi} = \ket{p, \alpha, \nu, \tau}\sigma_{p,\nu}^\alpha \sigma_{p+1, \tau}^\alpha \ket{\psi},
\end{eqnarray}
which can be realized by the general sequential implementation of ~\select~by $T$-count of $48S^2N_{\rm site} - 4$. One may alternatively consider the product-wise implementation of \select~as in Fig.~\ref{fig:select_general}(b) by consuming $T$-count of $48SN_{\rm site} - 4$. However, we point out that there is no advantage in terms of $T$-count for $S=1$ case, while it requires additional $\log N_{\rm site}$ ancillary qubits.
Therefore, we perform our runtime estimation based on the former implementation.

\section{Complexity of basic quantum operations}\label{sec:cost_basic}
Here we provide the details on the basic quantum operations that are used for circuit construction and $T$-count analysis.

\begin{figure}[tbp]
    \centering
    \includegraphics[width=0.99\linewidth]{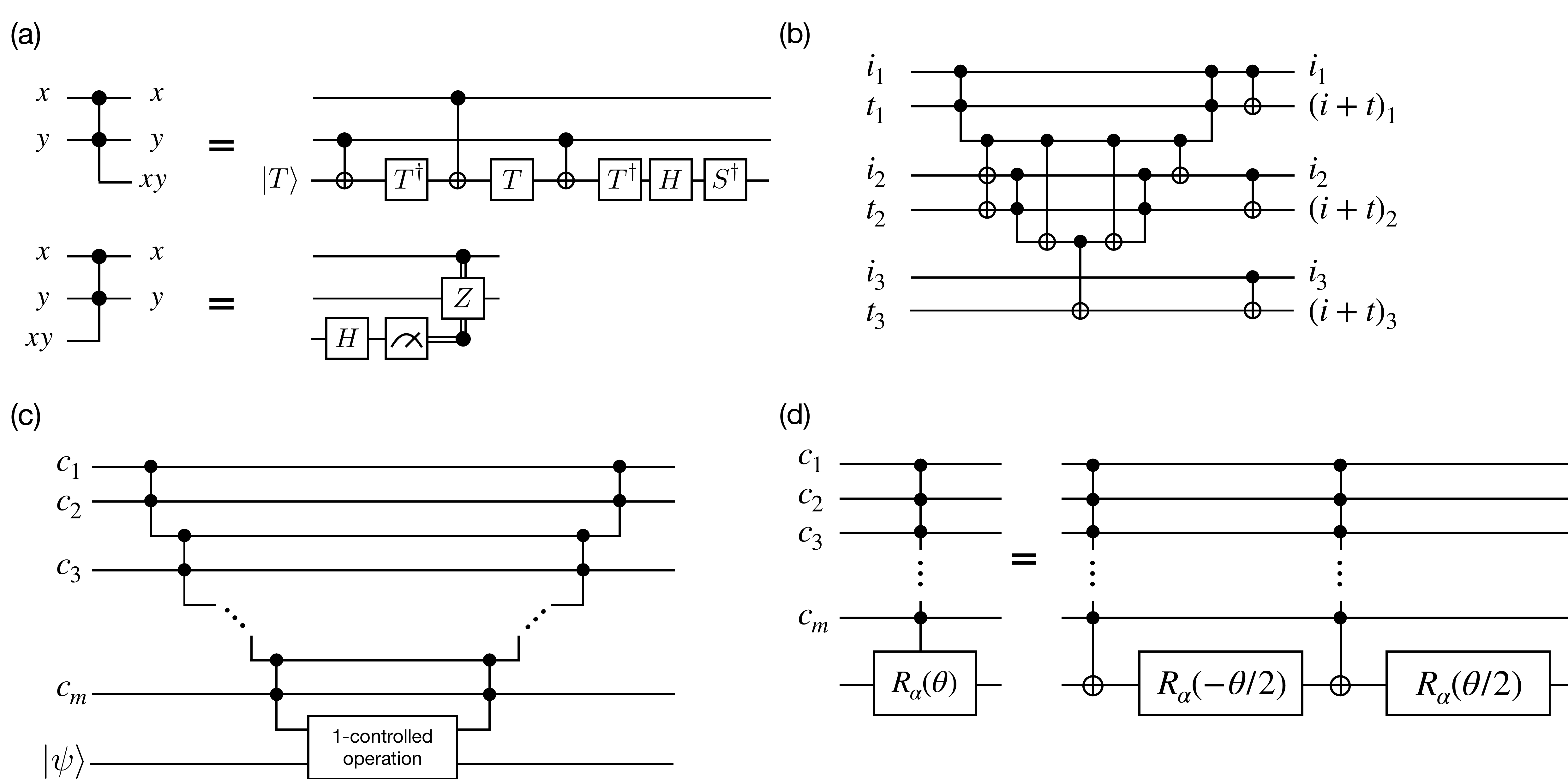}
    \caption{
    Graphical representation of basic quantum operations.
    (a) Computation and uncomputation of logical AND operation. (b) 3-qubit logical adder. (c) $m$-controlled operation (d) $m$-controlled rotation expressed using $m$-controlled NOT and single-qubit rotation. 
    }\label{fig:basic_operation}
\end{figure}

\begin{figure}[tbp]
    \centering
    \includegraphics[width=0.99\linewidth]{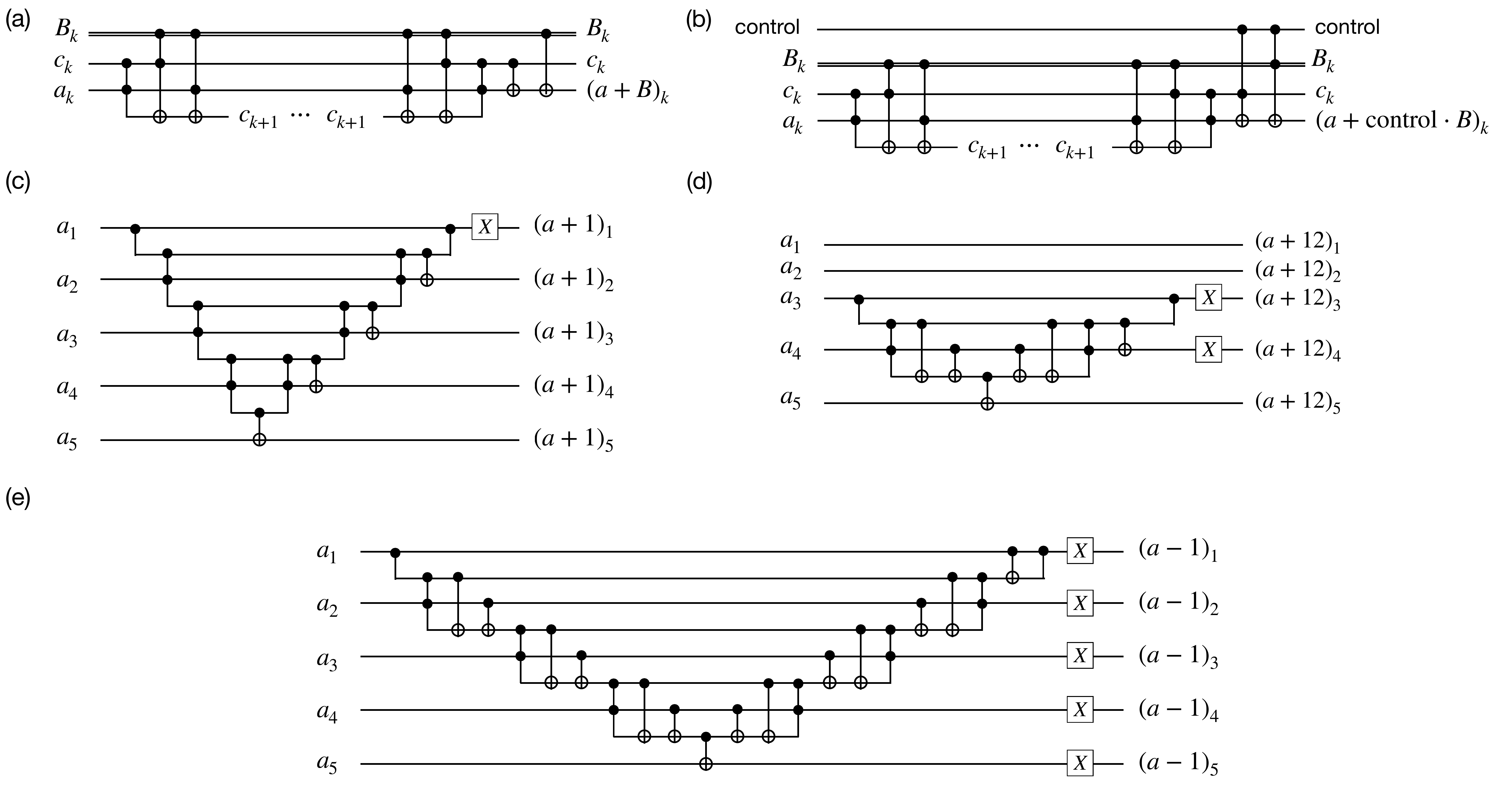}
    \caption{
    \black{Graphical representation of constant adder. Each panel represents
    (a) building block of constant adder, (b) building block of controlled constant adder, (c) 5-bit constant adder $a+1~({\rm mod~2^5})$, (d) 5-bit constant adder of $a+12~({\rm mod~2^5})$, (e) 5-bit constant subtracter of $a-1~({\rm mod~2^5})$.
    }
    }\label{fig:constant_adder}
\end{figure}

One of the most important operation that are commonly used in various basic gates presented in Table~\ref{tab:tcount_basic_operation} is the logical AND operation, whose explicit Clifford+$T$ construction is provided in Fig.~\ref{fig:basic_operation}(a). Gidney showed that, by consuming ancillary qubits to store the temporal results of addition, we can implement $n$-qubit logical adder with $T$-count of $4n+O(1)$ and controlled adder with $8n + O(1)$~\cite{gidney_halving_2018}~(Fig.~\ref{fig:basic_operation}(b)). We further push the use of temporary ancilla qubit to reduce the $T$-count for other controlled operations, such as $m$-controlled addition, SWAP, or rotation~(See Fig.~\ref{fig:basic_operation}(c),~(d)).
Also, we implement the {\tt UNIFORM} operation following the construction scheme presented in Ref.~\cite{babbush_encoding_2018}.
Finally, as a variant of the adder, we introduce a constant adder that performs some constant binary addition as shown in Figs.~\ref{fig:constant_adder}.

\begin{table}[]
    \centering
    \begin{tabular}{c|c}
         Operation& $T$-count  \\ 
         \hline \hline
         \black{General} Addition/Subtraction~\cite{gidney_halving_2018}  ({\tt SHIFT}\_{\tt INDEX})& $4n-4$\\
         \black{Addition/Subtraction of $2^k L$} & $4(n-k-1)-4$\\
         \hline 
        \black{Single-qubit rotation~\cite{Kliuchnikov2023shorterquantum}}& $\Gamma \lceil \log_2 (1/\delta_{\rm SS})\rceil + \Xi$\\
        {\tt UNIFORM}$_{2^k L}$~\cite{babbush_encoding_2018} & $8\log L + 2 \Gamma \log(1/\delta_{\rm SS}) + 2\Xi-4$\\
         \hline
         Arbitrary state synthesis~\cite{shende_synthesis_2006} & $2^{n+1}-2$ arbitrary rotations\\
         \hline \hline
         Controlled operation & $T$-count\\
         \hline
         $m$-controlled addition/subtraction & $4(m-1) + 8(n-1)$\\
         \black{$m$-controlled addition/subtraction of $2^k L$} & $4(m-1) + 8(n -k -2)$\\         
         \hline
         $m$-controlled NOT & 
         $
         \begin{cases}
         0 & (m=0,1) \\
         4(m-1) & (m\geq2) 
         \end{cases}
         $\\                  
         \hline
         $m$-control SWAP & $
         \begin{cases}
         0 & (m=0)\\
         4m & (m\geq 1)
         \end{cases}
         $\\         
         \hline
         \black{$m$-controlled rotation}& 
         $
         8(m-1) +2\Gamma\log (1/\delta_{\rm SS}) + 2\Xi
         $\\
         \hline
         $m$-controlled ${\tt UNIFORM}_{2^k L}$ & $4(m-1) + 2k + 10 \log L + 2\Gamma \log(1/\delta_{\rm SS}) + 2\Xi -4$
    \end{tabular}
    \caption{$T$-count required to perform basic operations that involves $n$ qubits. Here, $\Gamma$ and $\Xi$ are constants whose mean values are numerically computed as 1.03 and 5.6, respectively~\cite{Kliuchnikov2023shorterquantum}, and $\delta_{\rm SS}$ denotes the accuracy of rotation synthesis.}
    \label{tab:tcount_basic_operation}
\end{table}

\section{Szegedy quantum walk with imperfect \prepare}
\label{sec:imperfect_prepare}
In Sec.~\ref{subsec:hamiltonian_simulation_posttrotter}, we have seen that the Szegedy-type quantum walk operator enables us to explore the energy spectra of the block-encoded Hermitian operator. 
In particular, the oracular unitaries can be implemented efficiently if the Hamiltonian exhibit spatially periodic structure with $N_{\balpha}=O(1)$ types of interaction coefficients.
However, this argument strongly depends on the boundary condition; when the model is defined with open boundary condition or cylindrical boundary condition, the interaction configurations differ at the boundary sites, which affects the complexity of the \prepare~oracle in particular. Although the $T$-gate complexity of \prepare~remains to be $\log(N/\epsilon)$ and therefore does not dominate the total cost even when we incorporate boundary effect, it is desirable to deal with minimal effort.

Here, we present a method that use homogeneous implementation of \prepare~and still extract simulation results for lattice systems defined under boundary conditions that are not necessarily periodic along all directions.
To be concrete, we find that the error in the operation can be readily absorbed by constant shift in the Hamiltonian. 

Let us first consider a perfect and also a slightly erroneous \prepare~operation that takes the ancillary system to the following states:
\begin{eqnarray}
    \prepare_0: \ket{0}^{\otimes a} \mapsto \ket{\mathL_0} &=&  \sum_{l=1}^L\sqrt{\frac{w_l}{\lambda}}\ket{l},\\\label{eqn:perfect_prepare}
    \prepare : \ket{0}^{\otimes a} \mapsto \ket{\mathL} &=& \beta_1  \ket{\mathL_0} + \beta_2 \ket{\perp},\label{eqn:errorneous_prepare}
\end{eqnarray}
where $\ket{\perp}$ is orthogonal to any $\ket{l}.$ Our goal is to perform unbiased simulation of Szegedy quantum walk so that the obtained results correctly extracts the eigenspectra of the Hamiltonian. 
One naive solution is to ``correct" the state from $\ket{\mathL_0}$ to $\ket{\mathL}$ via, e.g., amplitude amplification. Meanwhile, by making use of the specific structure of \select~considered in this work, we can eliminate the error with negligible cost.
To be specific, we assume that the \select~operates nontrivially only in the subspace $\Pi \mathH \Pi$ where $\Pi = (\sum_{l=1}^L I^{\otimes n} \otimes \ket{l}\bra{l})$ and $\mathH$ is the total Hilbert space (confirm that this is satisfied in our construction presented in Sec.~\ref{subsec:hamiltonian_simulation_posttrotter}.). 
Under such an assumption, we introduce a constant shift $\gamma$ to the Hamiltonian to absorb the error.
The overlaps are explicitly written as
\begin{eqnarray}
(\bra{\mathL_0} \otimes I) \select (\ket{\mathL_0} \otimes I)     &=& \frac{H_0}{\lambda} + \gamma(-I),\\
(\bra{\perp}\otimes I) \select(\ket{\perp}\otimes I) &=& I,\\
(\bra{\perp}\otimes I) \select(\ket{\mathL_0}\otimes I) &=& 0,
\end{eqnarray}
which leads to
\begin{eqnarray}
(\bra{\mathL} \otimes I) \select (\ket{\mathL} \otimes I) = \frac{|\beta_1|^2 H_0}{\lambda} + (-\gamma |\beta_1|^2 + |\beta_2|^2)I.\label{eq:select_rescaled}
\end{eqnarray}
Therefore, if we pick $\gamma = |\beta_2|^2/|\beta_1|^2$, we can simulate the rescaled version of Hamiltonian.

For concreteness, let us consider a case where we want to simulate a lattice system with cylindrical boundary condition using \prepare~that is intended for periodic boundary condition.
We argue that the overhead due to the rescaling is negligible.
This can be understood from the fact that the $T$-count for estimating the eigenenergy (at arbitrary precision) using the Taylorization or qubitization algorithm is proportional to $\lambda$. 
As can be seen from Eq.~\eqref{eq:select_rescaled}, the L1 norm of the coefficient is amplified under the erroneous signal state $\ket{\mathL}$ as $\lambda \rightarrow \lambda/|\beta_1|^2$.
Since the ratio of coefficients is given as 
$|\beta_2/\beta_1|^2 = \order((N^{\frac{d-1}{d}}/N)^2) \nonumber
= \order(N^{-2/d})$
 in $d$-dimensional lattice systems with local interactions, 
we straightforwardly obtain $|\beta_1|^2 \approx 1 $, and thus does not affect the overall cost up to the leading order.
Throughout this work, for $T$-count and runtime analysis we assume to employ \prepare~oracle designed to implement a signal state for periodic boundary condition.

\section{Runtime analysis of qubitization-based quantum phase estimation} \label{sec:distselect_analysis}
\subsection{Overview}
In this section, we explain the details of the runtime analysis for the quantum phase estimation that is based on the qubitization algorithm.
As we have shortly mentioned in Sec.~\ref{subsec:hamiltonian_simulation_posttrotter}, the runtime of quantum circuits represented with the Clifford+$T$ formalism is typically evaluated with the number of $T$-gates, known as $T$-count, in the compiled quantum circuits. This is because each application of $T$-gates requires a time-consuming procedure consisting of magic-state injection, distillation, and teleportation in typical quantum error-correcting codes stabilized by Pauli operators. 
This is why we have compared the Hamiltonian simulation subroutines in Table ~1 in the main text.
While the $T$-count can capture the time-scaling of quantum algorithms when magic-state preparations are the most time-consuming factors, this estimation loses several vital factors in time analysis, and the actual execution time can vary a few orders of magnitudes depending on the strategy of compilation and executions. Here, we review several standard strategies and show how we calculated the execution times.

Throughout this paper, we assume the following situations: 1) Physical qubits are allocated on the vertices of 2d grids, which we call a qubit plane. 2) Logical qubits are encoded with surface codes~\cite{dennis2002topological,bravyi1998quantum,fowler2018low,fowlerSurfaceCodesPractical2012}. 3) \black{Logical operations are constructed based on} twist-based lattice surgery~\cite{horsman2012surface,fowler2018low}. 4) The method introduced by Litinski~\cite{litinski2019game} is applied to the construction of magic-state distillation, and purpose-specific strategies are employed for the compilation of \select~and \prepare~oracles. 
Note that these assumptions may change or become unnecessary in the future by using devices with more flexible connectivity or employing other logical-operation strategies such as defect-pair braiding~\cite{fowlerSurfaceCodesPractical2012} and twist-free approach~\cite{chamberland2022universal}. Nevertheless, we focus on the above scenario since they are standard assumptions in relevant studies on the runtime analysis~\cite{beverland2022surface,lee_evenmore_2021}, and we leave the other scenario as future work.

\subsection{Surface code and lattice surgery}\label{subsec:surface_code}
In fault-tolerant quantum computing with surface codes, each logical qubit is allocated as a $d \times d$ block on the qubit plane, where $d$ is the distance of surface codes. The error rates of logical qubits decrease exponentially with respect to $d$, as far as the error rates of physical qubits are smaller than the value called threshold. 
The required $d$ is determined from physical error rates of given devices, logical-error suppression rate in error-correcting codes, and required logical error rates. During the computation, we repeat a set of Pauli measurements called stabilizer measurements, according to the generator set of a stabilizer group. A period for all the stabilizer Pauli measurements is called a code cycle. The recovery Pauli operations are estimated with a delay due to classical processing and stored in the unit logic called Pauli frame~\cite{fowler2018low,riesebos2017pauli}.

All the basic logical operations must be achieved fault-tolerantly. The lattice surgery~\cite{horsman2012surface} is known to be the most promising strategy to efficiently achieve multi-qubit logical operations in fault-tolerant quantum computing. In this strategy, the logical Pauli measurements on multiple logical qubits are realized by connecting corresponding blocks with a proper circumference, and all the logical Clifford operations are performed via the logical measurements using lattice surgery and logical $H$ and $S$ gates. 
The time required for the multi-qubit logical Pauli measurements with the lattice surgery is independent of the number of the target logical qubits but linearly scales to the code distance $d$. Thus, it is convenient to define another unit, code beat, as the latency of $d$ code cycles. We need $2$ code beats for logical $S$ and CNOT gates and $3$ code beats for logical $H$ gates. Thanks to the nature of CSS codes, the latency for destructive single-qubit Pauli-$X, Z$ measurements and the preparation of the logical eigenstate of Pauli-$X, Z$ are finished within a single code cycle. Since we can take their latency into account by increasing the code beat, we neglect their latency hereafter.
\black{
Since lattice-surgery operations require ancillary logical-qubit cells that connect target logical qubits, the throughput of lattice-surgery operations depends on topological restrictions such as the allocation of logical qubits and routing of ancillary-qubit paths. See Ref.\,\cite{horsman2012surface,fowler2018low,beverland2022surface} for the detailed rules of the lattice surgery.
}

The non-Clifford gates, such as $T$ and Toffoli gates, cannot be performed with the combinations of the Clifford operations, and are known to take a longer time compared with the other operations. These operations are achieved with the gate-teleportation technique that consumes quantum resources named the magic states. 
For instance, when one desires to operate a $T$-gate through its teleportation, we need a logical qubit prepared in a magic state $\ket{A} = 1/\sqrt{2}(\ket{0} + e^{i\pi/4} \ket{1})$. While the noisy preparation of magic states can be achieved with the magic-state injection protocol, its infidelity is unacceptable for fault-tolerant operations. Therefore, we must create a magic state from a protocol called magic-state distillation, which allows the generation of magic states with arbitrarily small infidelity. 
A standard magic-state distillation protocol is 15-to-1 distillation. This protocol creates a single clean magic state from 15 noisy magic states to reduce the infidelity of magic states from $p$ to $35p^3$, and succeeds with probability $1-15p$~\cite{fowler2018low}. 
The latency of a single trial of magic-state injection requires constant code cycles. Typically, we need two-level distillation for the required logical error rate, and each generation requires $15$ code beats \black{with the construction proposed in Ref.\,\cite{litinski2019game}}. Note that large code distances are not required at the first distillation level since we only need to protect noisy magic states during distillation. Thus, the latency may be smaller. We typically assign a region of the qubit plane just for repeating magic-state preparation, which we call a magic-state factory.
The teleportation of non-Clifford gates requires the feedback of a Clifford operation or the adaptive change of consequent Pauli measurement basis.
Thus, successive gate teleportation will be blocked for a period of error-estimation delay, which is called a reaction time. The reaction time is determined by the throughput of the error-estimation algorithm for surface codes, physical error rates, chosen code distances, and its circuit implementation.
\black{
It should be noted that, unlike Clifford operations, the most time-consuming part of non-Clifford gates is the magic-state preparation. This process can be performed in advance, and the preparation part can always be parallelized by assigning many magic-state factories. This makes basic optimization methods and bottleneck analysis for Clifford and non-Clifford gates different.
}

\subsection{Time estimation without topological restriction}
\black{
The runtime estimation of quantum algorithms is a key part of resource estimation and finding the crossover point of quantum advantage. In this section, we explain methods to estimate the runtime of quantum programs that are written in terms of Clifford+$T$ formulation.
As we detail below, the appropriate runtime estimation heavily relies on the hardware configuration and compilation settings. To estimate actual runtime with reasonable costs, there are two runtime-estimation strategies. One is heuristically estimating the runtime from several critical parameters such as $T$-count and depth, and the other is calculating the runtime with software simulation of fault-tolerant quantum computers.
In the following, we introduce the first strategy in two different situations, the $T$-count limited and reaction limited estimations, as ones of the most commonly employed estimation strategies. We subsequently point out their defects and explain how to overcome them by software simulation while explicitly considering the topological restriction of logical qubits.
}

If quantum circuits consume magic states with a constant rate, we can roughly estimate the execution time by counting the number of $T$-gate. On the other hand, the consumption rate of magic states is not necessarily constant and the execution time may not be linear to the number of $T$-gates.
Fortunately, there is a canonical form proposed by Litinski~\cite{litinski2019game} that ensures an almost constant consumption rate of magic states. In this formalism, we convert the Clifford+$T$ circuits to two stacks of layers; the first one is a sequence of $\pi/8$ Pauli rotations, and the latter is a sequence of Pauli measurements. The number of $\pi/8$ Pauli rotations is equal to the number of $T$-gates $N_T$. The execution of $\pi/8$ Pauli rotation consumes a single magic state, and is achieved with a single code beat and post-processing on the ancillary qubits. 
\black{Based on Litinski's formalism,} the estimation of the execution time with the Clifford+$T$ formalism in a qubit-restricted situation is simple. Assuming that we have a small qubit plane and the number of the magic-state factory is very restricted, we can expect that the latter Pauli measurement part is negligible compared to the first $\pi/8$ rotation parts. Thus, the execution time can be estimated as the $\tau_T N_T$, where $\tau_T$ is the average time for generating a magic state and teleporting it. This is called a $T$-count limited estimation. In the most pessimistic scenario, we can only place a single factory on the qubit plane and the consumption rate of $T$-gate is much more frequent than the generation rate of a magic state factory.
In this case, the execution time of quantum phase estimation can be obtained by multiplying the number of $T$-gates for \select~and \prepare~oracles.

We can also consider the opposite scenario; we have a large qubit plane and can create arbitrary many magic-state factories. In this case, we can effectively have an on-demand magic-state supply. Thus, we can teleport multiple $T$-gates as far as they can be executed in parallel. 
Therefore, we assume that each simultaneous application is achieved with $\tau_{\rm group} = {\rm min}(\tau_{\rm tel}, \tau_{\rm react})$, where $\tau_{\rm tel}$ is the latency of lattice-surgery operations for the magic-state teleportation and $\tau_{\rm react}$ is the reaction time. This quantity is directly related with the execution time as $D_T \tau_{\rm group}$, where $D_T$ is the number of $T$-depth. 
When $\tau_{\rm tel} > \tau_{\rm react}$, we can hide the latency $\tau_{\rm tel}$ by time-optimal construction~\cite{fowler2012time,litinski2019game}. With ancillary qubits, we can convert the iterations of gate teleportation to parallel teleportation and a sequence of reactions. When we can utilize $K$ times larger qubit space, we can force $\tau_{\rm group} \sim D_T \tau_{\rm tel}/K + D_T \tau_{\rm react}$. This situation is called a reaction-limited.
An actual scenario would be the intermediate of the above two, and the actual execution time varies according to the size of the available qubit plane size.

\begin{figure}[tbp]
    \centering
    \includegraphics[width=0.99\linewidth]{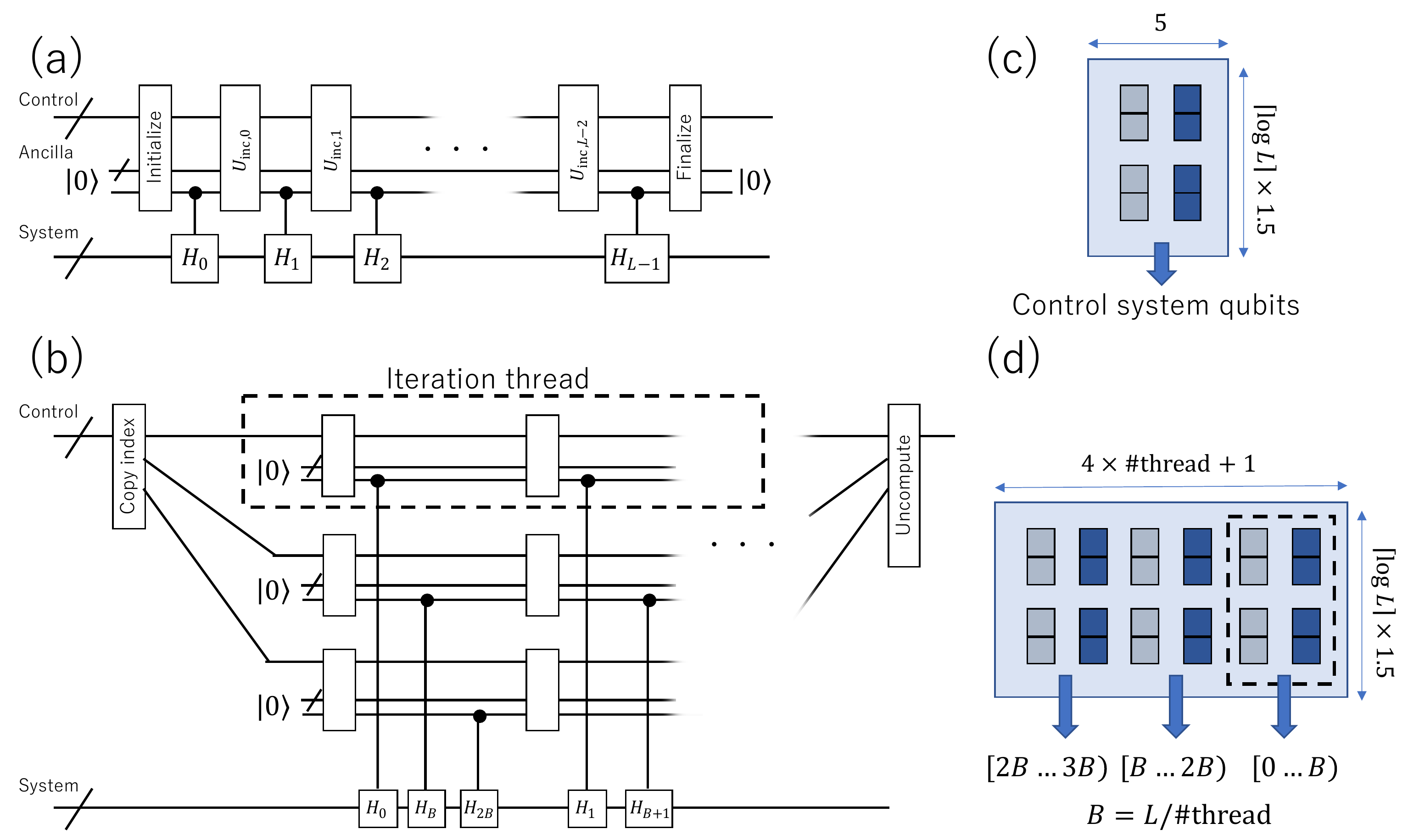}
    \caption{
    (a,b) Quantum circuits for \select~operations with sequential and parallelized constructions.
    (c,d) Qubit placement of \select~operations with sequential and parallelized constructions.
    }\label{fig:parallel_control}
\end{figure}

In the above strategy, the $T$-count and depth of \select~and \prepare~oracles are critical parameters. In the case of the ground-state energy estimation for condensed matter Hamiltonians, the coefficients of Hamiltonians are nearly uniform, and therefore the execution time is expected to be dominated by \select~modules (See Sec.~\ref{subsec:hamiltonian_simulation_posttrotter}). 
\black{
To implement \select~operation with a small number of $T$-gates, we use a method proposed in Ref.\,\cite{babbush_encoding_2018}. The action of \select~operations defined in Eq.\,(\ref{eq:select_def_again}) is divided into the product of small commutative unitary operations $\prod_{l=0}^{L-1} U_l$ where $U_l = \ket{l}\bra{l} \otimes H_l + (I-\ket{l}\bra{l}) \otimes I$.
To construct $U_l$, i.e., a gate controlled by $\lceil \log L \rceil$-qubit input register $\ket{l}$, we use $(\lceil \log L \rceil - 1)$ ancillary qubits and map $\ket{x}\ket{\psi}\ket{0}$ to $\ket{x}\ket{\psi}\ket{c_{x,l}}$, where $c_{x,l} \in \{0,1\}^{(\lceil \log L \rceil - 1)}$ and the last bit of $c_{x,l}$ is $1$ if $x=l$ and 0 if $x\neq l$. Then, we can achieve an action $\ket{x}\ket{\psi}\ket{c_{x,l}} \mapsto (U_l \ket{x}\ket{\psi})\ket{c_{x,l}} = \ket{x}(H_l'\ket{\psi})\ket{c_{x,l}}$ where $H_l' = H_l$ if $l=x$ and $H_l = I$ otherwise, using a Clifford operation $\ket{0}\bra{0}\otimes I + \ket{1}\bra{1} \otimes H_l$ where the control qubit corresponds to the last qubit of $\ket{c_{x,l}}$. 
To discuss the placement of logical qubits in the next subsection, we denote the register $\ket{x}$ and $\ket{c_{x,l}}$ as {\it control qubits}, and $\ket{\psi}$ as {\it system qubits}. Also, we denote the last qubit of $\ket{c_{x,l}}$ as {\it tail control qubit} since among the control qubits, only the tail control qubit interacts with system qubits in this iteration. 
}
\black{
The quantum circuit based on this iteration is shown in Fig.\,\ref{fig:parallel_control}~(a). At the first iteration, we construct $\ket{x}\ket{0} \mapsto \ket{x}\ket{c_{x,0}}$, of which the action is shown as {\tt Initialize}. In the $i$-th iteration, we need to update the state as $\ket{x}\ket{c_{x,i}} \mapsto \ket{x}\ket{c_{x,i+1}}$, which is achieved by a unitary operation named $U_{{\rm inc}, i}$. Finally, the ancillary qubits are disentangled as $\ket{x}\ket{c_{x,L-1}} \mapsto \ket{x}\ket{0}$ with {\tt Finalize}. This method enables the implementation of \select~operations with $T$-count of $4L-4$.
}

The above construction of \select~operation is a serial sequence of Toffoli gates for {\tt Initialize}, $U_{{\rm inc}, i}$, and {\tt Finalize}. Thus, its $T$-depth is almost equal to $T$-count. This means there is no room to reduce $T$-depth even when several magic-state factories are available. To enable a flexible performance tuning of \select~operations according to available magic-state supply rates, we propose a technique to reduce the $T$-depth with a similar idea of loop-parallelization in classical computing (See Fig.~\ref{fig:parallel_control}~(b)).
Suppose that the target Hamiltonian consists of $L$ terms of Pauli operators acting on $n$ qubits. In this technique, we divide the task of \select~operations to $b$ parallelized executions using $b$ times the number of control qubits as follows. First, we clone the input control registers to GHZ states $\sum_x \alpha_x \ket{x} \mapsto \sum_x \alpha_x \ket{x}^{\otimes b}$, which is denoted as {\tt CopyIndex} operation, and we call each cloned control register as a {\it thread}.
Then, we divide the $L$ terms to $b$ blocks, where the $i$-th block contains terms indexed from $(i L/b)$ to $((i+1)L/b-1)$, and \select~operations is performed as $b$ independent iterations by $b$ threads. 
Finally, we uncompute the cloned registers. This construction effectively reduces the $T$-depth from $O(L)$ to $O(L/b + \log L)$ with an additional $O(b \log n)$ qubits. The brief construction of this parallelization is shown in Fig.\,\ref{fig:parallel_control}~(b).
Note that as far as we keep $b$ sufficiently small than $n$, the overhead of $O(b \log n)$ ancillary qubits would be negligible compared to the system size $n$. The detailed decomposition and parallelization efficiency will be presented in another paper.

\black{
Although the $T$-count limited or reaction limited estimation based on the above counting provides a rough estimation of runtime, it should be noted that they ignore the overhead by topological restriction of the qubit plane. Since each logical qubit is placed at a certain position on the qubit plane and we need to connect the target blocks for lattice surgery, we sometimes cannot execute operations due to the conflict of their paths. This means $T$-gates are not necessarily teleported in parallel even if they are located in parallel on the quantum circuit representation. To take them into account, we need to consider the allocation of logical qubits and their routing for lattice-surgery operations, which is discussed in the next section.
}

\subsection{Placement of logical qubits and routing for lattice surgery}
\begin{figure}[tbp]
    \centering
    \includegraphics[width=0.99\linewidth]{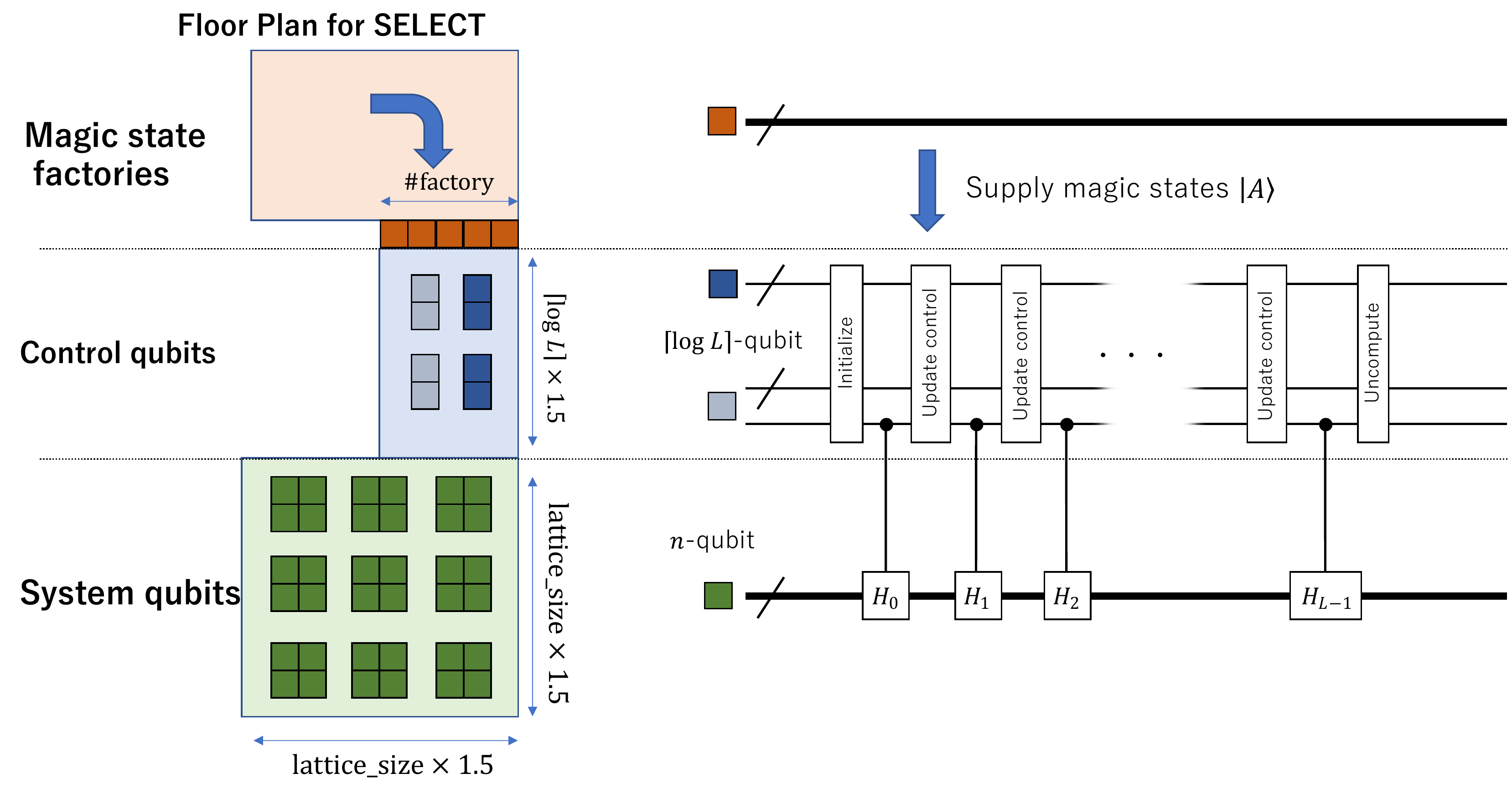}
    \caption{
    The floor plan of the qubit plane for the sequential construction of the \select~operation.
    }\label{fig:floor_plan}
\end{figure}

Once quantum circuits are translated to the sequence of intrinsic lattice-surgery operations, we need to place each logical qubit to a certain block on the qubit plane. While we should choose dense allocations to avoid the redundant usage of the qubit plane, too dense allocations may lead to path conflicts and the degradation of instruction throughput.
In this paper, we heuristically designed the placement of logical qubits for condensed matter Hamiltonians for 2d spins by leveraging the feature of spatially local interactions.
Our logical qubits mapping and correspondence to \select~circuits is illustrated in Fig.~\ref{fig:floor_plan}.

\black{
In our design, the qubit plane is divided into three parts; factory, control, and system. We assume that the design of magic state factories proposed by Litinski~\cite{litinski2019game} is used and they are simply located in the factory parts. So, we explain the detailed logical-qubit placement of the system and control parts.
}
\black{
The system part is used for locating system qubits in \select~operation. For the 2d Heisenberg models, the position of system qubits representing a spin at a certain coordinate is placed at the same coordinate of the system part to leverage the locality. For the 2d Fermi-Hubbard models, we heuristically determined the positions of system qubits according to those of fermions.
Since Clifford operations and lattice surgery need ancillary space around the logical-qubit cells, we created $O(\sqrt{n})$ corridors of ancillary space so that the vertical and horizontal edges of each logical-qubit cell touch them. 
Note that this is one of the most compact allocations while keeping Clifford operations available~\cite{chamberland2022universal,fowler2018low}. We use this compact allocation for the system part because of the observation that system qubits are less frequently accessed compared to the control part, and the compact allocation of the system part reduces the required number of physical qubits without losing the parallelism of \select~operations.
}

\black{
The control part is used for locating control qubits in \select~operations, i.e., input register $\ket{x}$ and ancillary register $\ket{c_{x,l}}$. Since the operations on this part are dominated by Toffoli-gates acting on the neighboring bits in the registers, we placed logical qubits for them as arrays in the order of head bits to tail bits. An example allocation of the control part is shown in Fig.\,\ref{fig:parallel_control}(c), where an array of dark cells corresponds to the input register and that of light cells to the ancillary register.
}

\black{
After we determine the placement for each part, we need to map the location of the three parts on the qubit plane. As shown in Fig.\,\ref{fig:floor_plan}, we allocate three parts in the order of factory, control, and system parts. The factory part is placed behind the control part since control qubits consume magic states while system qubits do not. Also, since only the tail control qubits in the control part interact with the system part, we let the system part touch the side of the control part where the tail control qubits are located. 
}

When we utilize the parallelization to the control part, we modify the qubit allocation of the control part as shown in Fig.\,\ref{fig:parallel_control}(d). 
If the Pauli terms are randomly assigned to each thread, tail control qubits of the threads try to touch system qubits at random positions, which induces the conflicts of Clifford operations and may become a bottleneck. This problem can be resolved by re-ordering the Pauli terms and appropriate block allocations. Since we now consider nearly uniform coefficients, the cost of \prepare~oracle is almost independent of the ordering of coefficients. Thus, we can change the order of Pauli terms to minimize the execution time for \select~operations. 
Suppose the list of sorted indices in the $i$-th Pauli operator is $(j_1, ... j_k)$. Then, we sort the Pauli terms with the set in ascending order, separate them into $b$ blocks, and assign them to each thread. 
When the thread number $b$ satisfies $\sqrt{n} > b$, $\sqrt{n}/b + k$ corridors are relevant to each thread. With this assignment, there are no potential conflicts as far as $k < \sqrt{n}/b$. Thus, we can expect this assignment will prevent the potential conflict of lattice-surgery paths.

\black{
Finally, we schedule the order of instructions and determine ancillary cells for logical operations. We translate Clifford+$T$ formalism into a 1D sequence of $S$, $H$, magic-state generation, one-body Pauli measurement, and two-body lattice-surgery operations. Each instruction becomes ready for execution when all the instructions that act on the target logical qubits earlier than the target instruction. Several instructions, such as feedback operations of $T$-gate teleportation, need to wait for relevant measurement outcomes to be error-corrected, i.e., there is a latency of reaction time. For logical $S$ and $H$ gates, logical qubits consume one of the neighboring and available ancillary cells. If there is no available ancillary cell, instructions wait for the next code beat. For lattice-surgery instruction, we find one of the shortest paths between target logical qubits with the breadth-first search algorithm and consume the ancillary cells on the path. If there is no path between logical qubits, instructions wait for the next code beat. 
}

\subsection{Parameter estimation and numerical evaluation}
Here, we estimate the relevant parameters according to the recent device technologies. We define $d$, $L$, and $m$ as a code distance, the number of terms in the target Hamiltonian, and the required digit of the ground energy, respectively. They are left as variables since they depend on the problem size. While a shorter code cycle is preferable to shorten the execution time, it is lower-bounded by the time for stabilizer measurements and error estimation algorithms. Currently, the latter is the dominant factor and is expected to be no shorter than $1~{\rm \mu s}$~\cite{holmes2020nisq+}. Thus, the code beat, which is defined by $d$ code cycles, is $d~{\rm \mu s}$. The time for two-level magic-state distillation $\tau_T$ is $15d~{\rm \mu s}$. The reaction time is typically assumed to be $10~{\rm \mu s}$. The $T$-count for the \select~operation is $4L$ and the $T$-depth is $4L/b$, where $b$ is a parallelization level. 

Summarizing the above, we can estimate the total execution time using several variables depending on the problem. In the case of $T$-count limited scenario, the execution time for \select~operation is about $15d~{\rm [\mu s]} \times 4L = 60dL~{\rm [\mu s]}$. In the case of $T$-depth limited case, the execution time for \select~operation is $10~{\rm [\mu s]} \times 4L/b = 40L/b~{\rm [\mu s]}$, which is $bd$ times smaller than the $T$-count limited case. Assuming the other parts, such as \prepare~, initial-state preparation, and quantum Fourier transform, are negligible compared to the \select, the whole execution time can be estimated by multiplying $2^m$ (or number of repetitions determined as $r=\pi \lambda/2\epsilon$ as discussed in Sec.~\ref{subsec:hamiltonian_simulation_posttrotter}) to that for \select. 

\begin{figure}[tbp]
    \centering
    \includegraphics[width=0.99\linewidth]{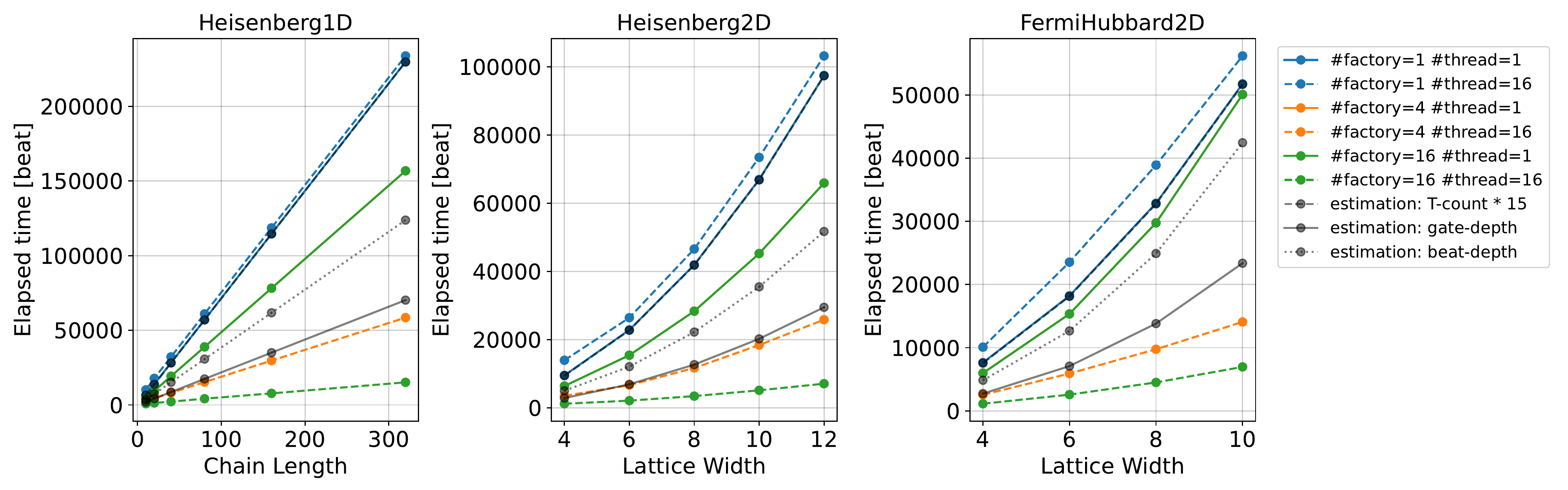}
    \caption{
    The required code beats for executing \select~operations for several Hamlitonian models.
    }\label{fig:select_latency}
\end{figure}

Finally, we verified our analysis with numerical calculation. We have synthesized the quantum circuits of \select~operations with and without parallelization, allocated logical qubits as shown in the floor plan, determined the routing of lattice surgery with depth-first search algorithms, and calculated the required code beat for \select~circuits with simulation. The observed coded beats to finish \select~operations are plotted as a function of problem sizes and the type of Hamiltonians in Fig.\ref{fig:select_latency}. 
In the numerical simulation, we used the following approximations to simplify the evaluation:
\begin{itemize}
    \item We assumed each magic-state factory generates a single magic state per 15 code beat. Note that this value may be improved by using more efficient magic-state distillation strategy, such as the construction proposed in Ref.\,\cite{fowler2018low,litinski2019game}.
    \item When Clifford operations are conditioned by previous measurement outcomes, the operations are error-corrected and become available after 1 code beat.
    \item While logical $S$ gates are required after $T$-gate teleportation with only probability of 0.5, we pessimistically assume that the logical $S$ gate is always required.
    \item Due to the implementation of \prepare, $L$ Hamiltonian terms are not necessarily assigned to the computational basis from $\ket{0}$ to $\ket{L-1}$ in the control qubits. Nevertheless, we assumed it is enough to iterate from $0$-th to $L-1$-th for simplicity. The conciliation between the cost of \select~and \prepare~is future work. 
\end{itemize}

According to the numerical results, when we do not perform parallelization and only have a single factory~(solid blue line), the runtime is close to the product of the rate of magic-state generation and $T$-count~(broken black line). When the number of factories increases to $4$, the runtime saturates at about half of $T$-count limited estimation~(solid green line). Compared to the depth of circuit~(black dotted line), it is about $1.2$ times larger. We expect that this discrepancy comes from the latency of reactions or path conflicts. When we employ parallelization, the runtime can be reduced to a smaller value while it requires a larger number of qubits and factories. With four threads and 16 factories, the runtime can be suppressed to $0.1$ times smaller values. Thus, we can conclude that when there are several magic-state factories, the parallelization is effective in further reducing the computational time with a small overhead of required logical qubits. The required code beats for all the configurations are summarized in Table.~\ref{table:select_time}.

\begin{table}[h]
 \centering
 \caption{
 The calculated runtimes for a single \select~operations for several Hamiltonian models.
 The execution times are shown with the unit of code beats, and $n_F$ means the number of magic-state factories.
 The runtime in terms of expected walltime is shown in Table~\ref{table:physical_qubit_count}.
 }\label{table:select_time}
 \begin{tabular}{l|c|c||c|c|c}
  Hamiltonian & lattice size & \#thread &
  \begin{tabular}{c}
         runtime~[beat]\\
             ($n_F=1$)
         \end{tabular}
  & 
  \begin{tabular}{c}
         runtime~[beat]\\
             ($n_F=4$)
         \end{tabular}
  & 
    \begin{tabular}{c}
         runtime~[beat]\\
             ($n_F=16$)
         \end{tabular}
  \\
  \hline
  \hline
  2d $J_1$-$J_2$ Heisenberg & $4\times 4$   & 1  & 9511 & 6389 & 6374 \\
  2d $J_1$-$J_2$ Heisenberg & $4\times 4$   & 16 & 13939 & 3514 & 1185 \\
  2d $J_1$-$J_2$ Heisenberg & $6\times 6$   & 1  & 22840 & 15433 & 15412 \\
  2d $J_1$-$J_2$ Heisenberg & $6\times 6$   & 16 & 26419 & 6634 & 2091\\
  2d $J_1$-$J_2$ Heisenberg & $8\times 8$   & 1  & 41929 & 28348 & 28327 \\
  2d $J_1$-$J_2$ Heisenberg & $8\times 8$   & 16 & 46583 & 11678 & 3424 \\
  2d $J_1$-$J_2$ Heisenberg & $10\times 10$ & 1  & 66958 & 45261 & 45239 \\
  2d $J_1$-$J_2$ Heisenberg & $10\times 10$ & 16 & 73465 & 18400 & 5127 \\
  2d $J_1$-$J_2$ Heisenberg & $12\times 12$ & 1  & 97498 & 65939 & 65917 \\
  2d $J_1$-$J_2$ Heisenberg & $12\times 12$ & 16 & 103225 & 25840 & 7065 \\
  \hline
  spin-1 Heisenberg chain & $10$  & 1  & 6559 & 4441 & 4407 \\
  spin-1 Heisenberg chain & $10$  & 16 & 10093 & 2547 & 780 \\
  spin-1 Heisenberg chain & $20$  & 1  & 13828 & 9391 & 9355 \\
  spin-1 Heisenberg chain & $20$  & 16 & 17779 & 4472 & 1308 \\
  spin-1 Heisenberg chain & $40$  & 1  & 28237 & 19227 & 19185 \\
  spin-1 Heisenberg chain & $40$  & 16 & 32179 & 8072 & 2172 \\
  spin-1 Heisenberg chain & $80$  & 1  & 57046 & 38887 & 38845 \\
  spin-1 Heisenberg chain & $80$  & 16 & 60983 & 15278 & 4157 \\
  spin-1 Heisenberg chain & $160$ & 1  & 114655 & 78208 & 78165 \\
  spin-1 Heisenberg chain & $160$ & 16 & 118585 & 29680 & 7728 \\
  spin-1 Heisenberg chain & $320$ & 1  & 229864 & 156848 & 156805 \\
  spin-1 Heisenberg chain & $320$ & 16 & 233789 & 58484 & 15116 \\
  \hline
  2d Fermi-Hubbard & $4\times 4$   & 1  & 7637 & 5981 & 5969 \\
  2d Fermi-Hubbard & $4\times 4$   & 16 & 10093 & 2549 & 1134 \\
  2d Fermi-Hubbard & $6\times 6$   & 1  & 18215 & 15362 & 15341 \\
  2d Fermi-Hubbard & $6\times 6$   & 16 & 23539 & 5914 & 2579 \\
  2d Fermi-Hubbard & $8\times 8$   & 1  & 32864 & 29772 & 29751 \\
  2d Fermi-Hubbard & $8\times 8$   & 16 & 38903 & 9759 & 4507 \\
  2d Fermi-Hubbard & $10\times 10$ & 1  & 51764 & 50097 & 50076 \\
  2d Fermi-Hubbard & $10\times 10$ & 16 & 56183 & 14079 & 6966 \\
 \end{tabular}
\end{table}

\section{Estimation of the number of physical qubits, code distances, and actual runtime}\label{sec:total_resource}
\black{Here, we explain how to estimate required number of physical qubits $N_{\rm ph}$ via determining the code distance $d$ of the surface code. In Sec.~\ref{subsec:rough_nphys}, we provide rough estimation on the physical qubit count to compare the space complexity between quantum algorithms used for the phase estimation. In Sec.~\ref{subsec:detailed_runtime}, we perform estimation that explicitly takes the mapping and connectivity of logical qubits into account,
so that we can combine all the results and estimate the actual runtime required for the qubitization-based phase estimation algorithm.
}

\subsection{Rough comparison between quantum algorithms} \label{subsec:rough_nphys}
\black{
First we describe how to obtain a rough estimate on the physical qubit count which yields Fig.~5 in the main text.
We choose the code distance so that the logical error rate $p_{\rm log}$ is smaller than the number of logical operations $N_{\rm op}$:
\begin{eqnarray}
    p_{\rm log} < 1/N_{\rm op}, \label{eq:logical_error}
\end{eqnarray}
where we estimate $p_{\rm log}=0.1(p/p_{\rm th})^{(d+1)/2}$ with the ratio of physical error rate over error threshold $p/p_{\rm th} = 0.1$. Note that this choice has been commonly employed in existing runtime estimation literature such as Refs.~\cite{litinski2019game, lee_evenmore_2021, goings_pnas_2022}.
In order to obtain rough estimate on  the number of logical operations, we focus solely on the $T$ gates. Based on the $T$-count that was also presented as Table~I. in the main text, we calculate  $N_{\rm op}$ as follows,
\begin{eqnarray}
N_{\rm op} = N_{\rm log} \times (T\text{-count}) \times (\text{code beat per $T$-gate}),
\end{eqnarray}
where $N_{\rm log}$ is the number of logical qubits used for the computation, i.e., the number of system qubits and the ancilla. Note that in the next section we perform more detailed analysis with a slightly different definition of $N_{\rm ph}$. The code beat for each $T$-gate is assumed to be 15~\cite{litinski2019game}.
By solving Eq.~\eqref{eq:logical_error}, we can determine the code distance $d$. Assuming that all the logical qubits are encoded with the framework of surface code with homogeneous code distance,  the number of physical qubits can be obtained as
\begin{eqnarray}
    N_{\rm ph} = 2d^2 N_{\rm log}.
\end{eqnarray}
}

\black{
As shown in Table~\ref{tab:phys_qubit_comparison}, although the post-Trotter methods requires ancilla qubits and hence consume larger $N_{\rm log}$, the significant reduction in the $T$-count in the qubitization algorithm results in fewer $N_{\rm ph}$ compared to Trotter-based methods. See also Fig. 5 in the main text for visualization of the spacetime cost for the lattice size of $10\times 10$.
}

\begin{table}[]
\begin{tabular}{|c|c|c|c|c|c|}\hline 
\ & \multicolumn{5}{c|}{($N_{\rm log}$, $d$, $N_{\rm ph}$) for 2d $J_1$-$J_2$ Heisenberg}                                                                           \\ \hline
System size & qDRIFT & random Trotter & Taylorization & \begin{tabular}{c} Qubitization\\(sequential)\end{tabular} & \begin{tabular}{c} Qubitization\\(product)\end{tabular} \\ \hline
$4\times4$ & 16, 29, 2.69e+04 & 16, 23, {\bf 1.69e+04} & 96, 25, 1.20e+05 & 24, 21, 2.12e+04 & 38, 21, 3.35e+04 \\ 
$6\times6$ & 36, 31, 6.92e+04 & 36, 25, {\bf 4.50e+04} & 146, 27, 2.13e+05 & 46, 23, 4.87e+04 & 64, 23, 6.77e+04 \\ 
$8\times8$ & 64, 33, 1.39e+05 & 64, 27, 9.33e+04 & 174, 27, 2.54e+05 & 74, 23, {\bf 7.83e+04} & 92, 23, 9.73e+04 \\ 
$10\times10$ & 100, 35, 2.45e+05 & 100, 29, 1.68e+05 & 232, 29, 3.90e+05 & 112, 25, {\bf 1.40e+05} & 134, 25, 1.68e+05 \\ 
$12\times12$ & 144, 35, 3.53e+05 & 144, 29, 2.42e+05 & 288, 29, 4.84e+05 & 156, 25, {\bf 1.95e+05} & 178, 25, 2.22e+05 \\ 
$14\times14$ & 196, 37, 5.37e+05 & 196, 31, 3.77e+05 & 340, 31, 6.53e+05 & 208, 27, {\bf 3.03e+05} & 230, 27, 3.35e+05 \\ 
$16\times16$ & 256, 37, 7.01e+05 & 256, 31, 4.92e+05 & 400, 31, 7.69e+05 & 268, 27, {\bf 3.91e+05} & 290, 27, 4.23e+05 \\ 
$20\times20$ & 400, 39, 1.22e+06 & 400, 31, 7.69e+05 & 568, 31, 1.09e+06 & 414, 29, {\bf 6.96e+05} & 440, 29, 7.40e+05 \\ 
\hline \hline
                              & \multicolumn{5}{c|}{($N_{\rm log}$, $d$, $N_{\rm ph}$) for 2d Fermi-Hubbard} \\ \hline
System size & qDRIFT & random Trotter & Taylorization & \begin{tabular}{c} Qubitization\\(sequential)\end{tabular} & \begin{tabular}{c} Qubitization\\(product)\end{tabular} \\ \hline                              
$4\times4$ & 32, 29, 5.38e+04 & 32, 25, 4.00e+04 & 131, 25, 1.64e+05 & 41, 21, {\bf 3.62e+04} & 49, 21, 4.32e+04 \\ 
$6\times6$ & 72, 31, 1.38e+05 & 72, 27, 1.05e+05 & 193, 27, 2.81e+05 & 83, 23, {\bf 8.78e+04} & 95, 23, 1.01e+05 \\ 
$8\times8$ & 128, 33, 2.79e+05 & 128, 29, 2.15e+05 & 249, 27, 3.63e+05 & 139, 25, 1.74e+05 & 151, 23, {\bf 1.60e+05} \\ 
$10\times10$ & 200, 35, 4.90e+05 & 200, 31, 3.84e+05 & 356, 29, 5.99e+05 & 213, 25, {\bf 2.66e+05} & 229, 25, 2.86e+05 \\ 
$12\times12$ & 288, 35, 7.06e+05 & 288, 31, 5.54e+05 & 444, 29, 7.47e+05 & 301, 27, 4.39e+05 & 317, 25, {\bf 3.96e+05} \\ 
$14\times14$ & 392, 37, 1.07e+06 & 392, 31, 7.53e+05 & 548, 31, 1.05e+06 & 405, 27, {\bf 5.90e+05} & 421, 27, 6.14e+05 \\ 
$16\times16$ & 512, 37, 1.40e+06 & 512, 33, 1.12e+06 & 668, 31, 1.28e+06 & 525, 27, {\bf 7.65e+05} & 541, 27, 7.89e+05 \\ 
$20\times20$ & 800, 39, 2.43e+06 & 800, 33, 1.74e+06 & 980, 33, 2.13e+06 & 815, 29, {\bf 1.37e+06} & 835, 29, 1.40e+06                                                                                                                                                                                         \\ \hline

\end{tabular}
\caption{\black{
Estimation on quantum resource to execute the phase estimation algorithm under various quantum subroutines. Here we show $N_{\rm log}$, $d$, and $N_{\rm ph}$, which indicate the logical qubit count, code distance for the surface code, and the physical qubit count given by $N_{\rm ph} = 2d^2 N_{\rm log}$. Note that, as opposed to the estimation in Sec.~\ref{subsec:detailed_runtime}, here we exclude the logical qubits used for the magic state factory, each of which consumes 176 logical qubits in assuming standard level-2 distillation. }
}\label{tab:phys_qubit_comparison}
\end{table}

\subsection{Detailed estimation with actual runtime}\label{subsec:detailed_runtime}

Next we proceed to describe how to perform more detailed estimation on $N_{\rm ph}$ together with the actual runtime of the algorithm.
We assume that the qubits are allocated as depicted in the floor plan of Fig.~\ref{fig:floor_plan} which consists of system qubits, control qubits, and magic state factory.
By assuming that all logical qubits are encoded by framework of surface code with homogeneous code distance of $d$, the number of physical qubits  $N_{\rm ph}$ is provided as
\begin{eqnarray}
N_{\rm ph} &=& \left(1.5^2 N_S + 1.5(4 b+1)\log L  + n_F A_F\right) \times (2d^2),
\end{eqnarray}
where $N_S$ is number of system qubits, $b$ is the number of parallelization threads, $n_F$ is the number of magic state factory, and $A_F$ is the floor area per single magic state factory.
Here, we follow the construction of Ref.~\cite{litinski2019game} and employ $A_F = 176$. Meanwhile, as was also noted in Sec.~\ref{subsec:surface_code}, we may reduce the code distance of logical qubits that are used for the first-level magic state distillation (e.g. two-fold), which results in significant reduction of physical qubits especially when number of magic state factories $n_F$ is large.

\black{In order to determine the code distance, now we}
estimate from the repetition count $r$ of qubitization oracle by neglecting the contribution from the subdominant \prepare~oracle as
\begin{eqnarray}
N_{\rm op} = (\text{\#logical qubits involved in \select}) \times (\text{\#code cycle per \select}) \times r,
\end{eqnarray}
\black{where $r$ is related with the total energy error $\epsilon$ via $\delta_{\rm PEA}= 0.9\epsilon /\lambda$ as $r = \lceil \pi/ (2\delta_{\rm PEA})\rceil$.}
By solving Eq.~\eqref{eq:logical_error} using these quantities, we find that it is sufficient to take $d$ around 21 or 25 around the quantum-classical crosspoint.

The total runtime can be estimated as 
\begin{eqnarray}
t \sim t_{\rm beat} \times N_{\rm beat} \times r,\label{eq:runtime_total}
\end{eqnarray}
where $t_{\rm beat} = t_{\rm cycle} \times d$ is the required time per code beat that is determined from the time per code cycle of surface code $t_{\rm cycle}=1\mu s$, and $N_{\rm beat}$ is the number of code beat per \select~oracle obtained in Sec.~\ref{sec:distselect_analysis}.
The overall runtime estimation results and quantum resource are summarized in Table~\ref{table:physical_qubit_count}, which is displayed in comparison with classical algorithms in Fig.~\ref{fig:crossover} in the main text. Here, we clearly observe that quantum advantage is achieved in runtime of hours, with use of physical qubits of $O(10^5)$.

Our results imply that code distance 21 is the minimum requirement to demonstrate the quantum advantage under the assumption of $p/p_{\rm th}=0.1$ and target of $\epsilon=0.01$~(See Fig.~7 in the main text for discussion with various hardware/algorithmic requirements.). In other words, we need a classical control unit of FTQC that can estimate/correct (non-logical) errors of $d=21$ logical qubits with $10^{-3}$ physical error rate and $1~{\rm \mu s}$ code cycle. However, according to Ref.\,\cite{ueno2022qulatis}, the current state-of-the-art implementation of lattice-surgery-compatible error decoders allows a code distance up to 11 with these assumptions. Therefore, achieving physical error rates ten-times smaller than the code threshold and increasing the number of qubits are not enough to demonstrate quantum computational advantage in the field of condensed matter physics. 

We expect several possible directions to overcome this difficulty.
First, as we have discussed in Fig.~7 in the main text, achieving smaller physical error rates can reduce the required code distance and number of physical qubits. The execution times are also reduced since it is proportional to code distances. 
Second strategy is to improve encoding strategies and error-estimation algorithms for quantum error correction, which effectively increase the code threshold or allowed code distances. 
Third, increasing physical qubits allows more allocations of magic-state factories, which enables further parallelization of \select~modules and reduction of the required code distances. 
Fourth, there is room for improving the workspace utilization for lattice surgery. We can use prepared logical Bell pairs~\cite{beverland2022surface} for more efficient scheduling, re-use dirty logical qubits~\cite{low2018trading}, or find more suitable Hamiltonian models to demonstrate advantage.
Finally, we can utilize longer code cycles to allow larger code distances. However, it should be noted that this approach also increases the runtime, and more resources are demanded to demonstrate an advantage. A possible direction is considering heterogeneous architecture, i.e., using superconducting qubits for the frequent-access region and ion qubits for low-access regions, but this requires careful consideration of the communication bandwidth between different qubit species.
We expect that one of these approaches cannot solely resolve the problems, and the co-design of the application, algorithm, system, and hardware is demanded in future. Our resource estimation is the first step towards such advanced designs.

\begin{table}[]
\begin{tabular}{l|c|c|c|c||c|c|}
Hamiltonian               & lattice size  & $(n_F, \#thread)$ & code distance $d$ & $N_{\rm ph}$ & Repetition count $r$ & Runtime {[}sec{]} \\ \hline \hline
2d $J_1$-$J_2$ Heisenberg & 4$\times$4 & (1, 1) & 19 & 1.96e+05 & 5.24e+03 & 9.46e+02\\
2d $J_1$-$J_2$ Heisenberg & 4$\times$4 & (16, 16) & 19 & 2.62e+06 & 5.24e+03 & 1.18e+02\\
2d $J_1$-$J_2$ Heisenberg & 6$\times$6 & (1, 1) & 21 & 2.86e+05 & 1.26e+04 & 6.03e+03\\
2d $J_1$-$J_2$ Heisenberg & 6$\times$6 & (16, 16) & 21 & 3.33e+06 & 1.26e+04 & 5.52e+02\\
2d $J_1$-$J_2$ Heisenberg & 8$\times$8 & (1, 1) & 23 & 4.18e+05 & 2.30e+04 & 2.22e+04\\
2d $J_1$-$J_2$ Heisenberg & 8$\times$8 & (16, 16) & 21 & 3.47e+06 & 2.30e+04 & 1.66e+03\\
2d $J_1$-$J_2$ Heisenberg & 10$\times$10 & (1, 1) & 23 & 5.12e+05 & 3.67e+04 & 5.64e+04\\
2d $J_1$-$J_2$ Heisenberg & 10$\times$10 & (16, 16) & 23 & 4.35e+06 & 3.67e+04 & 4.32e+03\\
2d $J_1$-$J_2$ Heisenberg & 12$\times$12 & (1, 1) & 25 & 7.28e+05 & 5.34e+04 & 1.30e+05\\
2d $J_1$-$J_2$ Heisenberg & 12$\times$12 & (16, 16) & 23 & 4.46e+06 & 5.34e+04 & 8.68e+03\\
\hline
spin-1 Heisenberg chain & 10 & (1, 1) & 19 & 4.68e+05 & 5.45e+03 & 6.79e+02\\
spin-1 Heisenberg chain & 10 & (16, 16) & 17 & 2.61e+07 & 5.45e+03 & 7.22e+01\\
spin-1 Heisenberg chain & 20 & (1, 1) & 21 & 9.49e+05 & 1.07e+04 & 3.10e+03\\
spin-1 Heisenberg chain & 20 & (16, 16) & 19 & 6.27e+07 & 1.07e+04 & 2.65e+02\\
spin-1 Heisenberg chain & 40 & (1, 1) & 23 & 2.04e+06 & 2.12e+04 & 1.37e+04\\
spin-1 Heisenberg chain & 40 & (16, 16) & 19 & 1.23e+08 & 2.12e+04 & 8.73e+02\\
spin-1 Heisenberg chain & 80 & (1, 1) & 23 & 3.82e+06 & 4.21e+04 & 5.52e+04\\
spin-1 Heisenberg chain & 80 & (16, 16) & 21 & 2.97e+08 & 4.21e+04 & 3.68e+03\\
spin-1 Heisenberg chain & 160 & (1, 1) & 25 & 8.72e+06 & 8.40e+04 & 2.41e+05\\
spin-1 Heisenberg chain & 160 & (16, 16) & 23 & 7.08e+08 & 8.40e+04 & 1.49e+04\\
spin-1 Heisenberg chain & 320 & (1, 1) & 27 & 2.00e+07 & 1.68e+05 & 1.04e+06\\
spin-1 Heisenberg chain & 320 & (16, 16) & 25 & 1.67e+09 & 1.68e+05 & 6.34e+04\\
\hline
2d Fermi-Hubbard & 4$\times$4 & (1, 1) & 21 & 2.72e+05 & 1.26e+04 & 2.02e+03\\
2d Fermi-Hubbard & 4$\times$4 & (16, 16) & 21 & 3.24e+06 & 1.26e+04 & 2.99e+02\\
2d Fermi-Hubbard & 6$\times$6 & (1, 1) & 23 & 4.29e+05 & 2.93e+04 & 1.23e+04\\
2d Fermi-Hubbard & 6$\times$6 & (16, 16) & 21 & 3.40e+06 & 2.93e+04 & 1.59e+03\\
2d Fermi-Hubbard & 8$\times$8 & (1, 1) & 23 & 5.70e+05 & 5.31e+04 & 4.01e+04\\
2d Fermi-Hubbard & 8$\times$8 & (16, 16) & 23 & 4.32e+06 & 5.31e+04 & 5.50e+03\\
2d Fermi-Hubbard & 10$\times$10 & (1, 1) & 25 & 8.86e+05 & 8.38e+04 & 1.05e+05\\
2d Fermi-Hubbard & 10$\times$10 & (16, 16) & 25 & 5.42e+06 & 8.38e+04 & 1.46e+04\\
\end{tabular}
 \caption{
 The estimated number of physical qubits, required code distance, and actual runtime for our target Hamiltonians.
 \black{Here, the runtime is based on the $T$-depth required to estimate the ground state energy with target precision $\epsilon=0.01$.}
 }\label{table:physical_qubit_count}
\end{table}

\end{document}